\documentstyle[12pt,epsfig]{article}

\newcommand{\lsim}{\raisebox{-0.13cm}{~\shortstack{$<$ \\[-0.07cm] $\sim$}}~}
\newcommand{\gsim}{\raisebox{-0.13cm}{~\shortstack{$>$ \\[-0.07cm] $\sim$}}~}

\newcommand{\ra}{\rightarrow}
\newcommand{\ee}{e^+e^-}
\newcommand{\s}{\smallskip}
\newcommand{\nn}{\noindent}
\newcommand{\non}{\nonumber}
\newcommand{\beq}{\begin{eqnarray}}
\newcommand{\eeq}{\end{eqnarray}}
\newcommand{\tb}{\tan\beta}

\newcommand{\mchi}{\mbox{$m_{\tilde\chi_1^0}$}}
\newcommand{\lsp}{\mbox{$\tilde\chi_1^0$}}
\newcommand{\om}{\mbox{$\Omega_{\tilde\chi_1^0} h^2$}}

\begin{document}
\baselineskip=16pt

\thispagestyle{empty}
\begin{flushright}
PM--01--27\\
TUM--HEP--421/01 \\
July 2001
\end{flushright}

\vspace{1cm}

\begin{center}

{\large\sc {\bf Constraints on the Minimal Supergravity Model}}

\vspace*{3mm}

{\large\sc {\bf and Prospects for SUSY Particle Production}}

\vspace*{3mm}

{\large\sc {\bf at Future e$^+$e$^-$  Linear Colliders}} 

\vspace{1cm}

{\sc A. Djouadi$^1$, M. Drees$^2$} and {\sc J.L. Kneur$^1$}

\vspace*{0.5cm}

$^1${\it Laboratoire de Physique Math\'ematique et Th\'eorique, UMR5825--CNRS,
\\ Universit\'e de Montpellier II, F--34095 Montpellier Cedex 5, France.}

\vspace*{3mm}

$^2${\it Physik Department, Technische Universit\"at M\"unchen, \\ 
James Franck Strasse, D--85748 Garching, Germany.}
\end{center}

\vspace{1cm}
\begin{abstract}
\nn We perform a complete analysis of the supersymmetric particle
spectrum in the Minimal Supergravity (mSUGRA) model where the soft
SUSY breaking scalar masses, gaugino masses and trilinear couplings
are unified at the GUT scale, so that the electroweak symmetry is
broken radiatively. We show that the present constraints on the Higgs
boson and superparticle masses from collider searches and precision
measurements still allow for large regions of the mSUGRA parameter
space where charginos, neutralinos, sleptons and top squarks as well
as the heavier Higgs particles, are light enough to be produced at the
next generation of $\ee$ linear colliders with center of mass energy
around $\sqrt{s} \sim 800$ GeV, with sizeable cross sections. An
important part of this parameter space remains even when we require
that the density of the lightest neutralinos left over from the Big
Bang, which we calculate using standard assumptions, falls in the
range favored by current determinations of the Dark Matter density in
the Universe. Already at a c.m. energy of 500 GeV, SUSY particles can
be accessible in some parameter range, and if the energy is increased
to $\sqrt{s} \simeq 1.2$ TeV, the $e^+e^-$ collider will have a reach
for high precision studies of SUSY particles in a range that is comparable to 
the discovery range of the LHC.
\end{abstract}

\newpage
\setcounter{page}{1}
\section*{1. Introduction} 

Supersymmetric theories (SUSY) \cite{SUSY} are the best motivated
extensions of the Standard Model (SM) of the electroweak and strong
interactions. They provide an elegant way to stabilize the huge
hierarchy between the Grand Unification (GUT) or Planck scale and the
Fermi scale, providing a natural framework to cancel the quadratic
divergences of the radiative corrections to the Higgs boson mass. The
most economical low--energy supersymmetric extension of the SM, the
Minimal Supersymmetric Standard Model (MSSM) \cite{MSSM}, allows for a
consistent unification of the three coupling constants of the SM gauge
group \cite{GaugeUni}. In addition, it can provide a natural solution
of the Dark Matter problem \cite{dmrev}, since it predicts the
existence of an electrically neutral, weakly interacting, massive and
absolutely stable particle; for large regions of parameter space the
thermal relic density of this particle agrees with the Dark Matter
density derived from cosmological arguments. The search for
Supersymmetric particles and for the required extended Higgs spectrum
is one of the main motivations for building high--energy colliders. \s

In the MSSM one assumes the minimal gauge group, i.e. the SM ${\rm
SU(3)_C \times SU(2)_L \times U(1)_Y}$ group; the minimal particle
content, i.e. three generations of fermions [without right--handed
neutrinos] and their spin--zero partners as well as two Higgs doublet
superfields to break the electroweak symmetry; and R--parity
conservation \cite{Fayet} which makes the lightest SUSY particle,
assumed to be the lightest neutralino $\tilde\chi_1^0$, absolutely stable.
In order to explicitly break supersymmetry [as required by experiment]
while preventing the reappearance of quadratic divergences, a
collection of soft terms is added to the Lagrangian \cite{soft}: mass
terms for the gauginos, mass terms for the scalar fermions, mass and
bilinear terms for the Higgs bosons, and trilinear couplings between
sfermions and Higgs bosons. In the general case, that is if one allows
for intergenerational mixing and complex phases, the soft SUSY
breaking terms will introduce a huge number (105) of unknown
parameters \cite{parameters}, in addition to the 19 parameters of the
SM. \s

This feature makes any phenomenological analysis in the general MSSM a
daunting task, if possible at all. In addition, almost all ``generic''
sets of these parameters are excluded by severe phenomenological
constraints, on flavor changing neutral currents (FCNC), additional
CP--violation, color and charge breaking minima, etc. Almost all FCNC
problems are solved at once if the MSSM parameters obey a set of
universal boundary conditions at the GUT scale. We will take these
parameters to be real, which solves all problems with CP
violation\footnote{If soft breaking parameters are universal at the
GUT scale, they are allowed to have large CP--violating phases only in
certain very narrow regions of parameter space where large
cancellations occur between various contributions to electric dipole
moments of the electron and neutron \cite{cancellations}. With all
high--scale soft breaking parameters being real, the model predicts
very small deviations from the SM in $B-$meson mixing and
CP--violating decays \cite{bdecays} now being explored at the
$B-$factories. However, we will see later that the model allows for
significant new contributions to $b \rightarrow s \gamma$ decays, and
related decay modes.}. The underlying assumption is that SUSY--breaking
occurs in a hidden sector which communicates with the visible sector
only through gravitational--strength interactions, as specified by
supergravity \cite{nilles}. Universal soft breaking terms then emerge
if these supergravity interactions are ``flavor--blind'' [like
ordinary gravitational interactions]. This is assumed to be the case
in the constrained MSSM or minimal Supergravity (mSUGRA) model
\cite{mSUGRA}. In this model the entire spectrum of superparticles and
Higgs bosons is determined by the values of five free parameters.
Since universal boundary conditions imply that the electroweak
symmetry is broken radiatively, which imposes one constraint on the
input parameters, one is left with only four continuous free
parameters and a discrete one. This makes comprehensive scans of
parameter space feasible. \s

Although other viable SUSY models exist, the mSUGRA model has become
the most frequently used benchmark scenario for supersymmetry, and has
been widely used to analyze the expected SUSY particle spectrum and
the properties of SUSY particles, and to compare the predictions with
available and/or expected data from collider experiments. Several
global or partial analyses of the present theoretical and experimental
constraints on the mSUGRA model have been performed in the literature;
see for instance Refs.~\cite{analyses,benchmark,Leszek}. \s
 
In this paper, we perform an independent analysis of the SUSY particle
spectrum in the mSUGRA model, taking into account theoretical
constraints\footnote{In order to further limit the parameter space,
one could require that the SUGRA model is not fine--tuned and the SUSY
breaking scale should not be too high, a constraint which can be
particularly restrictive since sparticles with masses beyond $\sim 1$
TeV would be excluded. However, the degree of fine--tuning which can
be considered acceptable is largely a matter of taste, so for the most
part we disregard this issue in our analysis.} and all available
experimental information \cite{PDG}: searches for the MSSM Higgs
bosons and SUSY particles at the LEP and Tevatron colliders,
electroweak precision measurements, the radiative $b \to s\gamma$
decay, etc. Special attention is paid to the implications of the
measurement of the anomalous magnetic moment of the muon recently
performed at Brookhaven \cite{BNL}, and to the $\sim 2\sigma$ evidence
for a SM--like Higgs boson with a mass $M_{\rm Higgs} \sim 115.6$ GeV
seen by the LEP collaborations \cite{hevidence}. We also discuss the
implication of requiring thermal relic neutralinos to form the Dark
Matter in the Universe. \s

We show that in a large part of the mSUGRA parameter space at least
one of these independent pieces of evidence for physics beyond the SM,
from the Dark Matter density, the $(g_\mu-2)$ measurement and the LEP2
Higgs boson--like excess, can be explained in mSUGRA. On the other
hand, only a small area of parameter space allows for these three
constraints to be fulfilled simultaneously. If all these indications
survive further scrutiny, the parameter space of the model would thus
already be tightly constrained. However, one has to keep in mind that
the statistical significance for the LEP Higgs signal and the
$(g_\mu-2)$ anomaly are at present still quite weak\footnote{ Recent
estimates of the uncertainties in the hadronic contributions to
$(g_\mu-2)$ might slightly push the theoretical prediction
\cite{SMgm2} towards the SM value and thus decrease the significance
of the discrepancy, see for instance Ref.~\cite{gerror}.}, while the
calculation of the thermal relic density relies on additional
assumptions that cannot be tested in collider experiments. \s

We then discuss prospects for producing SUSY particles and the heavier
Higgs bosons of the MSSM at future high--energy $e^+e^-$ linear
colliders\footnote{During the final stage of the present work, for
which preliminary results have been presented in Ref.~\cite{Orsay},
the paper ``Proposed Post--LEP Benchmarks for Supersymmetry"
\cite{benchmark}, which discusses some of the issues considered here
appeared on the web. A brief discussion of the prospects at future
$\ee$ colliders has also been given in Ref.~\cite{Francois}.
Prospects for the Tevatron Run II and for the LHC have been discussed
in Refs.~\cite{Tevatron,LHC}, respectively; of course, these earlier
studies used slightly weaker experimental constraints, as was
appropriate at the time of their writing.} with center of mass
energies $\sqrt{s}$ around 800 GeV as expected, for instance, at the
TESLA machine \cite{TESLA0}. We show that in large areas of the mSUGRA
parameter space the production rates for the lightest charginos and
neutralinos, as well as for sleptons, top squarks and the heavier
Higgs bosons are large enough for these particles to be discovered,
given the very large integrated luminosities, ${\cal L} \sim 500$
fb$^{-1}$, expected at this collider. Even at lower energies,
$\sqrt{s} \simeq 500$ GeV, charginos, neutralinos and tau sleptons can
be produced in some parameter range. If the energy is raised to
$\sqrt{s} \simeq 1.2$ TeV \cite{Wagner}, the $e^+e^-$ collider will
have a reach for probing the SUSY particle spectrum and the heavy
Higgs bosons which is comparable to the reach of the LHC. 

The remainder of this paper is organized as follows. In the next
section we briefly summarize the main features of the mSUGRA model and
the way it is implemented in our analysis. In section 3 the
experimental and cosmological constraints on the mSUGRA parameter
space are discussed. In section 4 we analyze the production of SUSY
particles and MSSM Higgs bosons at high energy $e^+ e^-$ colliders. 
Conclusions are given in section 5. For completeness, expressions
for the cross sections of all discussed particle production channels
in $e^+e^-$ collisions are collected in the Appendix.

\section*{2. The Physical Set--Up} 

\subsection*{2.1 The mSUGRA model} 

We will perform our analysis in the constrained MSSM or minimal
Supergravity model, where the MSSM soft breaking parameters obey a set
of universal boundary conditions at the GUT scale, so that the
electroweak symmetry is broken radiatively. For completeness and to
fix the notation, let us list these unification and universality
hypotheses and summarize the main features of the radiative
electroweak symmetry breaking (EWSB) mechanism. \s

Besides the unification of the gauge coupling constants, which is
verified given the experimental results from LEP1 \cite{GaugeUni} and
which can be viewed as fixing the Grand Unification scale $M_{\rm GUT}
\sim 2 \cdot 10^{16}$ GeV, the unification conditions are as follows:
\s

-- Unification of the gaugino [bino, wino and gluino] masses: 
\beq
M_1 (M_{\rm GUT})=M_2(M_{\rm GUT})=M_3(M_{\rm GUT}) \equiv m_{1/2}.
\eeq

-- Universal scalar [sfermion and Higgs boson] masses [$i$ is a 
generation index] 
\beq
M_{\tilde{Q}^i} (M_{\rm GUT}) &=& M_{\tilde{u}^i_R} (M_{\rm GUT}) =
M_{\tilde{d}^i_R}(M_{\rm GUT})  =M_{\tilde{L}^i} (M_{\rm GUT}) 
= M_{\tilde{l}^i_R} (M_{\rm GUT}) \non \\
&=& M_{H_u}(M_{\rm GUT}) =M_{H_d} (M_{\rm GUT}) \equiv  m_0.
\eeq

-- Universal trilinear couplings: 
\beq
A_u^{ij} (M_{\rm GUT}) = A_d^{ij} (M_{\rm GUT}) = A_l^{ij} (M_{\rm
GUT}) \equiv  A_0 \delta_{ij}.
\eeq

Besides the three parameters $m_{1/2}, m_0$ and $A_0$, the
supersymmetric sector is described at the GUT scale by the bilinear
coupling $B$ and the supersymmetric Higgs(ino) mass parameter
$\mu$. However, one has to require that EWSB takes place. This results
in two necessary minimization conditions of the two Higgs doublet
scalar potential which, at the tree--level, has the form \cite{HHG}
[to have a more precise description, one--loop corrections to the
scalar potential have to be included, as will be discussed later]:
\begin{eqnarray} 
V_{\rm Higgs} &=& \overline{m}_1^2 H_d^{\dagger} H_d + \overline{m}_2^2 
H_u^{\dagger} H_u + \overline{m}_3^2 (H_u \cdot H_d + {\rm h.c.}) \nonumber \\
 &+& \frac{g_1^2+g_2^2}{8}  (H_d^{\dagger} H_d - H_u^{\dagger} H_u)^2 + 
\frac{g_2^2}{2} (H_d^{\dagger}  H_u) (H_u^{\dagger}  H_d)
\label{vhiggs}, 
\end{eqnarray}
where we have used the usual short--hand notation:
\beq
\overline{m}_1^2= m^2_{H_d}+\mu ^2 \ , \ 
\overline{m}_2^2= m^2_{H_u}+\mu ^2 \ , \ 
\overline{m}_3^2= B\mu.
\eeq
The SU(2) invariant product of two doublets is defined as $\phi_1
\cdot \phi_2 = \phi_1^1 \phi_2^2 - \phi_1^2 \phi_2^1$, where the
superscripts are SU(2) indices. The two minimization equations
$\partial V_{\rm Higgs} / \partial H_d^0 = \partial V_{\rm Higgs} /
\partial H_u^0 = 0$ can be solved for $\mu^2$ and $B \mu$:
\begin{eqnarray} \label{eq:ewsb}
\mu^2 &=& \frac{1}{2} \bigg[ \tan 2\beta (m^2_{H_u} \tan \beta
- m^2_{H_d} \cot \beta) -M_Z^2 \bigg]; \non \\
B\mu &=& \frac{1}{2} \sin 2\beta \Bigg[ m^2_{H_u} + m^2_{H_d} + 2
\mu^2 \Bigg] .
\end{eqnarray}
Here, $M_Z^2=(g_1^2+g_2^2) \cdot (v_u^2 + v_d^2) /4$, and $\tan
\beta= v_u/v_d$ is defined in terms of the vacuum expectation values
of the two neutral Higgs fields. Consistent EWSB is only possible if
eq.~(\ref{eq:ewsb}) gives a positive value of $\mu^2$. The sign of
$\mu$ is not determined. Therefore, in this model one is left with
only four continuous free parameters, and an unknown sign:
\beq 
\tan \beta \ , \ m_{1/2} \ , \ m_0 \ , \ A_0 \ , \ \ {\rm sign}(\mu). 
\eeq 
All the soft SUSY breaking parameters at the weak scale are then
obtained through Renormalization Group Equations (RGE) \cite{RGE}. \s

The number of parameters could be further reduced by introducing an
additional constraint which is based on the assumption that the $b$
and $\tau$ Yukawa couplings unify at the GUT scale, as predicted in
minimal SU(5). This restricts $\tb$ to two narrow ranges around $\tb
\sim 1.5$ and $\tb \sim m_t/m_b$ \cite{Yunif}. The low $\tb$ solution
is ruled out since it leads to a too light an $h$ boson, in conflict
with searches at LEP2. However, Yukawa unification is not particularly
natural in the context of superstring theories, and minimal SU(5)
predictions are known to fail badly for the lighter generations. We
therefore treat all three third generation Yukawa couplings as
independent free parameters.

\subsection*{2.2 Calculation of the SUSY particle spectrum}

In this section, we briefly discuss our procedure for calculating the
SUSY particle spectrum in the constrained MSSM with universal boundary
conditions at the GUT scale, as well as related issues which are
relevant to our study. All results are based on the numerical FORTRAN
code {\tt SuSpect} version 2.0 \cite{suspect}, to which we refer for a
more detailed description. The algorithm essentially includes:

\begin{itemize}
\vspace{-2mm}
\item[--] Renormalization group evolution (RGE) of parameters between
the low energy scale [$M_Z$ and/or the electroweak symmetry breaking
scale] and the GUT scale.
\vspace{-2mm}
\item[--] Consistent implementation of radiative electroweak symmetry
breaking (EWSB). Loop corrections to the effective potential are
included using the tadpole method.
\vspace{-2mm}
\item[--] Calculation of the physical (pole) masses of the Higgs
bosons, scalar quarks and leptons as well as gluinos, charginos and
neutralinos.
\vspace{-2mm} 
\end{itemize}

\nn In more detail we proceed as follows. We first chose the
low--energy input values of the SM parameters. The gauge couplings
constants are defined in the ${\overline{\rm MS}}$ scheme at the scale
$M_Z$ [$\bar{s}_W^2 = 1- \bar{c}_W^2 \equiv
\sin^2\theta_W|^{\overline{\rm MS}}$]:
\beq 
g_1^2= \frac{4 \pi \alpha^{\overline{\rm MS}}_{\rm
em}(M_Z)} {\bar{c}^{2}_W } \ , \ g_2^2= \frac{4 \pi
\alpha^{\overline{\rm MS}}_{\rm em}(M_Z)} {\bar{s}^{2}_W } \ , 
\ g_3^2= 4 \pi \alpha^{\overline{\rm MS}}_s(M_Z).
\eeq
Their values have been obtained from precision measurements at LEP
and Tevatron \cite{PDG}:
\begin{eqnarray}
\alpha^{\overline{\rm MS}}_{\rm em}(M_Z)=1/127.938 \ , \ 
\alpha^{\overline{\rm MS}}_s(M_Z)= 0.1192 \ , \ 
\bar{s}^2_W = 0.23117.
\end{eqnarray}
The pole masses of the heavy SM fermions are \cite{PDG}: 
\begin{eqnarray}
M_{t}= 174.3 \ {\rm GeV} \ \ , \ \ 
M_{b} = 4.62 \  {\rm GeV} \ \ , \ \ 
M_{\tau} = 1.778 \ {\rm GeV} \ \ .
\end{eqnarray}
From the pole $b$--quark mass, one then obtains the $\overline{\rm
DR}$ mass, $\overline{m}_b (\overline{m}_b) \simeq 4.23$ GeV which is
then evolved, using two--loop ${\cal O}(\alpha_s^2)$ RGE, to obtain
the running mass at the scale $M_Z$, $\overline{m}_b(M_Z) \simeq 2.92$
GeV. Since the two--loop corrections to the difference between pole
and $\overline{\rm DR}$ top and bottom quark masses are not yet known, we
include, instead, the analogous two--loop corrections in the
$\overline{ \rm MS}$ scheme, which should be close to the
$\overline{\rm DR}$ ones. The difference should not be important in
view of the experimental errors in the determination of the two masses
\cite{PDG}, $M_t=174.3 \pm 5.1$ GeV and $\overline{m}_b(\overline{m}_b)
\simeq 4.24 \pm 0.11$ GeV. \s
 
Next, the $\overline{\rm DR}$--scheme values of the gauge and Yukawa
couplings are extracted from these inputs \cite{drbar}. The latter are
defined by [$v=174.1$ GeV]:
\beq 
\lambda_t (M_t) = \frac{\overline{m}_t (M_t)}{v \sin \beta} \ , \
\lambda_b (M_Z) = \frac{\overline{m}_b (M_Z)}{v \cos \beta} \ , \
\lambda_\tau (M_Z) = \frac{\overline{m}_\tau (M_Z) }{v \cos \beta} \ .
\eeq
All couplings are then evolved up to the GUT scale using two--loop
RGEs \cite{drbar,RGE2}. Here heavy (super)particles are taken to contribute
to the RGE only at scales larger than their mass, i.e. multiple
thresholds are included in the running of the coupling constants near
the weak scale. The GUT scale $M_{\rm GUT} \simeq 2 \cdot 10^{16}$ GeV
is defined to be the scale at which $g_1 = g_2 \cdot \sqrt{3/5}$. We
do not enforce $g_2 = g_3$ at the GUT scale and assume that the small
discrepancy [of the order of a few percent] is accounted for by
unknown GUT--scale threshold corrections \cite{gutthresh}. \s

In our numerical analyses we fix the MSSM parameters $\tan\beta$ [given
at scale $M_Z$] as well as $A_0$ and the sign of $\mu$, and then
perform a systematic scan over the high energy mSUGRA inputs $m_0$ and
$m_{1/2}$. Given these boundary conditions, all the soft SUSY breaking
parameters and couplings are evolved down to the electroweak
scale. Our default choice for this scale is the geometric mean of the
two top squark masses, $M_{\rm EWSB} = (m_{\tilde{t}_1}
m_{\tilde{t}_2})^{1/2}$, which minimizes the scale-dependence of the
one--loop scalar effective potential \cite{Vscale}. Since $\tb$ is
defined at scale $M_Z$, the vevs have to be evolved down from $M_{\rm
EWSB}$ to $M_Z$ \cite{Vscale}. \s

One--loop radiative corrections to the Higgs potential play a major
role in determining the values of the parameters $|\mu|$ and $B$ in
terms of the soft SUSY breaking masses of the two Higgs doublet
fields. We treat these corrections using the tadpole method. This
means that we can still use eq.~(\ref{eq:ewsb}) to determine
$\mu^2(M_{\rm EWSB})$; one simply has to add one--loop tadpole
corrections to $m^2_{H_d}$ and $m^2_{H_u}$ \cite{potential,Bagger}.
We include the dominant third generation fermion and sfermion loops,
as well as subdominant contributions from sfermions of the first two
generations, gauge bosons, Higgs bosons, charginos and neutralinos,
with the running parameters evaluated at $M_{\rm EWSB}$. As far as the
determination of $\mu^2$ and $B\mu$ is concerned, this is equivalent
to computing the full one--loop effective potential at scale $M_{\rm
EWSB}$. Since $|\mu|$ and $B$ affect masses of some (s)particles
appearing in these corrections, this procedure has to be iterated
until stability is reached and a consistent value of $\mu$ is
obtained; usually this requires only three or four iterations for an
accuracy of ${\cal O}(10^{-4})$, if one starts from the values of
$|\mu|$ and $B$ as determined from minimization of the RG--improved
tree--level potential at scale $M_{\rm EWSB}$. \s

At this stage, we check whether the complete scalar potential has
charge and/or color breaking (CCB) minima, which can be lower than the
electroweak minimum. These can e.g. appear in the top squark
sector\footnote{CCB minima involving first and second generation
sfermions are usually separated from the desired EWSB minimum by high
potential barriers, so that the EWSB minimum is still stable on
cosmological time scales \cite{claudson}.} for large values of the
trilinear coupling $A_t$. In order to avoid them, we impose the
(simplest) condition \cite{CCB}:
\beq
{\rm CCB}: && A^2_t < 3 \, (m^2_{\tilde t_L} +m^2_{\tilde t_R}+ \mu^2 +
m^2_{H_u} ). \label{CCBcons}
\eeq
Of course, we also reject all points in the parameter space which lead
to tachyonic Higgs boson or sfermion masses\footnote{Later on, we will
be more restrictive and discard the situations where SUSY particles
have masses which are lower than the mass of the neutralino $\tilde\chi_1^0$
which will be assumed to be the lightest SUSY particle}:
\beq
{\rm No\,  Tachyon}: && M_A^2 >0 \ \ , M^2_h > 0 \ \ , \ \ m^2_{\tilde
f} >0 . 
\eeq 
The EWSB mechanism is assumed to be consistent when all these
conditions are satisfied. \s

We then calculate all physical particle masses. The procedure is
iterated at least twice until stability is reached, in order to take
into account: (i) Realistic (multi--scale) particle thresholds in the
RG evolution of the dimensionless couplings via step functions in the
$\beta$ functions for each particle threshold. (ii) Radiative
corrections to SUSY particle masses, using the expressions given in
Ref.~\cite{Bagger}, where the renormalization scale is set to $M_{\rm
EWSB}$.  \s

We first evaluate the SUSY--radiative corrections to the heavy fermion
masses, $\bar{m}_t,\ \bar{m}_b$ and $\bar{m}_\tau$, following
Ref.~\cite{Bagger}.\footnote{In our procedure some of the leading
logarithmic terms are already included in the RG evolution of the
Yukawa couplings via the step functions. Therefore, care has to be
taken to avoid double counting when extracting the relevant radiative
corrections from the expressions given in Ref.~\cite{Bagger}.} This
includes SUSY--QCD corrections for the $t,b$ quarks [from
squark--gluino loops] and the dominant electroweak corrections for the
$b$ and $\tau$ masses [chargino--sfermion loops which are enhanced by
terms $\propto \mu \tb$]. As suggested in Ref.~\cite{viennese}, we use
the ``MSSM'' quark masses [essentially the Yukawa coupling times vev]
in the squark mass matrices. Our iteration then resums all SUSY--QCD
corrections to the quark masses of order $\left( \alpha_s \tb
\right)^n$. This is important at large $\tb$, where these corrections
can be quite sizable. The various sectors of the MSSM are then treated
as follows: \smallskip

-- In the sfermion sector, the soft scalar masses as well as the
trilinear couplings for the third generation are obtained using
one--loop RGE, and are frozen at the scale $M_{\rm EWSB}$. In the
third generation sfermion sector [$\tilde{t},\tilde{b}, \tilde{\tau}
$], mixing between ``left'' and ``right'' current eigenstates is
included, where the radiatively corrected running fermion masses at scale
$M_{\rm EWSB}$ are employed in the sfermion mass matrices. The
radiative corrections to the sfermion masses are included according to
Ref.~\cite{Bagger}, i.e. only the QCD corrections for the
superpartners of light quarks [including the bottom squark] plus the
leading electroweak corrections to the top squarks; the small
electroweak radiative corrections to the slepton masses have been
neglected. \s

-- In the gaugino sector, the SUSY breaking gaugino masses are
obtained using the two--loop RGEs and are also frozen at $M_{\rm
EWSB}$. The mass matrices for charginos and neutralinos are
diagonalized using analytical formulae \cite{diag}. The one--loop QCD
radiative corrections to the gluino mass are incorporated
\cite{drbar}, while in the case of charginos and neutralinos the
radiative corrections \cite{inocorr} are included in the gaugino and
higgsino limits, which is a very good approximation according to
Ref.~\cite{Bagger}. \s

-- In the Higgs sector, the running mass of the pseudoscalar Higgs
boson is obtained from the soft SUSY breaking Higgs masses [again
frozen at $M_{\rm EWSB}$] and the full one--loop tadpole corrections
\cite{Bagger}. This mass is then used as input, together with $M_t, \
\tb$ and some MSSM parameters [$A_t,A_b, \mu$ and the soft SUSY
breaking third generation squark masses], to obtain the pole masses of
the pseudoscalar Higgs boson $A$, the two CP--even $h$ and $H$ bosons
and the charged $H^\pm$ Higgs particles. This last step is similar to
the program {\tt HDECAY} \cite{hdecay}, which calculates the Higgs
spectrum and decay widths in the MSSM. The complete radiative
corrections due to top/bottom quark and squark loops within the
effective potential approach, leading NLO QCD corrections [through
renormalization group improvement] and the full mixing in the stop and
sbottom sectors are incorporated using the analytical expressions of
Ref.~\cite{SUBH}. We have verified that the results obtained for the
Higgs spectrum, in particular for the lightest $h$ boson mass, are
nearly the same as those obtained from the complete results of the
Feynman diagrammatic approach implemented in the program {\tt
FeynHiggs} \cite{herrors}. \s

Our results for some representative points of the mSUGRA parameter
space have been carefully cross--checked against other existing
codes. We obtain very good agreement, at the one percent level, with
the program {\tt SOFTSUSY} \cite{SOFTSUSY} which has been released
recently\footnote{We thank Ben Allanach for his gracious help in
performing this detailed comparison.}. We also find rather good
agreement for the SUSY particle masses computed by the program {\tt
ISASUGRA} \cite{ISASUGRA}, once we chose the same configuration [soft
SUSY breaking masses frozen at $M_Z$, some radiative corrections to
sparticle masses are not included, etc..]. The value of the lightest
Higgs boson mass is in less good agreement, presumably due to the more
sophisticated treatment of the Higgs potential in {\tt SuSpect}; we
will see later that a precise calculation of the $h$ boson mass is an
important ingredient of our analysis.

\section*{3. Constraints on the mSUGRA parameter space} 

\subsection*{3.1 Experimental Constraints}

\subsubsection*{i) Lower bounds on the SUSY particles masses}

A wide range of searches for SUSY particles has been performed at LEP2
and at the Tevatron, resulting in limits on the masses of these
particles \cite{PDG}. The pair production of the lightest chargino at
LEP2, $\ee \to \tilde\chi_1^+ \tilde\chi_1^-$, would probably have
been the cleanest SUSY process. In general it has the largest SUSY
production cross section at $\ee$ colliders, after the experimental
cuts needed to suppress the backgrounds, and the information that it
provides is one of the most important in the context of the mSUGRA
model. The negative outcome of searches for charginos at LEP2, up to
energies of $\sqrt{s} \simeq 208$ GeV, gives the approximate bound
$m_{\tilde\chi_1^\pm} \gsim 104$~GeV \cite{LEPres}.\footnote{This
bound is not valid if $|\mu| \ll M_2$, i.e. for a very higgsino--like
chargino which is almost degenerate with the LSP leading to a small
release of missing energy. We therefore exclude slightly too much in
the ``focus point'' region, see below. The true bound is also reduced
somewhat in scenarios with light sneutrino, since $\tilde \nu$
exchange in the $t-$channel reduces the cross section, and since
$\tilde\chi_1^\pm \rightarrow \tilde\nu + \ell^\pm$ decays can be
difficult to detect if the $\tilde\chi^\pm_1 - \tilde\nu$ mass
difference is small; this can happen for very small $m_0$ in mSUGRA,
but such scenarios are tightly constrained by slepton searches and
SUSY loop effects. An accurate treatment of this bound is not
important for the main topic to be investigated in this paper, the
reach for future $\ee$ colliders with energy much above the LEP
range.} \s

In mSUGRA the gaugino masses are unified at the GUT scale, leading to
the approximate relation $M_2 \simeq 2M_1 \sim M_3/3$ at the weak
scale. The bound on the lightest chargino mass thus translates into a
lower bound on the LSP mass, $m_{\tilde\chi_1^0} \gsim 50$ GeV [in the
gaugino--like region; in the higgsino--region, the bound is higher]
and also on the gluino mass, $m_{\tilde{g}} \sim M_3 \gsim 300$
GeV. In the case of the LSP, the bound can be improved by using
searches for neutralino production at LEP2, $\ee \to \tilde\chi_1^0
\tilde\chi_2^0$; however, these neutralino searches are relevant only for
low values of $\tb \lsim 2$ which are already excluded by Higgs boson
searches, as will be discussed later. In the case of gluinos, this
indirect bound is similar to the one obtained from direct searches at
the Tevatron, $m_{\tilde{g}} \gsim 300$ GeV, which is valid if
$m_{\tilde{q}} \simeq m_{\tilde{g}}$ \cite{D0}; the direct search
limits for $m_{\tilde q} \gg m_{\tilde g}$ are significantly weaker.
\s

The bound on $M_2$ also translates into a bound on the masses of first
and second generation squarks. The RG evolution of these masses [up to
small contribution from the D--terms] gives the approximate relation
$m^2_{\tilde{q}} \sim m_0^2 + 6 m_{1/2}^2$, which leads to
$m_{\tilde{q}} \gsim 250$ GeV, again of similar size as the bound from
direct searches at the Tevatron \cite{D0}. For third generation
squarks the RGE are more complicated and mixing between the
eigenstates is important, due to the large values of the Yukawa
couplings, so that the bounds from direct searches are relevant. We
use the bound from LEP2 \cite{LEPres}, which is almost independent of
the decay modes, and is applicable down to squark$-\lsp$ mass
splittings of a few GeV; Tevatron search limits \cite{cdfstop} are
stronger in some cases, but more dependent on details of squark decay,
and disappear for mass splittings below $\sim 30$ GeV. \s

By far the tightest slepton search limits also come from LEP2
\cite{LEPres}.  Here the coefficients of the $m_{1/2}^2$ term in the
RG evolution of $m^2_{\tilde l}$ are small so that [at least for the
SU(2) singlet ``right--handed'' states] $m_{\tilde{l}} \sim
m_0$. These LEP bounds are generally a few GeV below the kinematical
limit, except for some small regions of the parameter space with small
mass difference to the LSP.\footnote{Contrary to chargino pair
production, the cross sections for sfermion pair production in $\ee$
collisions is suppressed by a $\beta^3$ factor near threshold so that
it is rather tiny near the kinematical limit.} Since sneutrinos might
decay invisibly into $\nu \tilde\chi_1^0$, only indirect bounds can be
placed on their masses. Limits from searches for charged SU(2)
doublet sleptons, whose masses are related to sneutrino masses by
SU(2), are rather model--independent. In mSUGRA additional indirect
limits on sneutrino masses follow from searches for SU(2) singlet
sleptons and charginos, via the resulting constraints on $m_0$ and
$m_{1/2}$. \s

To summarize, in our analysis we impose the following bounds on
sparticle masses:
\beq
m_{\tilde\chi_1^+} \geq 104~{\rm GeV} & , 
& m_{\tilde{f}} \geq  100 \  {\rm GeV \ with} 
\ \tilde{f}= \tilde{t}_1, \tilde{b}_1, \tilde{l}^\pm, \tilde{\nu} \non \\
m_{\tilde{g}} \geq 300~{\rm GeV} & , & m_{\tilde{q}_{1,2}} \geq  260 \ 
{\rm GeV  \ with} \ \tilde{q}= \tilde{u}, \tilde{d}, \tilde{s} , \tilde{d}.
\eeq

\subsubsection*{ii) Constraints from the Higgs boson masses} 

The search for SUSY Higgs bosons was the main motivation for
extending the LEP2 energy up to $\sqrt{s}\simeq 208$ GeV. In the SM, a
lower bound \cite{hlimit,hevidence}
\beq
M_{H^0} \ >\ 113.5 \ {\rm GeV} \ {\rm at \ the \ 95\% C.L.}
\label{hbound}
\eeq
has been set by investigating the Higgs--strahlung process, $\ee \to
ZH^0$. In the MSSM, this bound is valid for the lightest CP--even
Higgs particle $h$ if its coupling to the $Z$ boson is SM--like,
i.e. if $g^2_{ZZh}/g^2_{ZZH^0} \equiv \sin^2(\beta- \alpha) \simeq 1$
where $\alpha$ is the mixing angle in the CP--even neutral Higgs
sector. This occurs in the decoupling regime where $M_A^2 \gg M_Z^2$.
For much lower values of $M_A$ and for large $\tb$, one has $M_h \sim
M_A$ and $\sin^2(\beta-\alpha) \ll 1$. In this case, the bound
(\ref{hbound}) applies to the mass of the heavy neutral CP--even Higgs
boson $H$. However, tighter constraints on the parameter space can
usually be derived from the search for the associated production of
CP--even and CP--odd Higgs particles, $\ee \to hA$, which has a cross
section proportional to $\cos^2(\beta-\alpha)$. At LEP2, a combined
mass exclusion in the MSSM,
\beq
M_h \, \simeq \, M_A \, \gsim \, 93.5 \ {\rm GeV}
\eeq
at the 95\% Confidence Level, has been set in this case. This limit is
lower than that from the Higgs--strahlung process, due to the less
distinctive signal and the $\beta^3$ suppression near threshold for
spin--zero particle pair production.  \s

Deriving a precise bound on $M_h$ for arbitrary values of $M_A$ [not
just for $M_A^2 \gg M_Z^2$ or $M_A \simeq M_h$] and hence for all values
of $\sin^2(\beta -\alpha)$ is rather complicated, since one has to
combine results from two different production channels, which have
different kinematical behavior, cross sections, backgrounds, etc. In
our analysis, we will use an interpolation formula for the excluded
value of $M_h$:
\beq 
M_h >  93.5 + 15x  + 54.3x^2 -48.4x^3- 25.7x^4  +24.8x^5 -0.5\, {\rm GeV},
\eeq 
with $x=\sin^2(\beta-\alpha)$. This formula fits the exclusion plot in
the $[M_h, \sin^2(\beta-\alpha)]$ plane given by the ALEPH
collaboration\footnote{The ALEPH search was performed in the energy
range up to $\sqrt{s}=202$ GeV leading to the bounds $M_h \gsim 107.7$
GeV in the decoupling limit and $M_h, M_A \gsim 91.5$ GeV for
$\sin^2(\beta- \alpha) \sim 0$. We have extended these end points to
the values $M_h \sim 113.5$ and 93.5 GeV obtained at $\sqrt{s} \simeq
208$ GeV, while leaving the general form of the exclusion contour
unchanged.} in Ref.~\cite{Aleph}. At this point we assume a very small
theoretical error of 0.5 GeV in the calculation of $M_h$ and $M_A$,
and have lowered the bounds accordingly; this is the typical
difference between our value of $M_h$ and the one obtained in the full
diagrammatic approach \cite{herrors}. The total theoretical error on
the calculated Higgs boson masses is probably much larger, even if all
input parameters were known exactly. The intrinsic uncertainty due to
unknown higher--order effects is usually estimated to be about two or
three GeV. In addition, the error on the top quark mass leads to an
uncertainty of the predicted value of $M_h$ of a few
GeV.\footnote{The finite gluino contributions to the two--loop
radiative corrections \cite{herrors}, which we did not take into
account, might also change $M_h$ by one or two GeV.} The impact of
these uncertainties will be discussed later.  \s

We will also study the implications of the $2.1 \sigma$ evidence for a
SM--like Higgs boson with a mass $M_H = 115.6$ GeV seen\footnote{This
used to be interpreted as $\sim 2.9 \sigma$ evidence for a Higgs boson
with mass $M_H = 115.0$ GeV. However, a recent re--evaluation led to a
lower significance, and correspondingly slightly higher preferred
Higgs mass.} by the LEP collaborations \cite{hevidence}. In view of
the theoretical and experimental uncertainties, we interpret this
result as favoring the range
\beq
113 \ {\rm GeV} \ \lsim \ M_h \ \lsim 117 \ {\rm GeV} \ {\rm and} \
\sin^2(\beta - \alpha) \geq 0.95.
\eeq
The second requirement ensures that the cross section for Higgs
production, $\sigma(\ee \to hZ)$, is similar to that of the SM [the
excess is assumed to come from the Higgs--strahlung
process].\footnote{The ``discovered" Higgs boson could also be the
heavy CP--even $H$ boson produced in the strahlung process with
$\cos^2(\beta-\alpha)$ close to unity, if the $h$ and $A$ bosons have
masses $M_h \sim M_A \geq 93.5$ GeV. However, in mSUGRA only a very
small set of input parameters can give this configuration. In fact, we
did not find any parameter set where this possibility is realized and
all other constraints are satisfied.}

\subsubsection*{iii) Constraints from electroweak precision observables} 

Loops of Higgs and SUSY particles can contribute to electroweak
observables which have been precisely measured at LEP, SLC and the
Tevatron. In particular, the radiative corrections to the
self--energies of the $W$ and $Z$ bosons, $\Pi_{WW}$ and $\Pi_{ZZ}$,
might be sizeable if there is a large mass splitting between some
particles belonging to the same SU(2) doublet; this can generate a
contribution which grows as the mass squared of the heaviest
particle. The dominant contributions to the electroweak observables,
in particular the $W$ boson mass and the effective mixing angle
$s_W^2$, enter via a deviation of the $\rho$ parameter \cite{drho0}
from unity, which measures the relative strength of the neutral to
charged current processes at zero momentum transfer, i.e. the breaking
of the global custodial SU(2) symmetry:
\beq    
\Delta M_W \simeq \frac{c_W^2 M_W}{2(c_W^2-s_W^2)} \Delta \rho \ \ , \ \
\Delta s_W^2 \simeq - \frac{2c_W^2 s_W^2}{c_W^2-s_W^2} \Delta \rho \non \\
\rho = (1-\Delta \rho)^{-1} \ ; \ \Delta \rho = \Pi_{ZZ}(0)/M_Z^2 - 
\Pi_{WW}(0)/M_W^2.
\eeq
Let us briefly summarize the possible contributions in the MSSM
\cite{Bagger,drhosusy1,drhosusy2}. A close inspection of the
contributions of the Higgs and chargino/neutralino sectors shows that
they are very small, $\Delta\rho \lsim 10^{-4}$.  In the former case,
the contributions are logarithmic, $\sim \alpha \log (M_h/M_Z)$, and
are similar to those of the SM Higgs boson [they are identical in the
decoupling limit]. In the latter case they are small because the only
terms in the chargino and neutralino mass matrices which could break
the custodial SU(2) symmetry are proportional to $M_W$. Since first
and second generation sfermions are almost degenerate in mass, they
also give very small contributions to $\Delta \rho$.\footnote{It has
recently been pointed out \cite{elwfit} that light sleptons and light
charginos could slightly improve the quality of global fits to
electroweak precision data. However, such configurations are excluded
in mSUGRA, chiefly by the Higgs boson search limits.} Only the third
generation sfermion sector can generate sizable corrections to the
$\rho$ parameter, because left--right mixing and [in case of the stop]
the supersymmetric contribution $\propto m_f^2$ leads to a potentially
large splitting between the sfermion masses. In particular, the
contributions of the $(\tilde{t}, \tilde{b})$ and $(\tilde{\tau},
\tilde{\nu})$ iso--doublets \cite{drhosusy2} might become unacceptably
large. They are given by:
\beq \Delta \rho (\tilde{t}, \tilde{b}) & = & 
\frac{3 G_F}{8\pi^2\sqrt 2} \, \bigg[ -
c^2_t s^2_t \;f(m^2_{\tilde t_1},m^2_{\tilde t_2}) -c^2_b s^2_b
\;f(m^2_{\tilde b_1},m^2_{\tilde b_2}) \non \\ &+& c^2_t \; [ c^2_b
\;f(m^2_{\tilde t_1},m^2_{\tilde b_1})+ s^2_b \;f(m^2_{\tilde
t_1},m^2_{\tilde b_2})] + s^2_t \; [c^2_b \;f(m^2_{\tilde
t_2},m^2_{\tilde b_1}) +s^2_b \;f(m^2_{\tilde t_2},m^2_{\tilde b_2}) ]
\; \bigg] ; \non \\ 
\Delta \rho (\tilde{\tau}, \tilde{\nu}) & = &
\frac{G_F}{8\pi^2\sqrt 2} \, \bigg[ - c^2_\tau s^2_\tau
\;f(m^2_{\tilde \tau_1},m^2_{\tilde \tau_2}) + c^2_\tau \;
f(m^2_{\tilde \tau_1},m^2_{\tilde \nu}) + s^2_\tau \; f(m^2_{\tilde
\tau_2},m^2_{\tilde \nu}) \; \bigg] ; \non \\ && \hspace*{2cm} f(x,y) =
x+y-2 \frac{x y}{x-y} \log  \frac{x}{y},
\label{rhostop}
\eeq
where $s_i$ and $c_i$ are the sine and cosine of the mixing angles. In
the stop sector large values of the trilinear coupling $A_t$ can be
dangerous, since they lead to values $m_{\tilde{t}_1} \ll
m_{\tilde{t}_2}, m_{\tilde{b}_1}$. In the stau/sbottom sectors large
values of the parameters $\tb$ and $\mu$ lead to sizable splitting
between the $\tilde{b}, \tilde{\tau}$ mass eigenstates.  \s

We have required the contributions of the third generation sfermions
to stay below the acceptable level of
\begin{eqnarray}
\Delta \rho (\tilde{f}) \lsim 2.2 \cdot 10^{-3}, 
\end{eqnarray}
which approximately corresponds to a 2$\sigma$ deviation from the SM
expectation~\cite{LEPrho}. Two--loop QCD corrections to $\delta
\rho(\tilde{t}, \tilde{b})$ have been calculated \cite{drho2} and can
be rather large, increasing the contribution by up to 30\%. In our
analysis we include the leading components of these corrections,
i.e. the full corrections due to gluon exchange and the correction due
to gluino exchange in the heavy gluino limit [which should be a good
approximation in our case]. However, in our numerical analyses we find
that this constraint is usually superseded by the Higgs boson search
constraint.

\subsubsection*{iv) The $b \to s \gamma$ constraint} 

Another observable where SUSY particle contributions might be large is
the radiative flavor changing decay $b \to s\gamma$ \cite{bsg0}. In
the SM this decay is mediated by loops containing charge 2/3 quarks
and $W$--bosons. In SUSY theories additional contributions come from
loops involving charginos and stops, or top quarks and charged Higgs
bosons.\footnote{We neglect contributions from loops involving gluinos
or neutralinos, which are much smaller than the chargino loop
contribution in mSUGRA type models \cite{bsg0}.}  Since SM and SUSY
contributions appear at the same order of perturbation theory, the
measurement of the inclusive decay $B \ra X_s \gamma$ given by the
CLEO and Belle Collaborations \cite{bsgCLEO} [as well as by the ALEPH
collaboration \cite{bsgALEPH}, albeit with a larger error] is a very
powerful tool for constraining the parameter space. \s

In our analysis we use the value quoted by the Particle Data Group
\cite{PDG}:
\begin{eqnarray} \label{brbsgex}
{\rm BR}(b \ra s \gamma) = (3.37 \ \pm 0.37 \ \pm 0.34 \ \pm 
0.24^{+0.35}_{-0.16} \ \pm 0.38) \cdot 10^{-4}. \label{CLEO}
\end{eqnarray}
The first three errors are, respectively, the statistical error, the
systematical error and the estimated error on the model used to
extract information about quark decays from $B$ meson decays. The
fourth error is due to the extrapolation from the data, where the
photon energy is cut off at 2.1 GeV, to the full range of possible
photon energies which contribute to the integrated partial width
\cite{bsgamma4}. The fifth error is an estimate of the theory
uncertainty. \s

Recently, the next--to--leading order QCD corrections to the decay
rate have been calculated in the MSSM \cite{bsg1,bsg2}. In the present
analysis, we will use the most up--to--date prediction of the $b\to
s \gamma$ decay rate \cite{bsg2}, where all known perturbative and
non--perturbative effects are included.\footnote{We thank the authors
of Ref.~\cite{bsg2}, in particular Paolo Gambino, for providing us
with their FORTRAN code and for their help in interfacing it with our
code for the MSSM spectrum.} This includes all the possibly large
contributions which can occur at NLO, such as terms $\propto \tan
\beta$, and/or terms containing logarithms of $m_{\rm
SUSY}/M_W$. Besides the fermion and gauge boson masses and the gauge
couplings discussed previously, we will use the values of the SM input
parameters required for the calculation of the rate given in
Ref.~\cite{bsg3}, except for the cut--off on the photon energy,
$E_{\gamma} > (1-\delta)m_b/2$ in the bremsstrahlung process $b \to
s\gamma g$, which we fix to $\delta=0.9$ as in Ref.~\cite{bsg2}. \s

To be conservative we add all the experimental and theoretical
uncertainties in eq.~(\ref{brbsgex}) linearly; note that most of these
errors are systematic or theory errors, which do not obey Gauss
statistics. We thus allow the predicted $b \to s\gamma$ decay
branching ratio to vary within the range\footnote{Our conservative
approach comfortably accommodates the uncertainty in BR($b \to s\gamma)$ 
related to the proper definition of heavy quark masses recently discussed 
in Ref.~\cite{bsgnew}. Another argument for a conservative
interpretation of the bound on BR$(b \ra s \gamma)$ is that small
modifications of the GUT scale boundary condition [specifically, small
non--vanishing values for $A_d^{23}$ or $A_d^{32}$] could greatly
alter the prediction for this branching ratio, without affecting any
of the other quantities discussed in this paper.}
\begin{eqnarray} \label{bsgbound}
2.0 \times 10^{-4} \leq {\rm BR}(b \ra s \gamma) \leq 5.0 \times 10^{-4}.
\label{CLEO2}
\end{eqnarray}

\subsubsection*{v) The contribution to the muon $g-2$:} 

Very recently, the Muon $(g-2)$ Collaboration in Brookhaven has
reported a new measurement of the anomalous moment of the muon
\cite{BNL}:
\begin{eqnarray}
(g_\mu-2) \equiv a_\mu^{\rm exp} = 11\, 659\, 202\, (14)(6) \, 10^{-10}, \
\end{eqnarray}
where the first uncertainty is statistical and the second
systematical. The value predicted in the SM, including the QED, QCD
and electroweak corrections is: $a_\mu^{\rm SM} = 11\, 659\, 160\ (7)
\, 10^{-10}$ \cite{davier} [see however Ref.~\cite{SMgm2,gerror} for
the size of the hadronic uncertainties]. This value differs from the
new average value by 2.6$\sigma$ \cite{BNL}. This leads to a $2\sigma$
range for interpreting the discrepancy as a New Physics contribution
of:
\begin{eqnarray}
11 \cdot 10^{-10} \leq a_\mu \, 
({\rm New\, Physics}) \, \leq 75 \cdot 10^{-10}.
\label{g-2}
\end{eqnarray}
The contribution of SUSY particles to $(g_\mu-2)$ is mainly due to
neutralino--smuon and chargino--sneutrino loops [if no flavor
violation is present]. In mSUGRA--type models, the contribution of
chargino--sneutrino loops usually dominates; it is given by
\cite{SUSYgm2}:
\beq
\Delta a_{\mu}^{\rm SUSY} &=& \frac{\alpha m_\mu}{4\pi} \sum_{k=1,2} \Bigg[
-\frac{3 m_{\tilde\chi_k^\pm} g_k^{L}g_{k}^R}{ m_{\tilde{\nu}}^2 (1-x_k)^3}
\left( 1 -\frac{4}{3}x_k +\frac{1}{3}x_k^2+\frac{2}{3} \ln x_k \right)
\non \\
&&+ \frac{m_{\mu} \left[ \left(g_k^L\right)^2 + \left( g_k^R
\right)^2 \right] }{ 3m_{\tilde{\nu}}^2 (1-x_k)^4}
\left( 1 +\frac{3}{2}x_k -3 x_k^2 +\frac{1}{2}x_k^3+3 x_k \ln x_k \right)
\Bigg],
\end{eqnarray}
with $x_k=m_{\tilde\chi_k^\pm}^2/m_{\tilde{\nu}}^2$. The
chargino--lepton--slepton couplings are determined by the matrices
$U,V$ which diagonalize the $\tilde\chi^\pm$ mass matrix:
$g_k^{L}=U_{k2}/(\sqrt{2}M_W s_W \cos \beta)$ and $g_k^R=-V_{k1}/s_W$
\cite{GH}. If the SUSY particles are relatively heavy, the
contribution of chargino--sneutrino loops can be approximated by
[$\tilde{m}$ is the mass of the heaviest particle]: $\Delta
a_{\mu}^{\tilde\chi^\pm \tilde{\nu}} \sim 1.3\cdot 10^{-5} \times (\tb
/ \tilde{m}^2)$, to be compared with the contribution of
neutralino--smuon loops, $\Delta a_{\mu}^{\tilde\chi^0 \tilde{\mu}}
\sim 1.1 \cdot 10^{-6} \times (\tb / \tilde{m}^2)$, which is an order
of magnitude smaller.  \s

Thus the SUSY contribution to $(g-2)$ is large for large values of
$\tb$ and small values of $m_0$ and $m_{1/2}$, e.g. reaching the level
of $40\cdot 10^{-10}$ for $\tb\simeq 50$ and $\tilde{m}\simeq 400$
GeV. Note also that the sign of the SUSY contribution is equal to the
sign of $\mu$, $a_\mu^{\rm SUSY} \propto (\alpha/\pi) \mu M_2 \tb
/\tilde{m}^4$. If the 2.6$\sigma$ deviation of the measured
$(g_\mu-2)$ from the SM prediction is to be explained by SUSY, the
sign of $\mu$ therefore has to be positive, $\mu>0$. Intriguingly,
this sign of $\mu$ is usually also favored by the constraint
(\ref{bsgbound}) on the BR$(b \ra s \gamma)$ \cite{bsg0,bsg2}.

\subsection*{3.2 Cosmological constraints from the LSP relic density}

In this section, we will discuss the contribution of \lsp\ particles
to the overall matter density of the Universe, following the standard
treatment \cite{DM1}, with the modifications outlined in
\cite{DM2}. This treatment is based on the following assumptions:
i) \lsp\ should be effectively stable, i.e. its lifetime should be
long compared to the age of the Universe. This assumption is natural
in models with conserved R--parity if the LSP resides in the visible
sector.
ii) The temperature of the Universe after the last period of entropy
production must exceed $\sim 10\%$ of \mchi. This assumption is also
quite natural in the framework of inflationary models, given that
analyses of structure formation determine the scale of inflation to be
$\sim 10^{13}$ GeV \cite{DM1}. \s

However, one should bear in mind that one can construct models where
one of these assumptions is violated, without changing the collider
phenomenology of the (mSUGRA) model we are considering. The
cosmological constraints we are about to derive therefore have a
different status than the constraints that follow from ``New Physics''
searches at colliders. It is nevertheless interesting to delineate the
regions of parameter space where the model can provide the
approximately correct amount of Dark Matter in the Universe under the
stated simple assumptions. \s

If these assumptions hold \lsp\ decouples from the thermal bath of SM
particles at an inverse scaled temperature $x_F \equiv \mchi / T_F$
which is given by \cite{DM1}
\beq \label{edm1}
x_F = \frac{ 0.38 M_{Pl} \langle v \sigma_{\rm ann} \rangle c (c+2)
m_{\tilde\chi_1^0} } {\sqrt{g_* x_F}}.
\eeq
Here $v$ is the relative LSP velocity in their center--of--mass frame,
$\sigma_{\rm ann}$ denotes the LSP annihilation cross section into SM
particles, $\langle \dots \rangle$ denotes thermal averaging, $M_{Pl}
= 2.4 \cdot 10^{18}$ GeV is the (reduced) Planck mass, $g_*$ is the
number of relativistic degrees of freedom (typically, $g_* \simeq 80$
at $T_F$), and $c$ is a numerical constant, which we take to be
0.5. One typically finds $x_F \simeq 20$ to 25. Today's LSP density in
units of the critical density is then given by \cite{DM1}
\beq \label{edm2}
\om = \frac {2.13\cdot10^8 / {\rm GeV}} { \sqrt{g_*} M_{Pl} J(x_F)},
\eeq
where the annihilation integral $J$ is defined as
\beq \label{edm3}
J(x_F) = \int_{x_F}^\infty \frac {\langle v \sigma_{\rm ann}
\rangle(x) } {x^2} dx.
\eeq
The quantity $h$ in eq.~(\ref{edm2}) is the present Hubble constant in
units of 100 km$/$(sec$\cdot$Mpc). Eqs.~(\ref{edm1})--(\ref{edm3})
describe an approximate solution of the Boltzmann equation which has
been shown to describe the exact numerical solution very accurately
for all known scenarios. \s

Since neutralinos decouple at a temperature $T_F \ll \mchi$, in most
cases it is sufficient to use an expansion of the LSP annihilation
cross section in powers of the relative velocity between the LSPs:
\beq \label{edm4}
v \sigma_{\rm ann} \equiv v \sigma(\lsp \lsp \rightarrow {\rm
SM\, particles}) = a + b v^2 + {\cal O}(v^4).
\eeq
The entire dependence on the parameters of the model is then contained
in the coefficients $a$ and $b$, which essentially describe the LSP
annihilation cross section from an initial S-- and P--wave,
respectively. The computation of the thermal average over the
annihilation cross section, and of the annihilation integral
eq.~(\ref{edm3}), is then trivial, allowing an almost completely
analytical calculation of \om\ [eq.~(\ref{edm1}) still has to be solved
iteratively, but this iteration converges very quickly]. Expressions
for the $a$ and $b$ terms for all possible two--body final states are
collected in \cite{DN3}. In these expressions we use running quark
masses at scale $Q = \mchi$, in order to absorb leading QCD
corrections. Of course, we also use the loop--corrected Higgs boson
spectrum. \s

In generic scenarios the expansion eq.~(\ref{edm4}) reproduces exact
results to $\sim 10\%$ accuracy, which is quite sufficient for our
purpose. However, it has been known for some time \cite{DM2} that this
expansion fails in some exceptional cases, all of which can be
realized in some part of mSUGRA parameter space: \s

-- If the mass splitting between the LSP and the next--to--lightest
superparticle NLSP is less than a few times $T_F$, co--annihilation
processes involving one LSP and one NLSP, or two NLSPs, can be
important. This can usually still be treated using the formalism of
eqs.~(\ref{edm1})--(\ref{edm4}) if the \lsp\lsp\ annihilation cross
section is replaced by a weighted sum of terms describing the
annihilation of two superparticles into SM particles \cite{DM2}. We
include the ${\cal O}(v^0)$ ``$a-$''terms for co--annihilation with a
charged slepton ($\tilde{e}_R, \ \tilde{\mu}_R$ or $\tilde{\tau}_1$)
\cite{DM3} or scalar top $\tilde t_1$ \cite{DM4}. In these cases
co--annihilation can reduce the relic density by one order of
magnitude \cite{DM3} or more \cite{DM4}. If \lsp\ is higgsino--like,
we include co--annihilation with both $\tilde\chi_2^0$ and
$\tilde\chi_1^\pm$ \cite{DM5}, assuming SU(2) invariance to
estimate co--annihilation cross sections for final states with two
massive gauge bosons $V$ from $\sigma(\lsp\lsp \rightarrow V
V)$. Since LEP searches imply $\mchi > M_W$ for higgsino--like LSP, so
that $\sigma(\lsp\lsp \rightarrow W^+ W^-)$ is large, co--annihilation
in this case ``only'' reduces the relic density by a factor $\lsim
3$. In our numerical scans co--annihilation with $\tilde{\tau}_1$ is
important near the upper bound on $m_{1/2}$ for fixed $m_0$, which
comes from the requirement $m_{\tilde{\tau}_1} > \mchi$. Higgsino
co--annihilation can be important in the ``focus point'' region $m_0^2
\gg m_{1/2}^2$. Finally, co--annihilation with $\tilde t_1$ can be
important in some scenarios with large $|A_0|$. \s

-- The expansion eq.~(\ref{edm4}) breaks down near the threshold for
the production of heavy particles, where the cross section depends
very sensitively on the center--of--mass energy $\sqrt{s}$. In
particular, due to the non--vanishing kinetic energy of the
neutralinos annihilation into final states with mass exceeding twice
the LSP mass (``sub--threshold annihilation'') is possible. We use the
approximate treatment of Ref.~\cite{DM2} for annihilation into $W^+
W^-$ and $hh$ pairs, which can be important for relatively light
higgsino--like and mixed LSPs, respectively.\footnote{Since for $\mchi
\simeq M_Z$ annihilation into $W^+W^-$ pairs is already open, a
careful treatment of $ZZ$ final states is less important; note also
that far above threshold, the cross section for $W^+ W^-$ pairs is
about two times higher than that for $Z$ pairs. Similarly, we did not
encounter situations where a careful treatment of final states
containing at least one heavy Higgs boson $H, \ A$ or $H^\pm$ is
important.} The integral (\ref{edm3}) then has to be computed
numerically. In mSUGRA, these effects can be of some importance for
$m_0^2 \gg m^2_{1/2}$. Note, however, that the bound $\mchi \geq M_W$
for higgsino--like LSP implies that sub--threshold annihilation into
$W^+ W^-$ pairs is no longer an issue. \s

-- The expansion eq.~(\ref{edm4}) also fails near $s-$channel poles,
where the cross section again varies rapidly with $\sqrt{s}$. In the
MSSM this happens if twice the LSP mass is near $M_Z$, or near the
mass of one of the neutral Higgs bosons. In this case we follow the
general procedure described in Ref.~\cite{DM2}, using the numerical
treatment outlined in Ref.~\cite{DM6}. In mSUGRA, the $Z$ and $h$
poles occur at low $m_{1/2}$; most of the $Z$--boson pole region is
now excluded by chargino searches at LEP. For small and moderate $\tb$
the $A$ and $H$ poles are not accessible. However, for large $\tb$,
i.e. large bottom Yukawa coupling, {\em both} soft breaking squared
Higgs masses are reduced from their GUT--scale input values, so that
$M_A$ can become quite small \cite{DM7}. Note that the pseudoscalar
$A$ boson pole is accessible from an S--wave initial state, while the
CP--even $H$ boson pole is only accessible from the P--wave; the
contribution from $H-$exchange is therefore suppressed by a factor of
$v^2$. We will see below that the effect of resonant $A-$exchange can
be quite dramatic. \s

Recent evidence suggests \cite{DM8} that $\Omega_{\tilde\chi_1^0}
\simeq 0.3$ with $h^2 \simeq 0.5$. We define
\beq
0.1 \leq \ \om \ \leq 0.3
\eeq
as the ``cosmologically favored'' region. To be conservative we also
discuss the region of parameter space where
\beq
0.025 \leq \ \om \ \leq 0.5;
\eeq
the lower bound comes from the requirement that \lsp\ should at least
form galactic Dark Matter, and the upper bound is a very conservative
interpretation of the lower bound on the age of the Universe.

\subsection*{3.3 Results} 

Using the theoretical, experimental and cosmological constraints
discussed in the previous sections, we perform a full scan of the
$(m_{1/2}, m_0)$ plane\footnote{For each value of $\tb$ and $A_0$, we
vary $m_0$ from $10$ to 2500 GeV with a grid of 10 GeV and $m_{1/2}$
from 5 to 1250 GeV with a grid of 5 GeV. This makes 62.500 points for
each choice of $\tb$ and $A_0$. The maximal value $m_0=2.5$ TeV
corresponds to first and second generation sfermion masses
$m_{\tilde{f}} \gsim 2.5$ GeV, while $m_{1/2}=1.25$ TeV leads to
gluino masses around 2.75 TeV and squark masses above 2.3 to 2.5 TeV.}
for given values of the parameters $\tb$ and $A_0$, fixing the
higgsino parameter $\mu$ to be positive to comply with the $(g_\mu-2)$
constraint. The results are shown in Figs.~1--7, which show the
regions in the $(m_{1/2}, m_0)$ plane excluded or favored by the
various constraints discussed above. In Figs.~1--4, the SM input
parameters and the EWSB scale are as discussed in section 2.2. In
Figs.~5 and 6 we show the effects of the uncertainties of the top and
bottom quark masses, and of the residual scale uncertainty,
respectively. Fig.~7 shows the effects of the radiative corrections to
the fermion and SUSY particle masses. \s

Let us first exhibit the effects of the individual constraints on the
$(m_{1/2}, m_0)$ parameter space for $\tb=40$ and $A_0=0$; Fig.~1. The
most stringent theoretical constraint, shown in the top--left frame of
Fig.~1, is the requirement of proper symmetry breaking.  In the small
green area the pseudoscalar Higgs boson mass takes tachyonic
values. The region with tachyonic sfermion masses is indicated in dark
blue. In the yellow area the iteration to determine $|\mu|$ does not
converge to a value $\mu^2 > 0$.  The latter constraint plays an
important role and excludes, depending on the value of $\tb$ [and
$m_t, m_b, M_{\rm EWSB}$ as will be discussed later], many scenarios
with $m_0 \gg m_{1/2}$. The requirement that the LSP is indeed the
lightest neutralino rules out the region (in light blue) of small
values of $m_0$ where the less massive $\tilde{\tau}_1$ slepton is
lighter than $\tilde\chi_1^0$. Turning to the experimental constraints on
SUSY particle masses, the requirement that the lightest charginos are
heavier than $\sim 104$ GeV (brown area) extends the region of no
EWSB\footnote{For small values of $m_0$ the right--hand side of this
boundary does not depend on $m_0$; in this region, $\tilde\chi_1^\pm$ is
wino--like and its mass is approximately given by $m_{\tilde\chi_1^\pm} \sim
M_2 \sim 0.8 m_{1/2}$. For larger values of $m_0$, one enters the
``focus point" \cite{Focuspoint} region where $\tilde\chi_1^\pm$ is a
mixture of higgsino and gaugino states; for even larger values of
$m_0$, $\mu$ becomes smaller and the chargino is higgsino--like with
$m_{\tilde\chi_1^\pm} \sim |\mu|$, until one reaches the ``no EWSB'' region
where no consistent value of $\mu$ is obtained. Note that close to
this boundary, the conservative experimental bound $m_{\tilde\chi_1^\pm}
\gsim 84$ GeV would have been more appropriate, but the strip where
the constraint would have been different is very small.}, while the
requirement of heavy enough sleptons, $m_{\tilde l} \gsim 100$ GeV,
(dark area) slightly extends the region where sfermions are
tachyonic. \s

The constraint from the measurement of the $b \to s \gamma$ branching
ratio excludes only a small additional part of the parameter space
with low $m_0$ and $m_{1/2}$ values (medium grey area) leading to
light charginos and top squarks [the constraint would have been
stronger for $\mu<0$]. For our choice $\mu > 0$ mSUGRA generally
predicts this branching ratio to be smaller than in the SM. However,
we will see later that exceptions to this rule can occur for large and
negative values of $A_0$. The lightest Higgs boson mass constraint
$M_h>113$ GeV (in both the medium and the light grey areas) is only
effective if $m_0 \lsim 1$ TeV and $m_{1/2} \lsim 300$ GeV since we
are in a large $\tb$ scenario where $M_h$ can easily be sufficiently
large. Note that for the values of $\tb$ and $A_0$ used here, there
are no points, not already ruled out by the constraints on EWSB and
the SUSY particle mass bounds, which are excluded by the $\Delta \rho$
constraint, since the splitting between the top squarks remains
moderate. The CCB constraint, which is somewhat related, is also not
effective in this case, because $A_t(M_{\rm EWSB})$ remains moderate
compared to the masses of the stop eigenstates.  \s

\begin{figure}[htbp]
\vspace*{-.5cm}
\centerline{\large $\tan\beta=40\ , \ A_0 = 0 \ ,\  {\rm sign}(\mu)>0$}
\hspace*{-.7cm}{\large $m_0$}\\[-1.cm]
\begin{center}
\hspace*{-.2cm}\mbox{\epsfig{figure=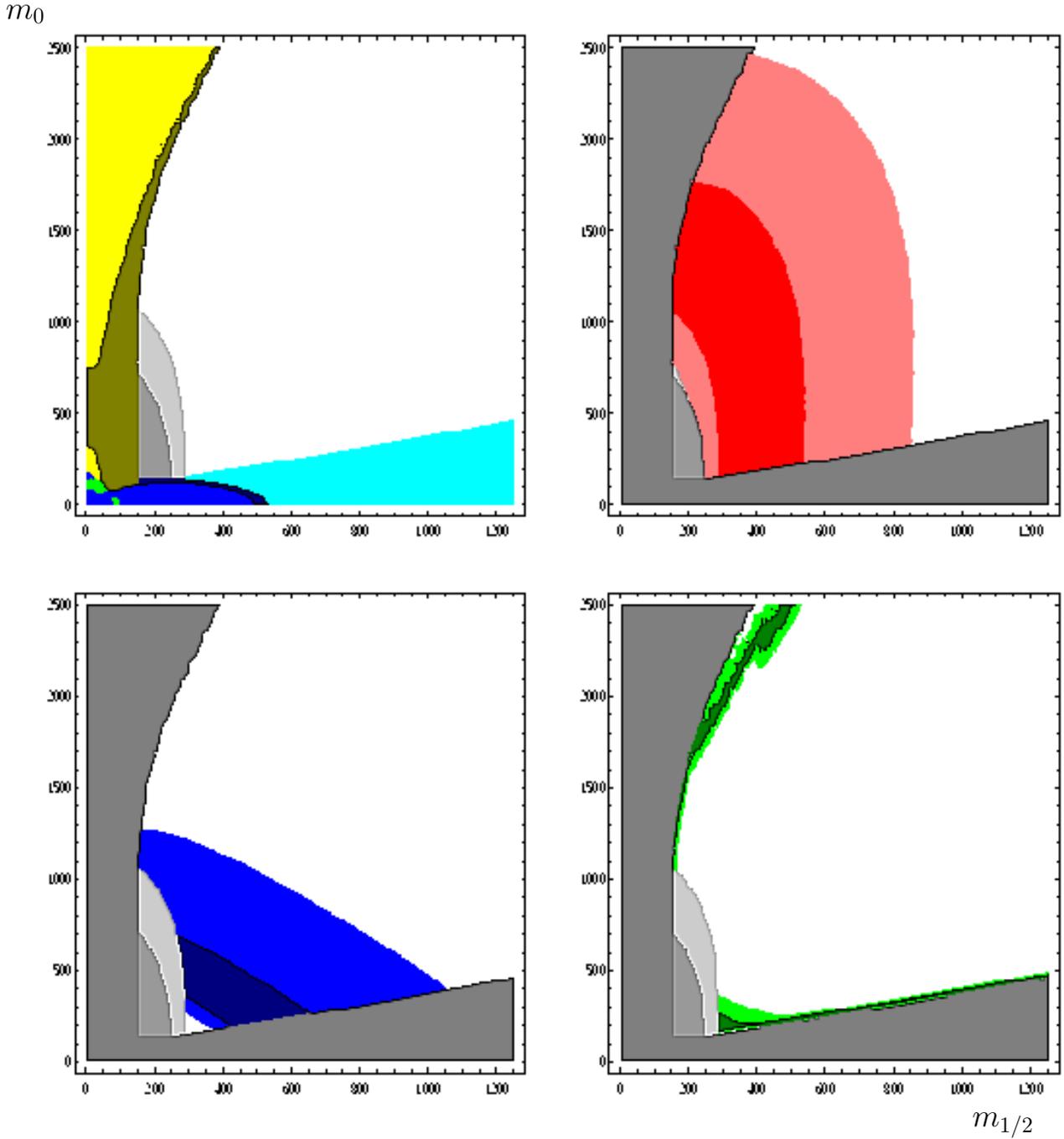,width=16.5cm}}
\end{center}
\vspace*{-.5cm}
\hspace*{14.5cm} {\large $m_{1/2}$}
\vspace*{-2mm}
\caption[]{Constraints on the $(m_{1/2}, m_0)$ mSUGRA
plane. Top--Left: individual constraints from non--convergent $\mu$
(yellow region), tachyonic $M_A$ (green), tachyonic sfermions (blue),
light sfermions (dark), light charginos (brown), $\tilde\chi_1^0$
non--LSP (light blue), BR($b \to s \gamma)$ (medium grey) and light
$h$ boson (light and medium grey). The three other plots are for the
$1\sigma$ (dark colors) and $2\sigma$ (light colors) ``evidence" for,
the Higgs boson (but with larger error bars, Top--Right),
$(g_\mu-2)$ (Bottom--Left) and the Dark Matter (Bottom--Right).}
\end{figure}

Let us now come to the positive indications for SUSY which are shown
in colors in the three other panels of Fig.~1. The ``Obelix menhir"
(dark red) in the top--right frame corresponds to the region where the
lightest $h$ boson mass lies in the range 113 GeV $ \leq M_h \leq 117$
GeV. It extends from $m_{1/2}$ values below 550 GeV for $m_0 \lsim 1$
TeV down to $m_{1/2} \sim 300$--$200$ GeV for larger $m_0$, with upper
contour at $m_0=1.7$ TeV. For larger values of $m_0$ and $m_{1/2}$ the
top squarks are very heavy and push $M_h$ beyond the 117 GeV
limit. One notices that the $M_h$ constraint is satisfied in a large
region of mSUGRA parameter space, since a variation of the $h$ boson
mass of a few GeV leads to a variation of $m_{1/2}$ of several hundred
GeV. This is due to the logarithmic dependence on the stop masses,
which in turn are mainly driven by $m_{1/2}$. Since the theoretical
error on $M_h$ [from higher order loop corrections, as well as from a
shift of $M_t$ by a few GeV within the experimental error, etc..] are
expected to be of the order of a few GeV, we display for illustration
the effect of including an additional uncertainty of $\pm 2$ GeV on
the $h$ boson mass which varies then in range 111 GeV $ \leq M_h \leq
119$ GeV (light red area). As can be seen, the impact is very large
and values $m_0 \sim 2$ TeV and $m_{1/2} \lsim 0.8$ TeV could be
reached [this point has also been realized recently, see
Ref.~\cite{Heiglein} for instance]. Some caution is therefore needed
when analyzing the consequences of a ``115 GeV $h$ boson'' for the
allowed mSUGRA parameter space. \s

In the blue regions of the bottom--left frame of Fig.~1 the
contribution of SUSY particles to the anomalous magnetic moment of the
muon is within two standard deviations (light blue) and one standard
deviation
(dark blue) from the central value of the measurement made by the
Brookhaven experiment, eq.~(\ref{g-2}). The $2\sigma$ area extends
from values $m_0 \lsim 1.2$ TeV for small $m_{1/2}$ to the boundary
where the neutralino $\tilde\chi_1^0$ is not the LSP for large values,
$m_{1/2} \sim 1$ TeV, except for a little corner with $m_{1/2}
\sim m_0 \sim 250$ GeV, where the SUSY contribution exceeds the
$2\sigma$ upper bound. In this area, charginos and smuons have
relatively small masses and can give too large a contribution to
$(g_\mu-2)$. The $1\sigma$ constraint is significantly more severe:
a large amount of the upper part of the $2\sigma$ area [where
charginos/smuons are too heavy to contribute] and a smaller area of
the lower part [where charginos and smuons are too light and generate
too large a contribution] are cut away. \s

Finally, the light green bands in the bottom--right frame correspond
to the regions where the LSP neutralino cosmological relic density is
in the required range, $0.025 \leq \om \leq 0.5$. The narrow band
slightly above the non \lsp \ LSP boundary is the region where
$\tilde{\tau}_1$ is almost degenerate with the neutralino $\tilde\chi_1^0$,
and $\tilde{\tau}_1$\lsp~ as well as $\tilde{\tau}_1 \tilde{\tau}_1$
co--annihilation is efficient enough to reduce the relic density. The
region below $m_{1/2} \lsim 400$ GeV and small $m_0$ is the bino--like
LSP region, where both the LSP and the $\tilde{\tau}_1$ are light
enough for the annihilation \lsp \lsp $\to \tau^+ \tau^-$ cross
section, through $t$--channel $\tilde{\tau}_1$ exchange, to be
sizeable. The area near the EWSB boundary is again the focus--point
region where the \lsp\ has a significant higgsino component, i.e. has
sizeable couplings to massive gauge bosons, making the annihilation
into $WW,ZZ$ efficient. Very close to the ``no EWSB'' region the \lsp\
is higgsino--like and is therefore almost degenerate with the lightest
chargino and next--to--lightest neutralino, in which case higgsino
co--annihilation takes place. The relic density can then even fall
below its lower bound. Note that we did not yet reach the regime
where the $s$--channel pseudoscalar Higgs boson poles play an
important role, although we have a relatively large value of $\tb$. We
also show the regions where the cosmological relic density is more
constrained, $0.1 \lsim \om \lsim 0.3$ (dark green). The areas become
narrower, in particular the bino--like \lsp~and the focus point
regions, but there is no qualitative change from what was discussed
above.  \smallskip

\begin{figure}[htbp]
\centerline{\large $\tan\beta=40\ , \ A_0 = 0 \ ,\  {\rm sign}(\mu)>0$}
\hspace*{-.4cm}{\large $m_0$}\\[-1.5cm]
\begin{center}
\hspace*{-.2cm} \epsfig{figure=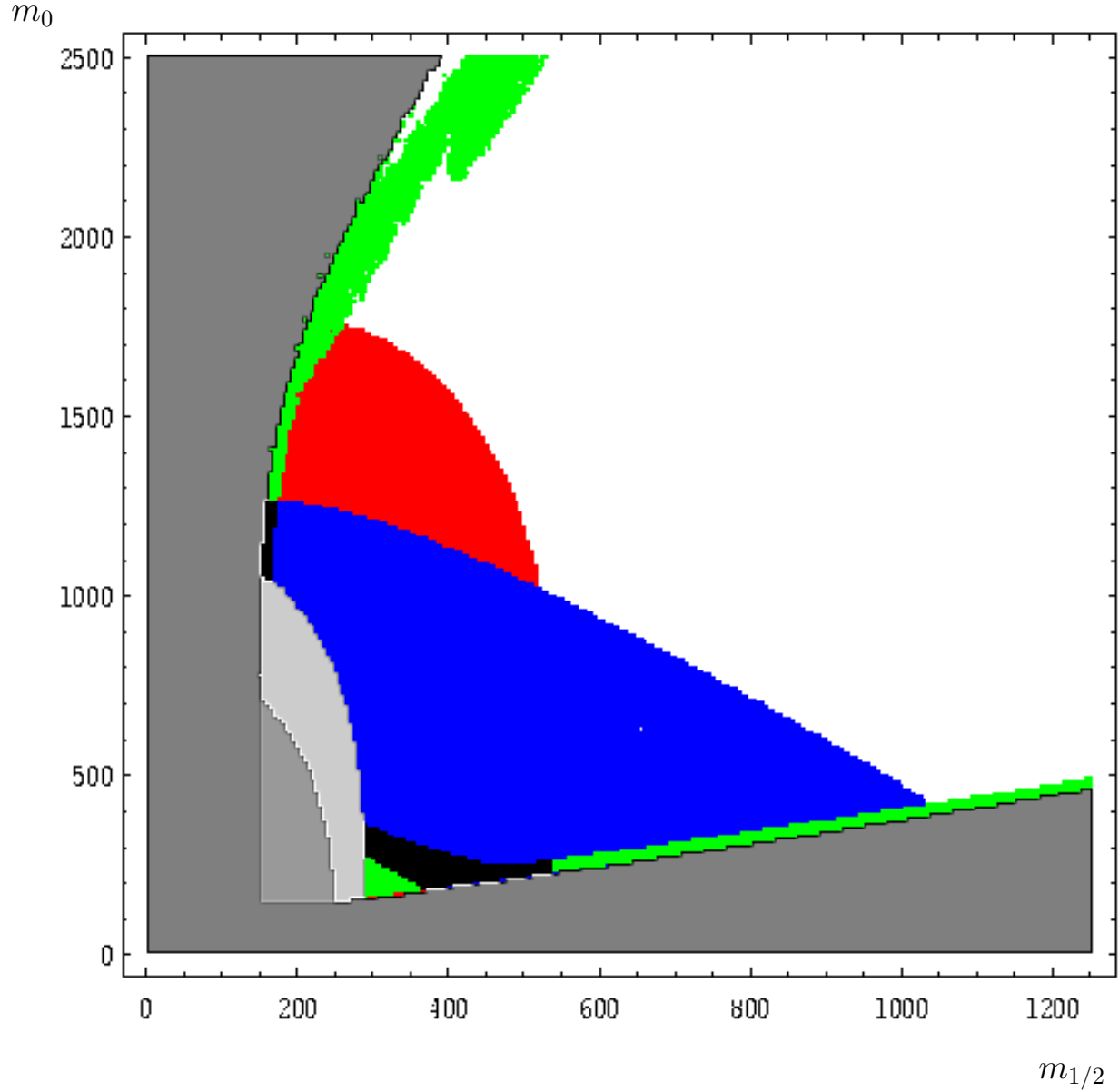,width=15cm,angle=-90}\\
\end{center}
\hspace*{14.5cm} {\large $m_{1/2}$}
\vspace*{-2mm}
\caption[]{Constraints on the $(m_{1/2}, m_0)$ mSUGRA plane for $\tan
\beta = 40$, $A_0=0$ and sign$(\mu)>0$. The grey areas are those
excluded by the requirement of EWSB breaking and limits on SUSY
particle masses (darker grey), BR($b \to s \gamma)$ (medium grey) and
$M_h > 113$ GeV (light and medium grey). The colors are for the
``evidence" for the Higgs boson (red), the $(g_\mu-2)$ (blue) and Dark
Matter (green).}
\end{figure}

Fig.~2 summarizes the situation when all the constraints as well as
the $2 \sigma$ ``evidence" from the LEP2 lightest $h$ boson, the
$(g_\mu-2)$ deviation and the neutralino cosmological relic density
are superimposed. [Note that some of these colored areas overlap as
can be inferred from Figs.~1; we refrained from allocating different
colors for these common regions, since one can deduce them by
continuing the boundaries up to the dark region which is the
intersection of all three areas]. As can be seen, the region excluded
by theoretical and experimental considerations is still relatively
modest. There are large areas of the parameter space where one can
accommodate a $\sim 115$ GeV $h$ boson and a SUSY explanation of the
$(g_\mu-2)$ deviation. On the other hand, the area where the
neutralino LSP is a good Dark Matter candidate is fairly small for
this value of $\tb$. The areas (in black) where all of the three
requirements are met are rather tiny and include only a part of the
region with a light bino LSP neutralino and a very small part of the
focus point region, the remaining pieces being removed by the
$(g_\mu-2)$ or Higgs boson constraints. Using the more constrained
scenario with a narrower range for \om \ and $1 \sigma$ errors for
$(g_\mu-2)$, would have collapsed the overlap region to a narrow strip
in the $\lsp - \tilde \tau_1$ co--annihilation region, with 420 GeV
$\leq m_{1/2} \leq$ 520 GeV. The lower bound on $m_{1/2}$ then comes
from the upper bound on $(g_\mu - 2)$, and the upper bound on
$m_{1/2}$ results from the upper bound on $M_h$. \s

\begin{figure}[htbp]
\noindent \hspace*{.9cm} $\tan\beta=5, \, A_0 = -1\, {\rm TeV}, \,  
{\rm sign}(\mu)>0$ \hspace*{1.4cm}
$\tan\beta=10, \, A_0 = 0, \,  {\rm sign}(\mu)>0$\\
\hspace*{-.7cm}{\large $m_0$}\\[16cm]
\hspace*{16.5cm} {\large $m_{1/2}$}\\[-17.7cm]
\begin{center}
\hspace*{-.2cm}\mbox{\epsfig{figure=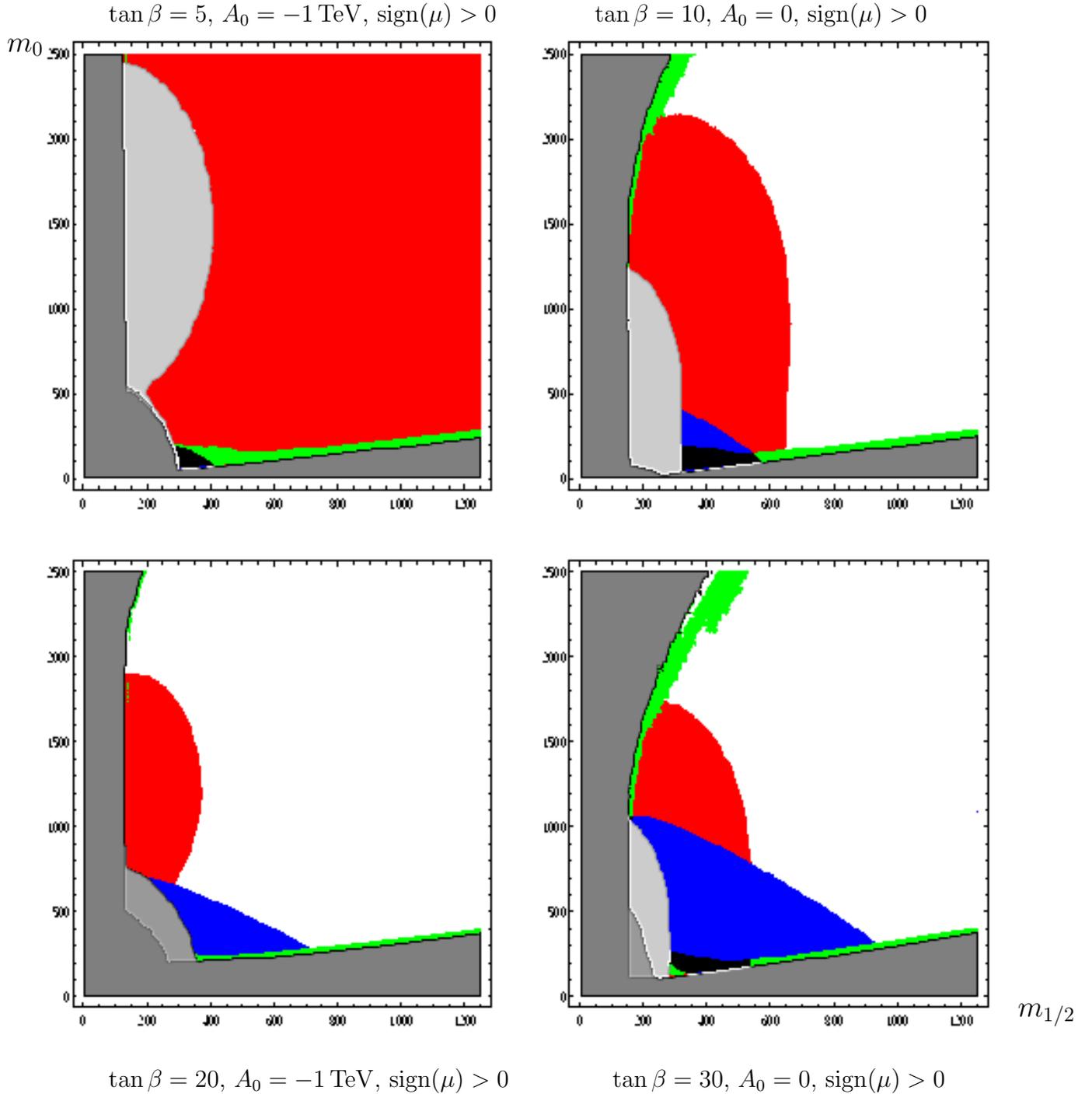,width=16.5cm}}\\[-.8cm]
\end{center}
\hspace*{.9cm} 
$\tan\beta=20, \, A_0 = -1\, {\rm TeV}, \,  {\rm sign}(\mu)>0$ 
\hspace*{1.5cm}
$\tan\beta=30, \, A_0 = 0 , \,  {\rm sign}(\mu)>0$\\
\caption[]{Constraints on the $(m_{1/2}, m_0)$ mSUGRA plane for values of 
$\tan \beta <40$, $A_0=0$ or -1 TeV and sign$(\mu)>0$. The grey areas 
are those excluded by the requirement of EWSB breaking and limits on SUSY 
particle masses (darker grey), BR($b \to s \gamma)$ (medium grey) and 
$M_h > 113$ GeV (light and medium grey). The colors are for the ``evidence" for
the Higgs boson (red), $(g_\mu-2)$ (blue) and Dark Matter (green).}
\end{figure}

\setcounter{figure}{3}
\renewcommand{\thefigure}{\arabic{figure}a}
\begin{figure}[htbp]
\noindent \hspace*{.9cm} $\tan\beta=45, \, A_0 = -1\, {\rm TeV}, \,  
{\rm sign}(\mu)>0$ \hspace*{1.5cm}
$\tan\beta=50, \, A_0 = 0, \,  {\rm sign}(\mu)>0$\\
\hspace*{-.7cm}{\large $m_0$}\\[16cm]
\hspace*{16.5cm} {\large $m_{1/2}$}\\[-17.7cm]
\begin{center}
\hspace*{-.2cm}\mbox{\epsfig{figure=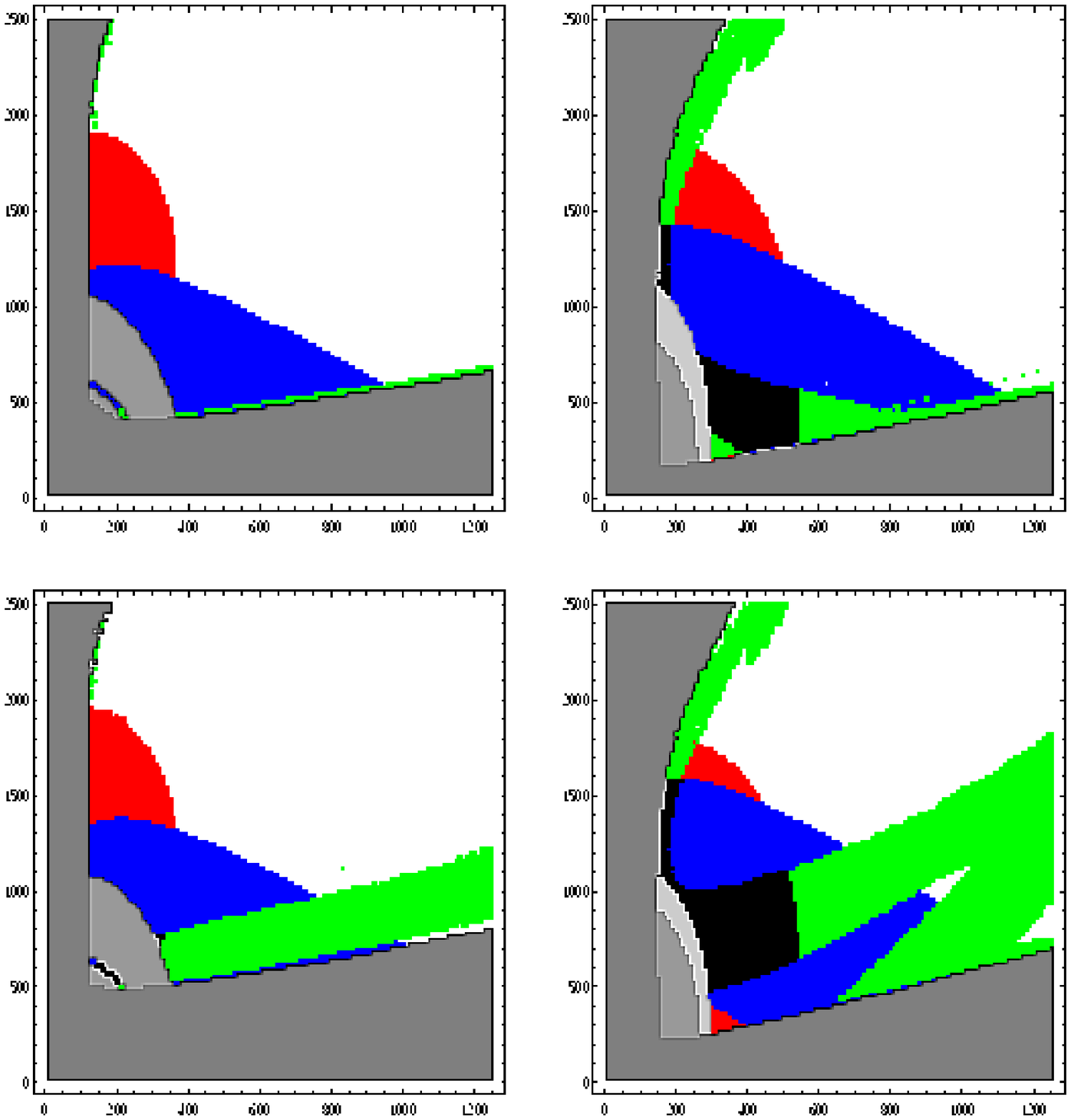,width=16.5cm}}
\end{center}
\noindent \hspace*{.9cm} 
$\tan\beta=55, \, A_0 = -1\, {\rm TeV}, \,  {\rm sign}(\mu)>0$ 
\hspace*{1.5cm}
$\tan\beta=60, \, A_0 = 0 , \,  {\rm sign}(\mu)>0$\\[.3cm]
\caption[]{Constraints on the $(m_{1/2}, m_0)$ mSUGRA plane for values of 
$\tan \beta >40$, $A_0=0$ or -1 TeV and sign$(\mu)>0$. The grey areas 
are those excluded by the requirement of EWSB breaking and limits on SUSY 
particle masses (darker grey), BR($b \to s \gamma)$ (medium grey) and 
$M_h > 113$ GeV (light and dark grey). The colors are for the ``evidence" for
the Higgs boson (red), and the $(g_\mu-2)$ (blue) 
and Dark Matter (green).}
\end{figure}

\setcounter{figure}{3}
\renewcommand{\thefigure}{\arabic{figure}b}
\begin{figure}[htbp]
\noindent \hspace*{.8cm} $\tan\beta=45, \, A_0 = -1\, {\rm TeV}, \,  
{\rm sign}(\mu)>0$ \hspace*{1.5cm}
$\tan\beta=50, \, A_0 = 0, \,  {\rm sign}(\mu)>0$\\
\hspace*{-.7cm}{\large $m_0$}\\[16cm]
\hspace*{16.5cm} {\large $m_{1/2}$}\\[-17.7cm]
\begin{center}
\hspace*{-.2cm}\mbox{\epsfig{figure=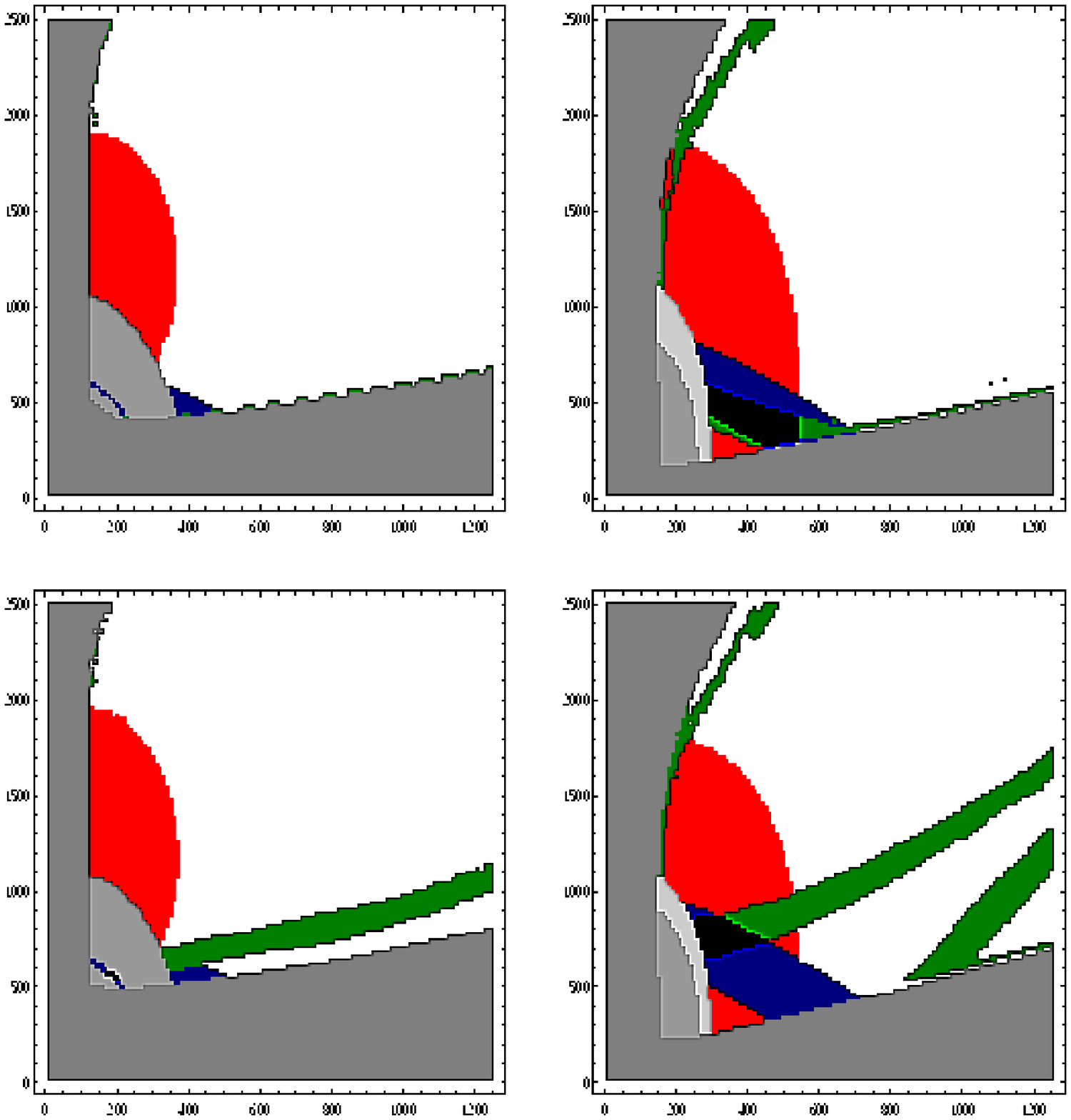,width=16.5cm}}
\end{center}
\noindent \hspace*{.8cm} 
$\tan\beta=55, \, A_0 = -1\, {\rm TeV}, \,  {\rm sign}(\mu)>0$ 
\hspace*{1.5cm}
$\tan\beta=60, \, A_0 = 0 , \,  {\rm sign}(\mu)>0$\\[.3cm]
\caption[]{As Fig.~4a, except that we have used the $1\sigma$ range
for $(g_\mu-2)$ and the tighter constraint (31) on \om.}
\end{figure}
\setcounter{figure}{4}

In Figs.~3 and 4 we show the same $(m_0, m_{1/2})$ plane for smaller
and larger values of $\tb$, respectively, choosing $A_0 =0$ or $-1$
TeV. The most striking features are as follows: \s

(i) The region where EWSB does not take place decreases with
decreasing $\tb$. In fact, for $\tb \sim 5$ there is no EWSB problem,
and the lower limit on $m_{1/2}$ comes from the requirement
$m_{\tilde\chi_1^\pm} \simeq 0.75 m_{1/2} \gsim 104$ GeV. In contrast, for
$\tb \gsim 50$ the EWSB constraints rules out a larger part of the
parameter space. For very large values, $\tb > 60$, the requirement of
EWSB becomes extremely strong and most of the parameter space is ruled
out. The ``no EWSB'' region is controlled in part by the top and
$b$--quark Yukawa couplings since the RG evolution of $M_{H_u}$ and
$M_{H_d}$ is very sensitive to modifications of $\lambda_t$ and
$\lambda_b$, respectively. Note that EWSB with $\tb > 1$ requires
$m^2_{H_u} < m^2_{H_d}$ at the weak scale, which in turn requires
$\lambda_b \lsim \lambda_t$, leading to an upper bound on $\tb$. The
SUSY radiative corrections to $m_b$ play an important role here. For
positive values of $\mu$, the radiatively corrected $b$--mass, and thus
$\lambda_b(M_Z)$, is smaller and the problem is postponed to larger
$\tb$. For $\mu<0$, the situation worsens and one reaches the
no--EWSB regime for smaller values of $\tb$. A resummation of these
corrections, as performed in our analysis \cite{viennese}, is
therefore crucial for an accurate treatment of scenarios with very
large $\tb$. The no--EWSB region is also smaller for sizable
(and negative) values of the mixing parameter $A_0$ which enters the
RGEs of the Higgs boson masses, the stop/sbottom masses and the SUSY
radiative corrections to $m_b$ and $m_t$. \s

(ii) The region where the \lsp\ is not the LSP decreases with smaller
$\tb$, since the mixing in the $\tilde{\tau}$ sector is proportional
to $\mu \tb$, so that the splitting between the two stau masses
$m_{\tilde{\tau}_1}$ and $m_{\tilde{\tau}_2}$ becomes smaller. The net
effect is that $\tilde{\tau}_1$ becomes heavier, for given values of
$m_0$ and $m_{1/2}$. On the other hand, taking $A_0 = -1$ TeV
increases the value of $|\mu|$ as determined by EWSB, due to the same
RGE effect that decreases the ``no EWSB'' region. This increases
$\tilde \tau$ mixing and thus reduces $m_{\tilde \tau_1}$, thereby
extending the region where \lsp\ is not the LSP.\s

(iii) The $b \to s\gamma$ constraint is more restrictive for larger
$\tb$ values, although it is still not very constraining in the region
of parameter space favored by the ``115 GeV Higgs boson'' and the
deviation in $(g_\mu - 2)$. It also becomes stronger\footnote{Note,
however, the narrow blue strips at small $m_0$ and small $m_{1/2}$ in
the left panels in Figs.~4. Here $b \ra s \gamma$ decays are
``accidentally'' suppressed by cancellations between various new
contributions. In the region below this strip the predicted branching
ratio falls above the upper bound of eq.(22), while in the excluded
region above the strip the predicted branching ratio is too small.}
when $A_0$ is sizable (and negative), since this reduces
$m_{\tilde{t}_1}$. For example, in the bottom--left panel in Fig.~3
the $b \ra s \gamma$ constraint excludes the entire region capable of
explaining the $(g_\mu - 2)$ anomaly, so that no overlap region exists
even at the 2 $\sigma$ level. The $\Delta \rho$ [and CCB] constraint
also becomes more relevant for the small $\tb$ and large $-A_0$, again
due to the possibly large mass splitting between the two top
squarks. However, the region where $\Delta \rho$ is too large is
already ruled out by the $b \to s\gamma$ [and $M_h$] constraint. \s

(iv) For small $\tb$ one needs a large mixing in the stop sector [and
thus a sizable and negative $A_0$] to maximize the $h$ boson mass and
to satisfy the bound $M_h>113$ GeV. This is true, for example, for
$\tb = 5$ [top--left frame in Fig.~3]. Even with large stop mixing,
$M_h$ cannot exceed 117 GeV, so that the whole parameter space is
filled by the $M_h \sim 115 \pm 2$ GeV ``evidence". For increasing
$\tb$ [which leads to an increase of $M_h$], these areas become
smaller and smaller, until one reaches values $\tb \gsim 20$ where the
maximal value of $M_h$ reaches a plateau, as is well known. Note that
the mixing changes the form of the red area since $M_h$ is maximal for
$|A_t| \simeq \sqrt{6 m_{\tilde{t}_1}m_{\tilde{t}_2}}$ [the so--called
maximal mixing scenario]; for a fixed value of $A_t$, $M_h$ decreases
for larger or smaller stop masses which are controlled by $m_0$ and
$m_{1/2}$. It is curious to note that for $\tb = 5$, $A_0 < 0$ is
crucial for allowing a small overlap region where all positive
``indications'' can be explained [at the $2\sigma$ level, at least],
whereas for $\tb = 20$ taking $A_0 = -1$ TeV removes the overlap
region that is present for $A_0 = 0$; as noted in (ii), the constraint
that \lsp\ is the LSP also plays an important role here.\s

(v) The area where the contribution of SUSY particles allows an
explanation of the deviation of the $(g_\mu-2)$ from the Brookhaven
result is also very sensitive to the value of $\tb$. For $\tb \sim 5$
and $A_0$ near zero, the contribution of chargino and smuon loops is
not sufficient, except in regions excluded by the other
constraints. [We stress again that no point for $\mu <0$ can comply
with this constraint]. For larger values of $\tb$, the $(g_\mu-2)$
domain becomes larger and for $\tb \gsim 40$, values of $m_{1/2}$ or
$m_0$ in excess of 1 TeV are still compatible [within $2 \sigma$]
with the central Brookhaven result. We also note that
$(g_\mu -2)$ is less sensitive to $A_0$ than the $b \ra s \gamma$
constraint is. In the former case the sensitivity is only due to the
increased value of $|\mu|$ required by EWSB, while the latter constraint
is directly sensitive to $A_t$ through the $\tilde t$ mass matrix, as
explained above. Nevertheless changing $A_0$ from 0 to $-1$ TeV
significantly reduces the values of $m_0$ and $m_{1/2}$ required to
explain the deviation in $(g_\mu-2)$, in order to compensate for the
increase of $|\mu|$.\s

(vi) The region where the \lsp \ LSP is required to make the Dark
Matter in the universe is also very sensitive to the values of $\tb$
and $A_0$. For small $\tb$, the regions with light bino--like LSP and
the mixed gaugino--higgsino LSP [focus--point] are smaller; the latter
is even absent for $\tb \sim 5$, or for $A_0 = -1$ TeV. If $\tb \gsim
20$, this latter choice of $A_0$ also brings the light bino LSP--like
region in conflict with the $b \to s \gamma$ constraint, which becomes
more severe when $A_0$ is reduced below 0, as discussed in
(iii). [Note that for $\tb \sim 5$, there is a region where the
required value of \om\ is attained due to $\lsp - \tilde t_1$
co--annihilation; however, this region is excluded by the $h$ boson
mass constraint.] For very large $\tb$ values, the bino--like region
becomes much wider, due to reduced $\tilde \tau_1$ mass and larger
Higgs exchange contributions. Moreover, the mixed or higgsino--like
region increases in size, since the focus--point scenario is easier to
realize at large $\tb$. For $\tb \gsim 50$, there are additional
domains where \om~is in the interesting range, the regions near the
pseudoscalar $A$ boson or scalar $H$ boson $s$--channel poles. As
discussed previously, for $\tb \gg 1$, $M_A$ [and thus also $M_H$]
become smaller, and their Yukawa couplings to $b$ quarks and $\tau$
leptons [proportional to $\tb$] are strongly enhanced. The resulting
large $\tilde\chi_1^0 \tilde\chi_1^0 \to b \bar{b}, \tau^+ \tau^-$
annihilation cross sections reduce the relic density to the required
level. Note that with our treatment of the QCD correction to the
bottom Yukawa coupling, these Higgs pole regions open up only for
values $\tb \gsim 50$ in agreement with the recent analyses performed
in Ref.~\cite{Leszek}. Moreover, the corrections to the physical Higgs
masses are of some importance here. The running $A$ and $H$ masses are
very close to each other in this region of parameter space, but these
corrections decrease $M_A$ and increase $M_H$\footnote{Note that $A$
cannot couple to two equal squarks, e.g. $\tilde t_1$ or $\tilde b_1$
pairs, while $H$ does have such couplings.}, leading to a mass
splitting of 10\% or more for $\tb = 60$. This is larger than the
width of these Higgs bosons, which amounts to typically 4\% of their
mass. $H-$exchange still does not lead to a separate favored region in
our scans, due to the $P-$wave suppression, but the large Higgs mass
splitting increases the width of the cosmologically favored region. As
a result, even the $1 \sigma$ overlap region becomes quite sizable for
$\tb \gsim 50$ and $A_0 = 0$, as shown in the right panels in
Fig.~4b. \s

We now discuss, for the choice $\tb = 40$ and $A_0=0$, the effect of
using different top and bottom quark mass input values, approximately
$1~\sigma$ and $2\sigma$ higher or lower than the central experimental
values, $M_t \sim 175 \pm 5$ GeV and $\overline{m}_b (\overline{m}_b)
=4.25 \pm 0.125$ GeV; see Fig.~5. For smaller top quark mass,
$M_t=170$ GeV, the region where EWSB does not occur becomes much
larger. The $M_h$ constraint becomes also much more severe; one needs
significantly heavier stops, and hence larger values of $m_0$ and/or
$m_{1/2}$, to obtain a sufficiently large value of $M_h$. The $(g_\mu-2)$
domain remains almost the same, since in the relevant region of
parameter space the value of $|\mu|$ required by EWSB is only slightly
reduced by this reduction of the top mass. On the other hand, the DM
region, and in particular the mixed higgsino--gaugino region, 
becomes wider. For larger $M_t$ the trend is reversed: the region
where EWSB does not occur and the one with large higgsinos--gaugino
mixing almost disappear. The Higgs mass constraints become less
severe, while the domain where 113 GeV $\leq M_h \leq$ 117 GeV is
reduced. \s

\renewcommand{\thefigure}{\arabic{figure}}
\begin{figure}[htbp]
\noindent \hspace*{.6cm}
$M_t=170  \, {\rm GeV} \ , \ \bar{m}_b(\bar{m}_b)=4.24  \, {\rm GeV}$ 
\hspace*{1.2cm}
$M_t=180 \, {\rm GeV} \ , \  \bar{m}_b(\bar{m}_b)=4.24  \, {\rm GeV}$\\
\hspace*{-.7cm}{\large $m_0$}\\[16cm]
\hspace*{16.5cm} {\large $m_{1/2}$}\\[-17.7cm]
\begin{center}
\hspace*{-.2cm}\mbox{\epsfig{figure=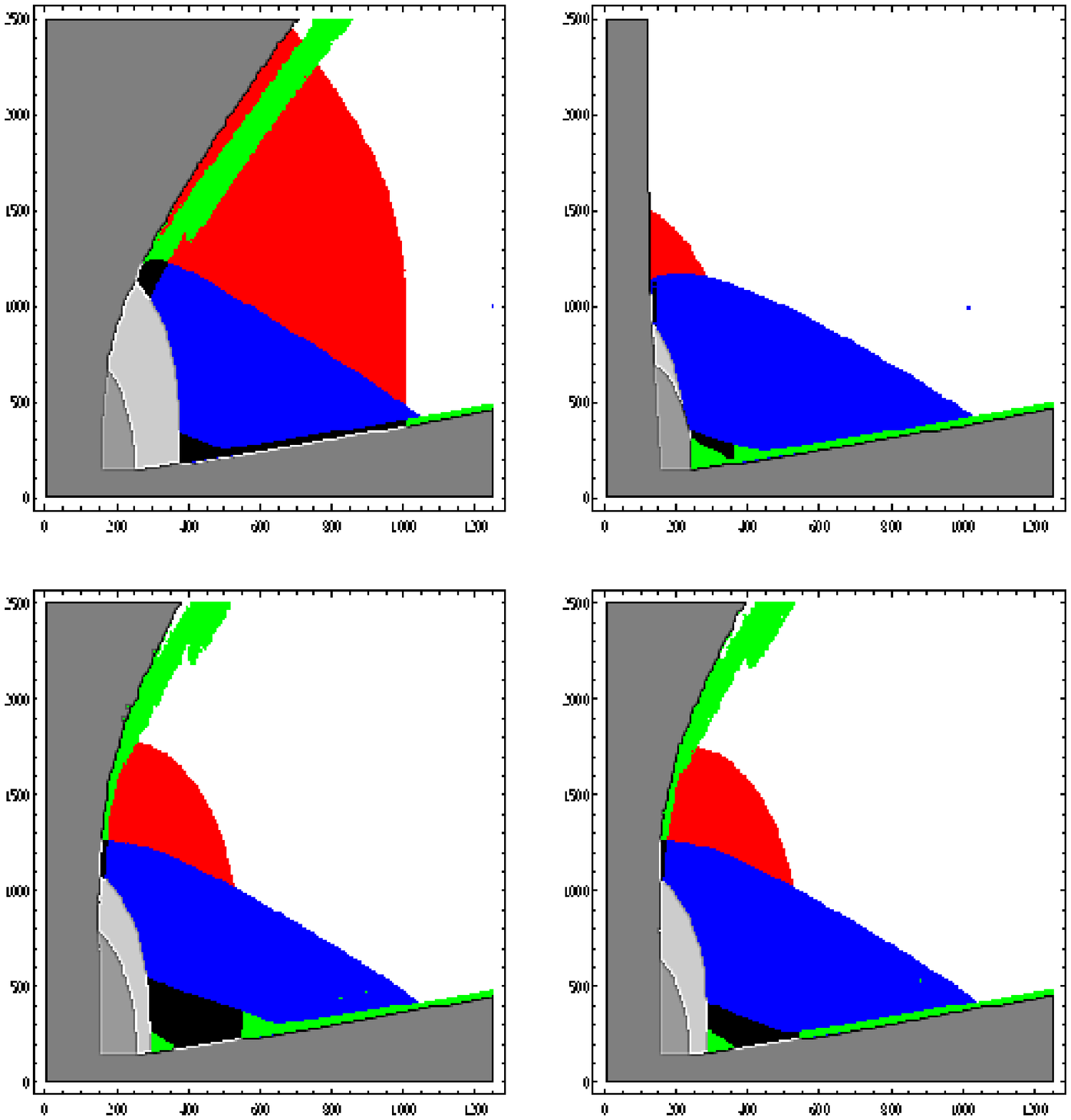,width=16.75cm,angle=-90}}
\end{center}
\noindent \hspace*{.6cm} 
$M_t=174.3 \, {\rm GeV}, \  \bar{m}_b(\bar{m}_b)=4.5  \, {\rm GeV}$ 
\hspace*{1.2cm}
$M_t=174.3 \, {\rm GeV}, \, \bar{m}_b(\bar{m}_b)=4.0  \, {\rm GeV}$\\[.3cm]
\caption[]{Constraints on the $(m_{1/2}, m_0)$ mSUGRA plane for $\tan \beta
=40$, $A_0=0$, sign$(\mu)>0$ and different values of the pole top quark mass
$M_t$ and $\overline{\rm MS}$ bottom quark mass. The notation is as in
Fig.~2.} 
\end{figure}
%
\begin{figure}[htbp]
\hspace*{1.3cm} $M_{\rm EWSB}= 0.5\sqrt{m_{\tilde{t}_1} m_{\tilde{t}_2} }$ 
\hspace*{4.2cm} $M_{\rm EWSB}=2\sqrt{m_{\tilde{t}_1}m_{\tilde{t}_2}}$\\
\hspace*{-.9cm}{\large $m_0$}\\[7.8cm]
\hspace*{16.5cm} {\large $m_{1/2}$}\\[-10.cm]
\begin{center}
\hspace*{-.2cm}\mbox{\epsfig{figure=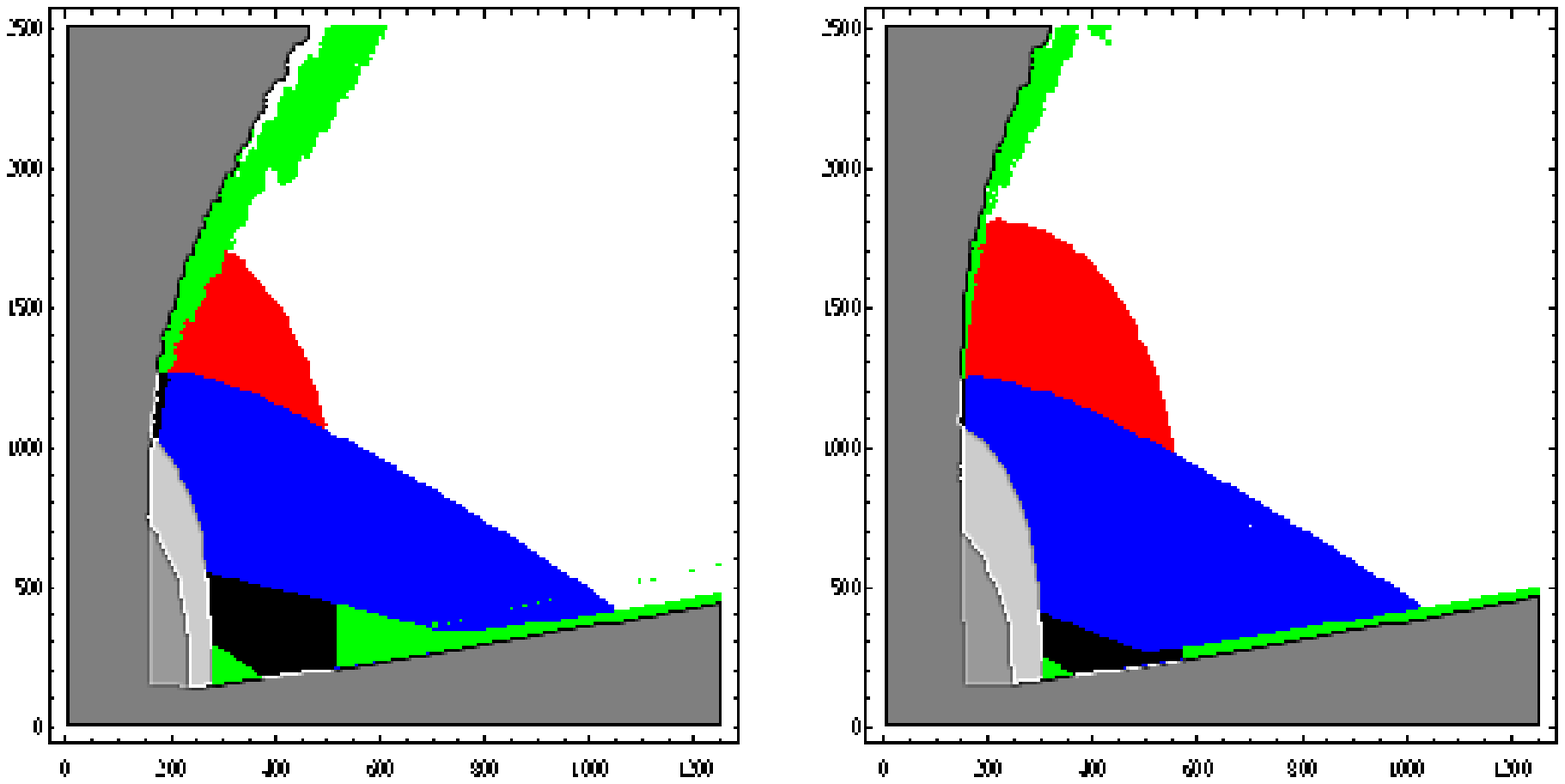,width=16.75cm}}\\[-1.8cm]
\end{center}
\vspace*{-.7cm}
\caption[]{Constraints on the $(m_{1/2}, m_0)$ mSUGRA plane for $\tan \beta
=40$, $A_0=0$, sign$(\mu)>0$ with different choices of the EWSB
scale. The notation is as in Fig.~2}
\end{figure}

Since for the features related to EWSB an increase of $m_b$ is more or
less equivalent to an increase of $\tb$, our results for $\bar{m}_b=4$
GeV and $\bar{m}_b=4.5$ GeV look similar to those where the central
value $\overline{m}_b (\overline{m}_b) \sim 4.25$ GeV is kept, but
with an input value $\tb \sim 37$ and $\sim 44$, respectively. One
sees that in this case, the changes are rather modest. These modest
changes however conspire to produce a significant reduction of the
overlap region when $\overline{m}_b (\overline{m}_b)$ is reduced from
from 4.5 to 4.0 GeV (bottom panels in Fig.~5). The effect of varying
$m_b$ would have been more striking for large $\tb \gsim 50$, where
EWSB and the cosmological relic density become much more sensitive to
the $b$--Yukawa coupling. For instance, the regime where it is
difficult to realize EWSB is reached for lower values of $\tb$, $\tb
\sim 55$, if $\overline{m}_b (\overline{m}_b) \sim 4.5$ GeV is used as
an input. \s

There is a residual scale dependence in our treatment of the MSSM
scalar potential, since only the one--loop corrections are fully
included. A standard RG improvement of the effective potential does
not completelly remove the scale dependence, due to the presence of
several a priori unrelated scales.~\footnote{More sophisticated
attempts to further reduce this residual scale dependence have been
proposed, see e.g. ref. \cite{CaCleQui}.} The effect of varying the
scale at which EWSB is realized [i.e. the scale at which the effective
one--loop scalar potential is evaluated and the running of the soft
SUSY breaking terms is frozen] by a factor of 2 in either direction is
displayed in Fig.~6. Here again, except for a small change in the
shape of the $M_h =115 \pm 2$ GeV domain, the most striking change
occurs for the area where the cosmological relic density is in the
interesting range. Increasing the EWSB scale to $M_{\rm EWSB}= 2
\sqrt{m_{\tilde{t}_1} m_{\tilde{t}_2}}$ (right panel) increases the
predicted value of $|\mu|$. This reduces the size of the
cosmologically favored mixed higgsino--bino region, but leaves the
light bino region largely unaffected.  On the other hand, the choice
$M_{\rm EWSB}= 0.5 \sqrt{m_{\tilde{t}_1} m_{\tilde{t}_2}}$ (left
panel) leads to a significant reduction of the predicted value of
$|\mu|$, and a corresponding decrease of the ``no EWSB'' area as well
as the cosmologically preferred area where the LSP has a significant
higgsino component. Moreover, the reduction of $|\mu|$ leads to a
reduction of the masses of the heavy Higgs bosons. This significantly
increases the cosmologically preferred region where the LSP is
bino--like, and hence also increases the overlap region, as can be
seen by comparing the left panel in Fig.~6 with Fig.~2. In addition a
new feature occurs: the opening of the region where the $s$--channel
pseudoscalar $A$ boson pole starts to play a role in the LSP
annihilation cross section. For the present choice of the scale [and
$\tb$] this region is still rather tiny, a small line parallel to the
co--annihilation region. If the scale is reduced to much lower values,
e.g. to $M_{\rm EWSB}=M_t$ or $M_Z$, this area would have been much
more sizeable. However, in this case, the requirement of proper
electroweak symmetry breaking will exclude large portions of the
parameter space. \s

\begin{figure}[htbp]
\hspace*{1cm} No SUSY RC to sparticle masses
\hspace*{2.cm}
No SUSY RC to (s)particle masses.\\
\hspace*{-.9cm}{\large $m_0$}\\[7.8cm]
\hspace*{16.5cm} {\large $m_{1/2}$}\\[-10.cm]
\begin{center}
\hspace*{-.2cm}\mbox{\epsfig{figure=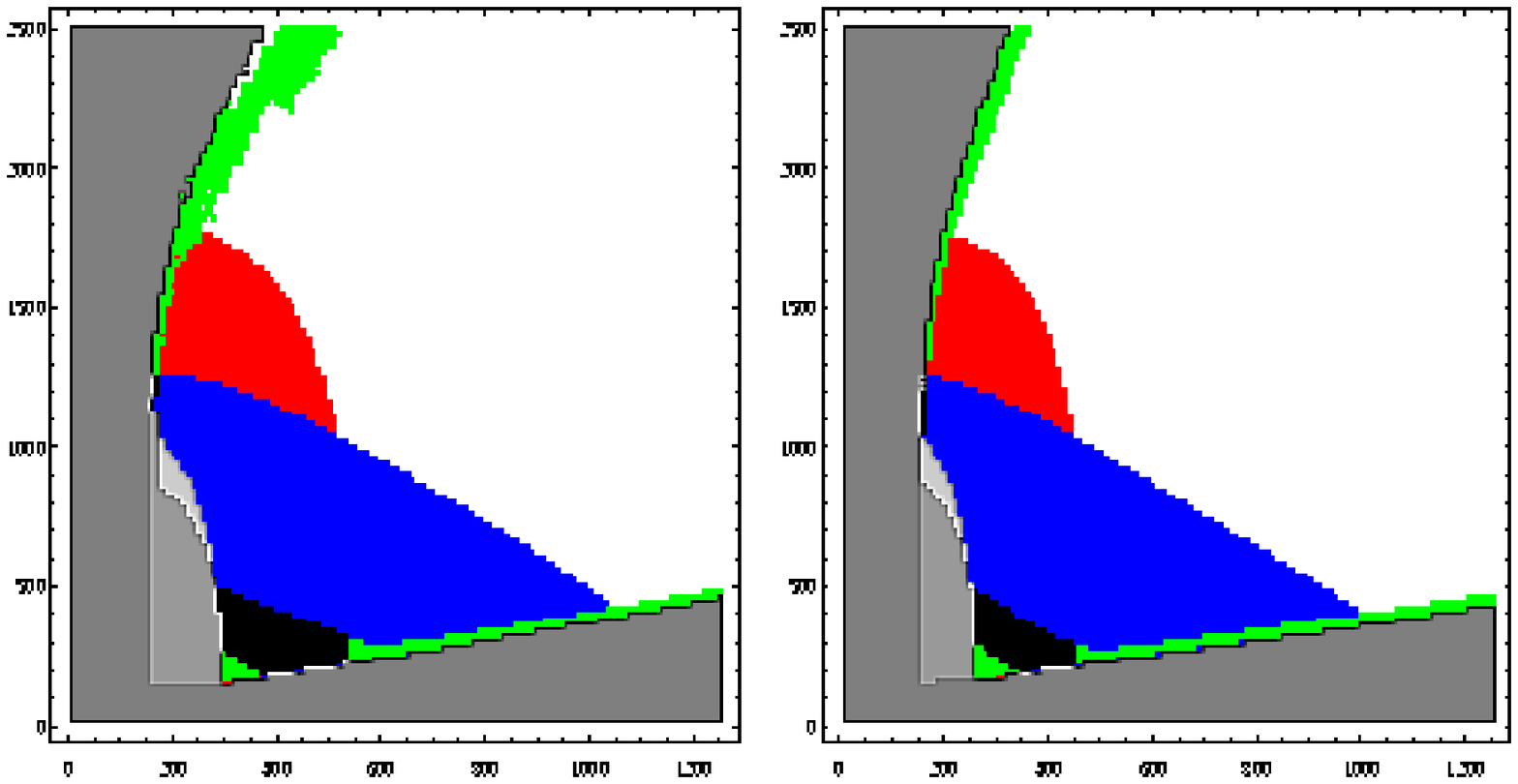,width=16.75cm}}\\[-1.8cm]
\end{center}
\vspace*{-.7cm}
\caption[]{Constraints on the $(m_{1/2}, m_0)$ mSUGRA plane for $\tan
\beta =40$, $A_0=0$, sign$(\mu)>0$ when certain radiative corrections 
are not included. In the left panel we have omitted corrections to all
superparitcle masses, and in the right panel we have in addition
neglected corrections to heavy fermion masses. The notation is as in
Fig.~2.} 
\end{figure}

Finally, we show in Fig.~7 the effect of the SUSY radiative
corrections on the parameter space. In the left panel we have switched
off all radiative corrections to the masses of SUSY particles
[neutralinos, charginos, gluinos and squarks]. This has almost no
effect on the Higgs boson and $(g_\mu-2)$ constraints, but the
``light" bino--like LSP DM region becomes slightly larger. The largest
effect is the increase of the region excluded by the $b \to s\gamma$
constraint: the radiative corrections tend to increase the stop and
chargino masses, so switching them off makes these sparticles lighter,
leading to larger contributions to the $b\to s\gamma$ decay width. \s

The impact of the SUSY radiative corrections to the top and bottom
quark masses [right panel] is slightly more important. These
corrections tend to increase $M_t$ relative to the running mass, so
switching them off increases the size of the top Yukawa coupling
$\lambda_t$. This shrinks the ``no EWSB'' region, as well as the
cosmologically favored region where the LSP is higgsino--like. The
value of $m_t$ that appears in the stop mass matrix, which is the
``MSSM'' running mass in the notation of Ref.~\cite{viennese}, also
increases. This shifts the right boundary of the area excluded by the
$b \ra s \gamma$ constraint to slightly lower values of
$m_{1/2}$. Moreover, the region where 113 GeV $\lsim M_h \lsim$ 117
GeV becomes slightly smaller.\footnote{Recall that our calculation of
$M_h$ and BR$(b \ra s \gamma)$ uses $M_t$ and $\overline{m}_b (
\overline{m}_b)$ as input quark masses; these are not affected by the
SUSY loop corrections. The effect of these corrections is thus
entirely through the changes of the sparticle spectrum. We also note
that switching off the SUSY loop corrections to quark masses is not
entirely consistent, since we use the routine of Ref.~\cite{bsg2} for
the calculation of BR$(b \ra s \gamma)$ which explicitly includes
these corrections. We nevertheless feel that Fig.~7 is a reasonable
illustration of the importance of these corrections.} Since the SUSY
radiative corrections tend also to decrease the bottom quark mass
$m_b$, their removal will lead to an increase of $\lambda_b$, and
hence to a more extended region where the cosmological relic density
is in the interesting range, as is the case for larger $\bar{m}_{b}$
values [Fig.~5] or larger values of $\tb$ [Fig.~4]. Finally, we note
that switching off all two--loop terms in the RGE would dramatically
extend the ``no EWSB'' region, as also pointed out in
Ref.~\cite{SOFTSUSY}. 

\section*{4. Sparticle and Higgs Production in e$^+$e$^-$ Collisions}  

In this section we will discuss the prospects for producing SUSY
particles and heavier Higgs bosons at future linear $\ee$ colliders
with center of mass energies between 500 GeV and 1.2 TeV, in the
context of the mSUGRA model. We first list the production processes
that we will analyze and then discuss the regions of the $(m_{1/2},
m_0)$ parameter space, for various input values of $\tb$ and $A_0$, in
which these processes are accessible. For completeness, analytical
expressions for the relevant total production cross sections are given
in the Appendix. \s

In this exploratory study we will assess the accessibility of certain
production modes simply through the corresponding total cross section,
without performing any background studies. However, in most cases the
clean experimental environment offered by $\ee$ colliders should
indeed permit discovery of a certain mode, given a sample of a few
dozen events. Difficulties might arise in some narrow regions of
parameter space, which we will point out in the following discussion.

\subsection*{4.1 Production processes}

In our study, we will consider the following production processes: 
\beq
{\rm Pair~production~of~the~lightest~charginos}: \ \  
\ee \to \tilde\chi_1^+ \tilde\chi_1^- .
\eeq
This proceeds through $s$--channel photon and $Z$ boson exchange as
well as $t$--channel sneutrino exchange. For higgsino--like charginos,
$M_2 \gg |\mu|$, or heavy sneutrinos, only the $s-$channel diagrams
contribute substantially. Note that the $Z$ boson couples more
strongly to the wino components of $\tilde \chi_1^\pm$ than to its
higgsino component, and the $t$--channel $\tilde{\nu}_e$ exchange
contributes with opposite sign as the $s-$channel diagrams. The cross
section is thus maximal for heavy sneutrino and $M_2 \ll |\mu|$. For
small $m_{\tilde{\nu}_e}$, which corresponds to small $m_0$, the
$t$--channel contribution can reduce the cross section
significantly. If the chargino is higgsino--like, the cross section
becomes insensitive to $m_{\tilde \nu}$, since the $\tilde{\nu}e
\tilde\chi_1^\pm$ couplings are zero in this limit. We thus see that
the production cross section for fixed chargino mass can vary
significantly across the parameter space, but it is always rather
large, so that masses close to the kinematical threshold should be
accessible. The only exception can occur in the very higgsino--like
region, where the $\tilde \chi_1^\pm - \lsp$ mass difference becomes
small. In the most extreme case one may have to rely on events with
additional hard photon to suppress backgrounds \cite{chen}. However,
we found that this can happen only in a very narrow strip near the
``no EWSB'' boundary, where the LSP relic density is below the
currently favored range. \s

The $\tilde \chi_1^\pm$ decay pattern depends on its mass difference
to the LSP, as well as to sleptons and charged Higgs bosons. If the
lighter chargino is the lightest charged sparticle [generally for $m_0
\gsim m_{1/2}$], it mostly decays into \lsp\ plus a real or virtual $W$
boson which in turn decays into $f \bar f'$ pairs with well--known
branching ratios. For smaller ratio $m_0/m_{1/2}$ real or virtual
slepton exchange contributions become important, in some cases leading
to a leptonic branching ratio near 100\%. There is a very narrow strip
in parameter space where $\tilde \chi_1^\pm \ra \tilde \nu_\ell
\ell^\pm \ (\ell = e, \mu, \tau)$ has a large branching ratio but the
charged lepton is very soft. Here one might again have to require the
existence of an additional hard photon in the event to suppress
backgrounds. However, in this case charged slepton pair production is
also accessible. There are also regions where $\tilde \chi_1^\pm$
decays predominantly into $\tau^\pm \nu_\tau \lsp$, either because the
$\tilde \tau_1$ mass is reduced compared to the other slepton masses
[this happens at large $\tb$ \cite{bcdpt}], or because L--R mixing
greatly enhances the $\tilde \chi_1^\pm \tilde \tau_1 \nu_\tau$
couplings relative to the corresponding couplings of $\tilde e_1
\simeq \tilde e_R$ and $\tilde \mu_1 \simeq \tilde \mu_R$; this latter
effect can become important already for moderate values of $\tb$, if
$m_{1/2}/2 \lsim m_0 \lsim m_{1/2}$ \cite{lykken}. Finally, for very
large values of $\tb$, charged Higgs boson exchange contributions can
also play a role, again leading to an enhanced branching fraction for
the $\tau$ mode. However, a large or even dominant branching ratio
into $\tau \nu_\tau \lsp$ is not expected to significantly degrade the
mass reach for $\tilde \chi_1^+ \tilde \chi_1^-$ production at $\ee$
colliders, which should be very close to the kinematical limit [as at
LEP]. To the contrary, the measurement of $\tilde \chi_1^\pm$ decay
branching ratios might allow one to extract information about
(s)particles that are too heavy to be pair--produced.
\beq {\rm
Production~of~the~lightest~neutralinos}: \ \ 
\ee \to \tilde\chi_1^0 \tilde\chi_2^0 . \eeq 
This process is mediated by $s-$channel $Z-$boson exchange and $t-$
and $u-$channel $\tilde{e}_L, \tilde{e}_R$ exchanges. In the gaugino
limit, $|\mu| \gg M_1 , M_2$, the $Z$ boson coupling to neutralinos
vanishes and only the $t$ and $u$--channel contributions are
present. The latter will be suppressed for high selectron masses,
i.e. for large $m_0$; however, $m_0 \gg m_{1/2}$ also generally
implies that $|\mu|$ is not so large, so the size of the $Z-$exchange
contribution increases in this region. In the extreme higgsino limit,
only the $Z$ boson exchange contribution will survive since the $e
\tilde{e} \tilde\chi_{1,2}^0$ couplings are ${\cal O}(10^{-5})$.
Except in the extreme higgsino limit, the cross section is much
smaller than the cross section for chargino pair production; however,
as will be shown later, the anticipated high luminosity should
ensure a detectable signal over most of the kinematically accessible
parameter space. \s

As well known \cite{olddec,bcdpt} the $\tilde \chi_2^0$ branching
ratios depend strongly on details of the SUSY particle spectrum. For
example, the branching ratio into $\ell^+ \ell^- \lsp$ can vary
between nearly 100\% and almost zero. The former occurs if $\tilde
\chi_2^0 \ra \ell^\pm \tilde \ell^\mp_1$ is the only 2--body decay
mode of $\tilde \chi_2^0$, while the latter situation is e.g. realized
if $\tilde \chi_2^0 \ra h \tilde \chi_1^0$ is dominant. However, at an
$\ee$ collider hadronic $\tilde \chi_2^0$ decays are as easily
detectable as decays into charged leptons. The only potentially
difficult scenarios are the extreme higgsino region, where the $\tilde
\chi_2^0 - \lsp$ mass difference is small [but stays about twice as
large as the $\tilde \chi_1^\pm \lsp$ mass difference], or scenarios
where $\tilde \chi_2^0$ decays almost exclusively into the invisible
mode $\nu \bar{\nu} \lsp$; however, this latter scenario is {\em never}
realized in mSUGRA, given current experimental constraints. \s

We will not discuss the production of pairs of the next--to--lightest
neutralinos, $\ee \to \tilde\chi_2^0 \tilde\chi_2^0$, since in mSUGRA
this process leads to the same reach at $\ee$ colliders as
$\tilde\chi^\pm_1$ pair production; the approximate equality
$m_{\tilde\chi_2^0} \simeq m_{\tilde\chi_1^\pm}$ holds in both the
higgsino and the gaugino limit. However, the neutralino production
cross section is smaller due to the absence of the photon exchange
channel, and because the $Z$ boson does not couple to neutral SU(2)
gauginos. Nor will we consider the production of heavier states, $\ee
\to \tilde\chi_i^0 \tilde\chi_j^0$ with $i$ or $j>2$ and $\ee \to
\tilde\chi_i^\pm \tilde\chi_j^\mp$ with $i$ or $j>1$, since these
channels cannot extend the overall discovery reach. Of course, one
would eventually like to also study these channels in detail in a
clean environment.
\beq
{\rm Pair~production~of~charged~sleptons}: \ \  
\ee \to \tilde{\ell}^+ \tilde{\ell}^- .
\eeq
Pairs of SU(2) doublet [``left--''] and singlet [``right--handed'']
selectrons are produced via $s$--channel photon and $Z$ boson exchange
and the $t$--channel exchange of the four neutralinos
$\tilde\chi_i^0$. Since the vector boson couplings to charged slepton
current eigenstates are full strength gauge couplings and L--R mixing
between selectrons is negligible [for our purposes], the first two
channels always contribute. Since the electron Yukawa coupling is
tiny, only the exchange of the gaugino--like neutralinos plays a role
here. To good approximation the cross section is therefore determined
by the sizes of the soft breaking gaugino masses $M_1$ and $M_2$. In
mSUGRA they are related via eq.(1), which leads to $M_1 \simeq M_2/2$
at the weak scale. The value of $|\mu|$ is not relevant
here. Moreover, mixed $\tilde e_L^\pm \tilde e_R^\mp$ production is
possible through the exchange of neutralinos in the $t-$ or
$u-$channel. However, in mSUGRA the overall mass reach in selectron
pair production will be determined by $\tilde e_R$ pair production,
since the $\tilde e_R - \tilde e_L$ mass difference can be sizable. \s

The production of second and third generation charged sleptons only
proceeds through $s-$channel $\gamma$ and $Z$ boson exchange. Since
the mixing in the $\tilde{\tau}$ sector can be large, it has to be
included in the $Z \tilde{\tau} \tilde{\tau}$ couplings; in this case
we will only consider the production of the lighter $\tilde{\tau}$
states, $\ee \to \tilde{\tau}_1 \tilde{\tau}_1$, which offers the
largest reach.\footnote{The $Z \tilde \tau_1 \tilde \tau_1$ coupling
vanishes for a specific value of the $\tilde \tau$ mixing angle,
$\cos^2 \theta_\tau = 2 \sin^2 \theta_W$. However, the photon exchange
contribution ensures that the $\tilde \tau_1$ pair production cross
section remains sizable even in this case.} Note that, with the
exception of $\tilde e_L^\pm \tilde e_R^\mp$ production, the cross
section for the production of sleptons near threshold is strongly
suppressed by $\beta^3$ factors, $\beta$ being the cms velocity of the
sleptons. Therefore only slepton masses up to several GeV below the
kinematical limit can be probed. \s

The sleptons will mostly decay into their partner leptons and the
gaugino--like neutralinos and [if accessible] charginos. Slepton pair
production at next--generation $\ee$ colliders will only be accessible
if $m_0$ is not very large, away from the ``focus--point'' region. The
gaugino--like neutralinos and charginos will then be the lighter
states, since mSUGRA predicts $|\mu| > M_2$ unless $m_0 \gg
m_{1/2}$. In particular, for the lighter slepton mass eigenstates the
by far dominant decay will be $\tilde{\ell}^\pm_1 \to \ell^\pm
\tilde\chi_1^0$. If phase space allows it, the heavier left--handed
sleptons will decay predominantly into wino--like charginos $\tilde
\chi_1^\pm$ or neutralinos $\tilde \chi_2^0$, plus a neutrino or
charged lepton, since these decays occur via SU(2) couplings which
exceed the U$(1)_{\rm Y}$ couplings responsible for $\tilde \ell_L \ra
\ell \lsp$ decays. In the case of $\tilde{\tau}$ sleptons, both states
might be able to decay via the charged current because of L--R
mixing. For large $\tb$ the $\tilde \tau_2$ decay pattern can be
rather complicated \cite{bartlstau}. The mass reach for $\tilde
\tau_1$ pair production is expected to be a little lower than that for
$\tilde \mu$ pair production, since the $\tau$ leptons produced in
$\tilde \tau_1 \ra \tau \lsp$ will themselves decay, which degrades
the visible energy. This becomes of some concern in a narrow strip of
parameter space close to the lower bound on $m_0$, i.e. near the
region excluded by the requirement $\mchi < m_{\tilde \tau_1}$. On
the other hand, $\tilde \tau_1$ pair production generally still gives
a larger reach in the $(m_{1/2}, m_0)$ plane, since for $\tb \gg 1$,
$m_{\tilde \tau_1}$ is significantly smaller than $m_{\tilde
\mu_R}$. Moreover, the measurement of the polarization of the produced
$\tau$ leptons could yield important information about the SUSY model
\cite{miho}.
\beq
{\rm Pair~production~of~sneutrinos}: \ \  \ee \to \tilde{\nu} 
\bar{\tilde\nu}.
\eeq
Muon and tau sneutrino pairs are produced only through $s$--channel
$Z$--boson exchange. Electron sneutrinos can also be produced
through $t$--channel diagrams with the exchange of the [gaugino--like]
charginos, which enhances the cross section significantly if $m_{1/2}$
is not too large. All $\tilde \nu$ pair production cross sections show
the familiar $\beta^3$ behavior near threshold. \s

In mSUGRA the sneutrinos [as well as the ``left--handed'' charged
sleptons] are usually heavier than the SU(2) gauginos. Sneutrinos can
thus generally decay through both neutral and charged currents into
leptons and gauginos, $\tilde{\nu} \to \nu \tilde\chi_i^0$ and
$\tilde{\nu} \to l^\pm \tilde\chi_i^\mp$. In the narrow range where
$m_{\tilde\chi_1^0} \le m_{\tilde{\nu}} \le m_{\tilde\chi_1^\pm},
m_{\tilde\chi_2^0}$, which can occur for $m_{1/2} \gg m_0$ if the
overall SUSY mass scale is rather low so that the negative $D-$term
contribution to $m^2_{\tilde \nu}$ plays a role, the only allowed
decay will be the invisible decay into the LSP and a neutrino.
However, in mSUGRA such scenarios are excluded by the $h$ mass
constraint. Usually this invisible decay is disfavored by the
smallness of the U$(1)_{\rm Y}$ gauge coupling.
\beq {\rm
Pair~production~of~stops~and~sbottoms}: \ \ 
\ee \to \tilde{t}_1 \bar{\tilde t}_1 \ , \ \tilde{b}_1 \bar{\tilde b}_1 .
\eeq 
In $\ee$ collisions, squarks can only be produced through $s$--channel
photon and $Z$ boson exchange. In our analysis, we will consider only
the pair production of the states $\tilde{t}_1$ and $\tilde{b}_1$
which are in general significantly lighter than the other squark
states, due the large Yukawa couplings of top and bottom quarks. In
both cases $L-R$ mixing can affect the coupling to $Z$ bosons, and
hence the total cross section \cite{Asquarks}, but the latter remains
sizable unless these processes are kinematically suppressed.  \s

The mixing in the stop sector and the large top quark mass might
generate some complications in the decays of $\tilde{t}_1$, since the
usual two--body squark decays $\tilde{q} \to q \tilde g, \ q
\tilde\chi_i^0$ and $q' \tilde\chi_i^\pm$ are kinematically closed for
sizable regions of parameter space. In this case, $\tilde t_1$ decays
are dominated by loop induced two--body decay channels [such as the
flavor changing decay into a charm quark and the LSP \cite{kenkob}],
or by tree--level three--body [$b W\tilde\chi_1^0, bH^+
\tilde\chi_1^0$ or $b l \tilde{l}$] or even four--body [$b \lsp f
\bar f'$ where $f, f'$ are massless fermions] decay modes
\cite{stopdecays}. In contrast, tree--level two--body decays $\tilde
b_1 \rightarrow b \tilde \chi_i^0$ decays are always possible at least
for $i=1$. As in the case of $\tilde \tau$, the decay patterns of the
heavier states $\tilde b_2$ and $\tilde t_2$ can again be rather
complicated \cite{bartlstau}.
\beq 
{\rm Pair~production~of~Higgs~bosons}: \ \ 
\ee \to AH \ {\rm and} \ H^+H^- .
\eeq 
If the pseudoscalar Higgs boson $A$ is relatively heavy, $M_A \gsim
200$ (130) GeV for moderate (large) values of $\tb$, we are in the
so--called decoupling regime where the running masses of the heavier
Higgs bosons $A,H$ are almost degenerate\footnote{We mentioned earlier
that even for large $M_A$, the $A$ and $H$ pole masses can differ by
up to 10\% if $\tb$ is very large.}; for even larger $M_A$, the
$H^\pm$ mass also becomes very similar. In that limit, which is
realized for most of the allowed mSUGRA parameter space, the lighter
$h$ boson mass reaches its maximal value, $M_h \lsim 130$ GeV, and $h$
has Standard Model like couplings to fermions and gauge bosons. The
pseudoscalar Higgs boson has no tree--level couplings to two gauge
bosons. In the decoupling limit, the couplings of $H$ to two massive
vector bosons are also strongly suppressed, as is the $Z A h$
coupling. The only significant $2 \to 2$ process for the production
of the heavier neutral Higgs bosons is thus associated $HA$ production
through $Z$ boson exchange in the $s$--channel. The charged Higgs
particles can be pair--produced photon and $Z$ boson exchange. These
cross sections are sufficiently large [although once again suppressed
by $\beta^3$ factors near threshold] to allow for the production of
detectable quantities of these particles up to almost the beam
energy. \s

If $\tb > \sqrt{\overline{m}_t( \overline{m}_t) / \overline{m}_b(
\overline{m}_b )}$ or $M_A < 2 M_t$ the heavy neutral Higgs bosons
will mainly decay into $b\bar b$ and $\tau^+\tau^-$ pairs [with
relative abundance $\sim 9:1$]. For smaller values of $\tb$, decays
into $t\bar t$ pairs dominate, if kinematically allowed.  The charged
Higgs bosons will dominantly decay into $tb$ and $\tau \nu$ final
states. In some cases decays into SUSY particles [charginos,
neutralinos and possibly sleptons and stops; only this last mode can
compete with decays into $t \bar t$, or with $b \bar b$ at large
$\tb$] can be possible \cite{Ohmann}. These decays will be more
difficult to analyze, but in most cases should be clean enough to be
detectable at $\ee$ colliders. [The main background will probably come
from the direct pair production of heavier superparticles.] \s

Of course, the light scalar $h$ boson can be produced [in association
with a $Z$ boson, and/or through gauge boson fusion] over the entire
mSUGRA parameter space. Detailed studies \cite{e+e-,TESLA} have shown
that detection of this particle is straightforward at an $\ee$
collider with c.m. energy $\sqrt{s} \gsim 250$ GeV even with moderate
luminosities, $\int {\cal L} dt \gsim 10$ fb$^{-1}$. With the high
luminosities expected at a linear collider, $\int {\cal L} dt \sim
500$ fb$^{-1}$, detailed studies of the profile of this particle can
be made already at $\sqrt{s} \lsim 500$ GeV: the mass can be measured
at the permille level, various couplings to fermions and gauge bosons
[which determine the production cross sections and decay branching
ratios] can be measured at the percent level, and even the Higgs
self--coupling can be measured at the level of $\sim 10-20\%$. In the
MSSM these measurements should reveal some deviations from Standard
Model predictions for $M_A \lsim 500$ GeV to 1 TeV. Nevertheless this
indirect evidence for the existence of heavier Higgs bosons will have
to be confirmed through their direct production. \s

\subsection*{4.2 Results}

In a first step, we will analyze (s)particle production at an $\ee$
collider with $\sqrt{s}=800$ and an annual luminosity of ${\cal
L}=500$ fb$^{-1}/$yr. This is e.g. expected for the second phase of the
TESLA machine \cite{TESLA2}. We will consider a given channel to be
visible if its total cross section exceeds $\sigma_{\rm min}=0.1$ fb,
which means that a sample of 50 signal events per year will be
required, with the assumed luminosity [or 100 events with a two years
running], to establish discovery.\footnote{Here, we will only discuss
direct production of the new particles. In some areas of the parameter
space, cascade decays of SUSY particles might allow for the detection
of states which are not accessible directly since the corresponding
cross sections are too small.} This number should usually be
sufficient for discovery in the clean environment offered by $\ee$
colliders \cite{e+e-,TESLA}. \s

We will also illustrate the potential of the first phase of the $\ee$
collider with $\sqrt{s} = 500$ GeV and the same luminosity. We will
then discuss the increase of the discovery potential if the energy of
the collider could be raised to $\sqrt{s}= 1.2$ TeV
\cite{Wagner}. This could be accomplished either by simply extending
the electron and/or positron accelerators, or by increasing the RF
power if the cavity modules generate higher gradients than originally
envisaged [as is the case for the most recent
TESLA cavities].\footnote{We thank Peter Zerwas for a discussion on
this point.} However, in order to compensate for the $1/s$ drop of
most background cross sections with raising c.m. energy, we will
require a smaller minimal cross section than in the previous cases,
$\sigma_{\rm min}=0.025$ fb. In order to obtain the same number of
signal events one would then either have to increase the luminosity
[as is the case for the TESLA machine \cite{TESLA2} where the
luminosity is expected to scale with the energy], or to extend the
running period of the machine.

Figs.~8--10 show the regions in the usual $(m_{1/2}, m_0)$ plane where
various superparticles and heavy Higgs bosons can be discovered at an
$\ee$ collider with a c.m. energy $\sqrt{s}=800$ GeV and a luminosity
$\int {\cal L} dt =500$ fb$^{-1}$, for the same values of $\tb$ and
$A_0$ taken in Figs.~1--3. The grey areas are again those excluded by
theoretical and experimental constraints. The colored regions
correspond to the following production processes: $\ee \to
\tilde\chi_1^0 \tilde\chi_2^0$ (green+red), $\ee \to \tilde\chi_1^+
\tilde\chi_1^-$ (red), $\ee \to \tilde{\ell}^+ \tilde{\ell}^-$ (blue),
$\ee \to \tilde{\nu} \bar{\tilde\nu}$ (purple), $\ee \to \tilde{t}_1
\bar{\tilde t}_1$ (dark blue), $\ee \to \tilde{b}_1 \bar{\tilde b}_1$
(dark green) and the heavy MSSM Higgs boson production $ \ee \to HA$
and $ H^\mp H^\pm$ (yellow). Note that some of these regions are
overlapping. For example, in Fig.~8 the chargino and neutralino areas
should be extended until the lower boundary where the LSP is not the
$\tilde\chi_1^0$, the $\tilde{\tau}$ region includes the sneutrino and
Higgs regions, and the sneutrino area includes the Higgs boson
region. Moreover, the region where $\lsp \tilde \chi_2^0$ production
is accessible always includes the entire $\tilde \chi_1^\pm$ pair
production region. \s

\begin{figure}[htbp]
\centerline{\large $\tan\beta=40\ , \ A_0 = 0 \ ,\  {\rm sign}(\mu)>0$}
\hspace*{-.4cm}{\large $m_0$}\\[-1.cm]
\begin{center}
\hspace*{-.2cm} \epsfig{figure=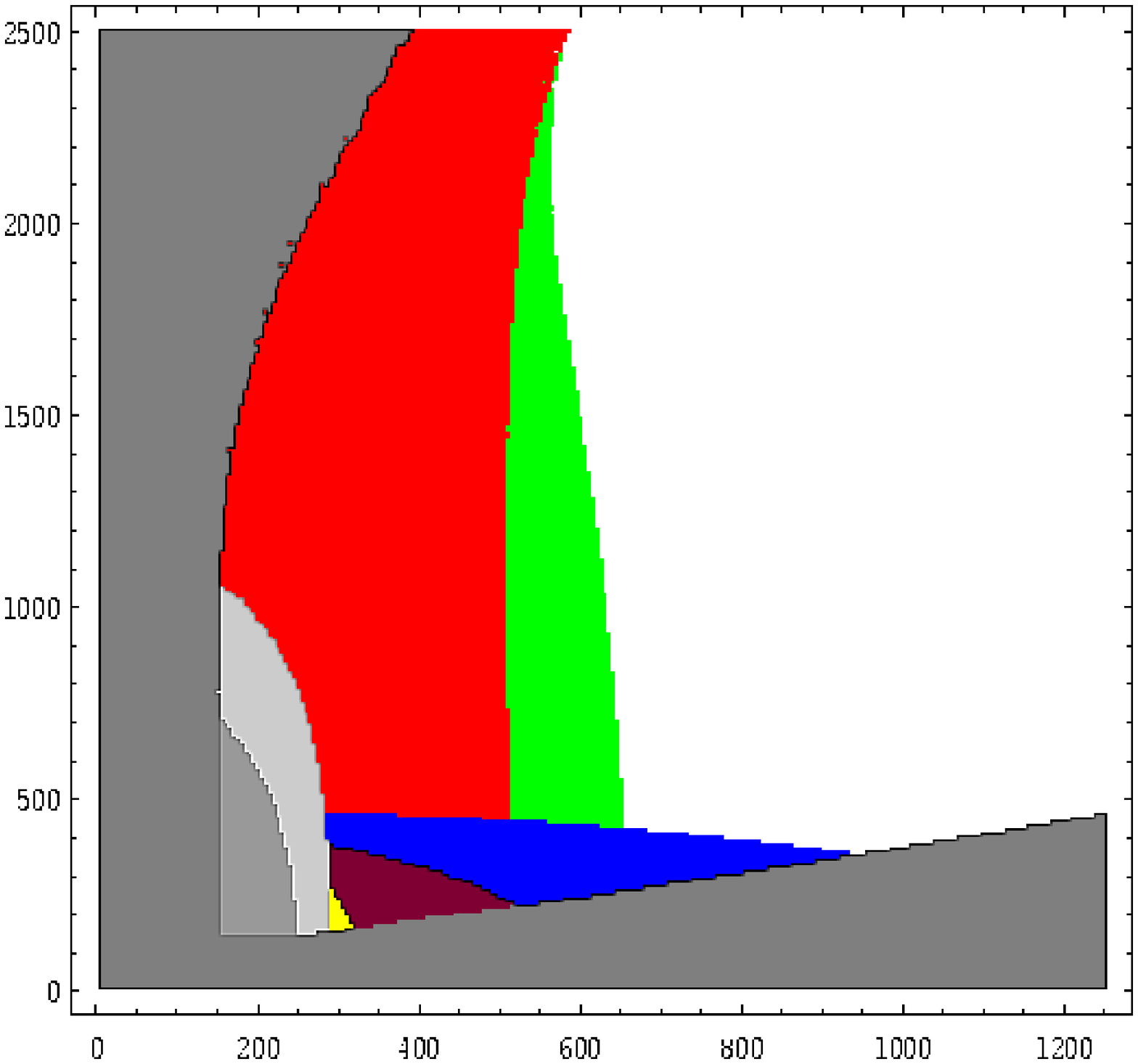,width=15cm}\\
\end{center}
\hspace*{14.5cm} {\large $m_{1/2}$}
\vspace*{-2mm}
\caption[]{The $(m_{1/2}, m_0)$ mSUGRA plane where SUSY and Higgs
particles can be produced at an $e^+e^-$ collider with a c.m. energy
$\sqrt{s}=800$ GeV. The grey areas are those excluded by theoretical
and experimental constraints. The colored regions are those where a given
cross section is large enough for the particles to be produced:
$\tilde\chi_1^0 \tilde\chi_2^0$ (green), $\tilde\chi_1^+
\tilde\chi_1^-$ (red), $\tilde{\ell}^+ \tilde{\ell}^-$ (blue), $\tilde{\nu}
\bar{\tilde\nu}$ (purple), $\tilde{t}_1 \bar{\tilde t}_1$ (dark blue),
$\tilde{b}_1 \bar{\tilde b}_1$ (dark blue) and the heavy MSSM $H,A,H^\pm$
bosons (yellow). Note that some of these regions are overlapping.}
\end{figure}

\begin{figure}[htbp]
\noindent \hspace*{.9cm} $\tan\beta=5, \, A_0 = -1\, {\rm TeV}, \,  
{\rm sign}(\mu)>0$ \hspace*{1.5cm}
$\tan\beta=10, \, A_0 = 0, \,  {\rm sign}(\mu)>0$\\
\hspace*{-.7cm}{\large $m_0$}\\[16.2cm]
\hspace*{16.5cm} {\large $m_{1/2}$}\\[-17.9cm]
\begin{center}
\hspace*{-.2cm}\mbox{\epsfig{figure=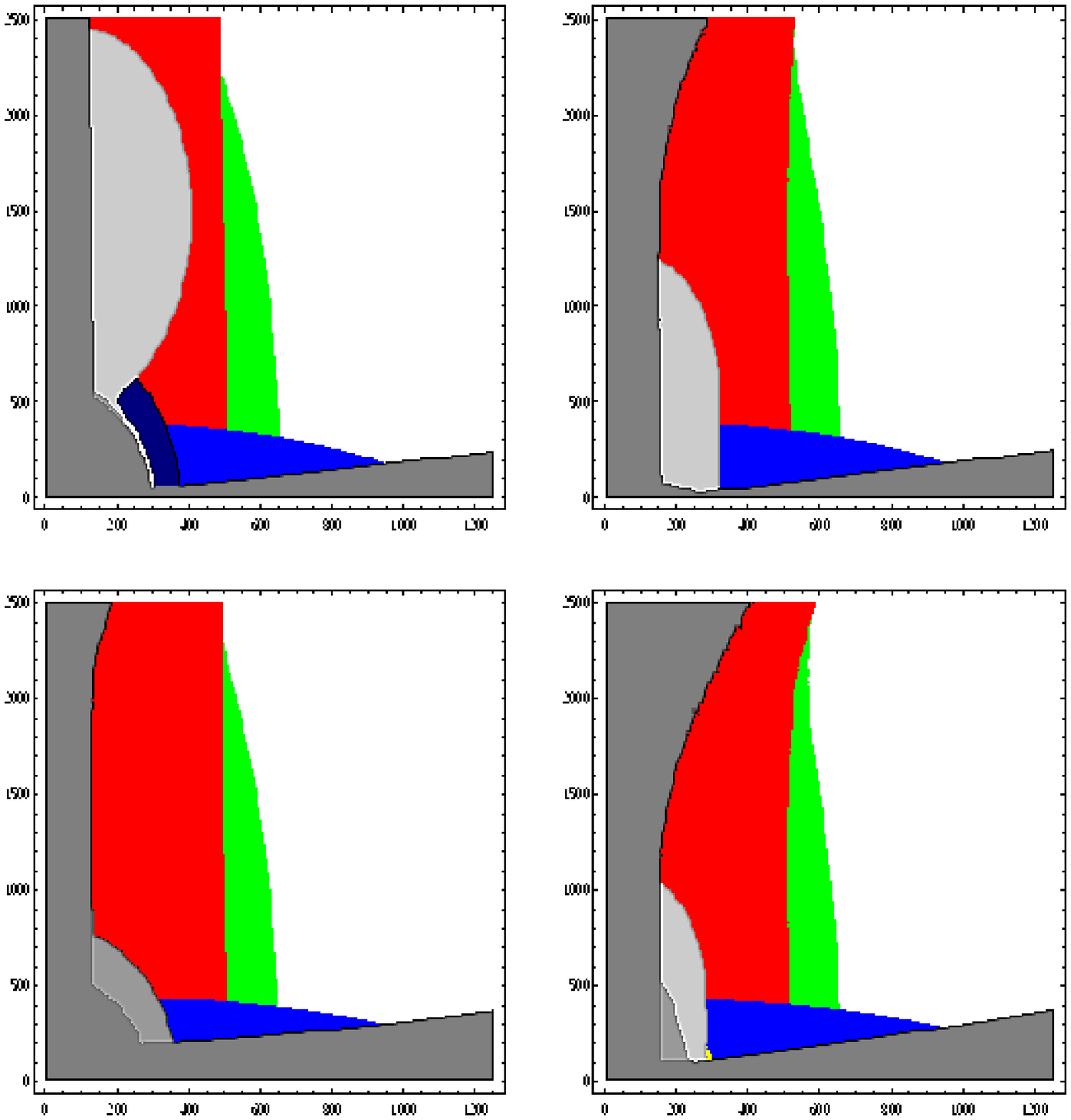,width=16.5cm}}
\end{center}
\hspace*{.9cm} $\tan\beta=20, \, A_0 = -1\, {\rm TeV}, \, {\rm sign}(\mu)>0$ 
\hspace*{1.5cm}
$\tan\beta=30, \, A_0 = 0 , \,  {\rm sign}(\mu)>0$\\[.3cm]
\caption[]{The $(m_{1/2}, m_0)$ mSUGRA plane where SUSY and Higgs particles
can be produced at an $e^+e^-$ collider with a c.m. energy $\sqrt{s}=800$ 
GeV for different values of $\tan \beta <40$ and $A_0$  with
sign($\mu) >0$. The notation is as in Fig.~8.}
\end{figure}

\begin{figure}[htbp]
\noindent \hspace*{.7cm} $\tan\beta=45, \, A_0 = -1\, {\rm TeV}, \,  
{\rm sign}(\mu)>0$ \hspace*{1.7cm}
$\tan\beta=50, \, A_0 = 0, \,  {\rm sign}(\mu)>0$\\
\hspace*{-.7cm}{\large $m_0$}\\[16.2cm]
\hspace*{16.5cm} {\large $m_{1/2}$}\\[-17.9cm]
\begin{center}
\hspace*{-.2cm}\mbox{\epsfig{figure=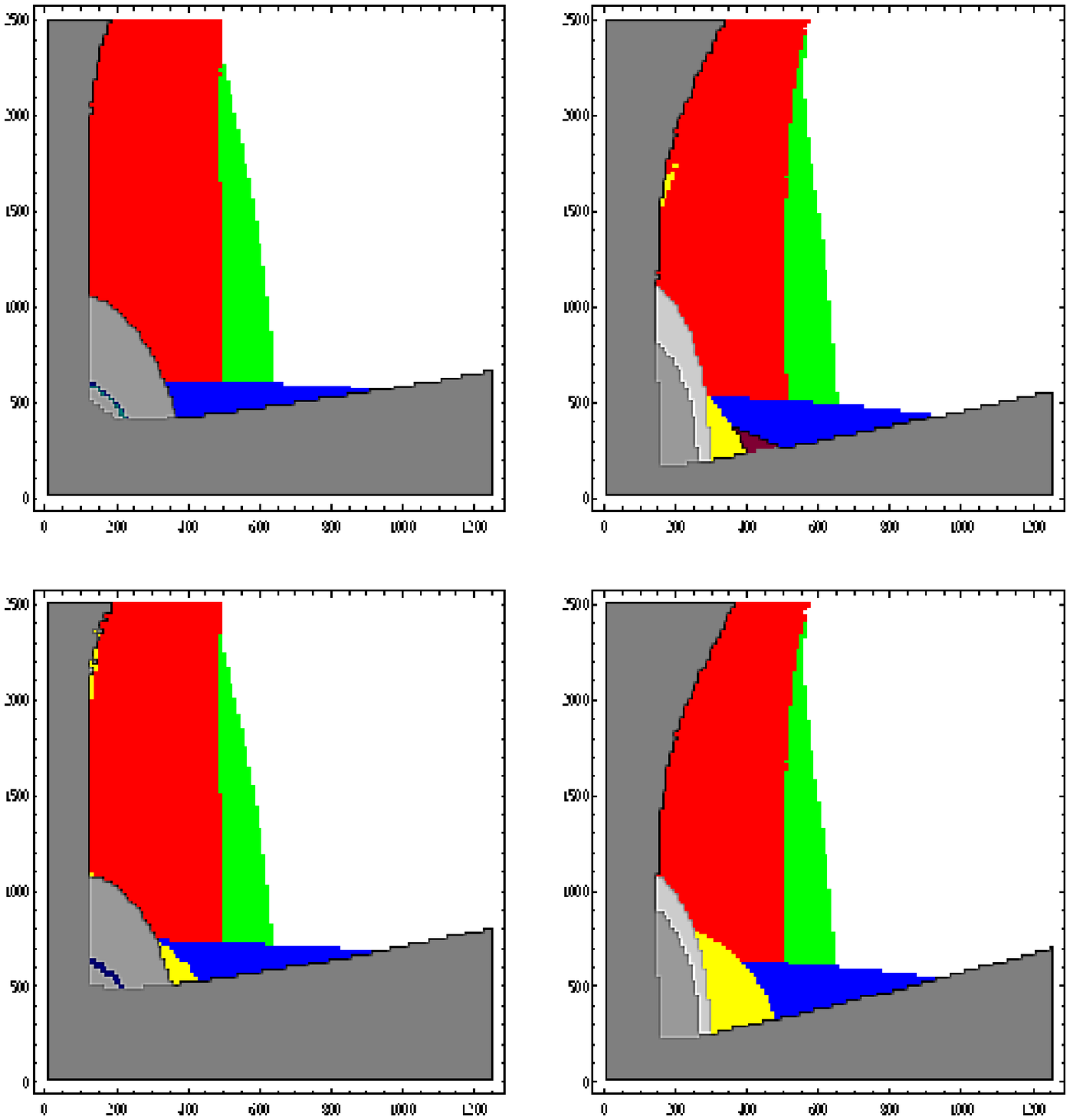,width=16.5cm}}
\end{center}
\noindent \hspace*{.7cm} $\tan\beta=55, \, A_0 = -1\, {\rm TeV}, 
\,  {\rm sign}(\mu)>0$ 
\hspace*{1.7cm}
$\tan\beta=60, \, A_0 = 0 , \,  {\rm sign}(\mu)>0$\\[.3cm]
\caption[]{The $(m_{1/2}, m_0)$ mSUGRA plane where SUSY and Higgs particles
can be produced at an $e^+e^-$ collider with a c.m. energy $\sqrt{s}=800$ 
GeV for different values of $\tan \beta >40$ and $A_0$  with
sign($\mu) >0$. The notation is as in Fig.~8.}
\end{figure}
As stated earlier, chargino pair production, $\ee \to \tilde\chi_1^+
\tilde\chi_1^-$, can be probed for mass values up to the kinematical
limit, $m_{\tilde\chi_1^\pm} \sim 400$ GeV for $\sqrt{s}=800$
GeV. This implies that for wino--like charginos, i.e. in the region
$m_0 \lsim 1.5$ TeV, values $m_{1/2} \sim 1.3 m_{\tilde\chi_1^\pm}
\sim 500$ GeV can be probed. For larger values of $m_0$, the chargino
becomes a mixture of higgsino and wino, since $\mu$ is relatively
small. This reduces its mass for fixed $m_{1/2}$, so that larger
values of $m_{1/2}$ can be probed. As long as the chargino is
wino--like, the region where $\tilde \chi_1^\pm$ pair production is
accessible is almost independent of $\tb$; however, some parts of
this region might be ruled out by the experimental constraints,
e.g. for $\tb \sim 5$ [top--left frame in Fig.~9] where a part has
been eaten by the $M_h \gsim 113$ GeV requirement. As noted in the
discussion of Fig.~3, choosing $A_0 = -1$ TeV increases the value of
$\mu$ required for EWSB, removing the higgsino and mixed regions, and
the maximal value of $m_{1/2}$ that can be probed in chargino pair
production becomes independent of $m_0$.  \s

Searches for the associated production of the lightest and
next--to--lightest neutralinos, $\ee \to \tilde\chi_1^0
\tilde\chi_2^0$, further increase the reach in $m_{1/2}$ in the
bino--like limit. In this region the $Z\tilde\chi_1^0 \tilde\chi_2^0$
coupling, which is proportional to the product of higgsino components
of \lsp\ and $\tilde \chi_2^0$, is small.  However, selectrons are not
too heavy, so selectron exchange in the $t-$channel produces a cross
section that should be detectable given the anticipated large
luminosity. The situation is thus quite different from the one at
LEP2, where the integrated luminosity is three orders of magnitude
smaller, in which case the gain from neutralino production over the
information obtained from chargino pair production is
marginal. However, for very large values of $m_0$ and $A_0 = 0$ the
lighter neutralinos become almost higgsino--like, and are therefore
almost degenerate with the lightest charginos. The reach in chargino
and neutralino production then becomes almost the same.\footnote{We
remind the reader that the visible energy released in the pair
production of light higgsino--like states can be quite small, making
the study of these channels difficult in the extreme higgsino region.}
Finally, for $A_0 = -1$ TeV the light neutralinos remain gaugino--like
even for $m_0 \gsim 2$ TeV. However, here the selectrons are so heavy
that the cross section for neutralino pair production becomes too
small even at a high--luminosity collider. In this region of parameter
space even the discovery reach could thus still be increased by yet
higher luminosities. \s

Charged sleptons can be produced for relatively small values of
$m_0$. $\tilde \tau_1$ pair production remains accessible for values
of $m_{0}$ from $\sim 400$ GeV down to the region where the
$\tilde\chi_1^0$ is not the LSP.\footnote{Recall that the region very
close to the lower bound on $m_0$ could be difficult to access
experimentally.} Sneutrinos are heavier than $\tilde \tau_1$, since
they are pure SU(2) doublets. Hence their pair production can be
probed in a smaller region of the parameter space. Most slepton pair
production channels are insensitive to $\tb$. On the other hand, the
region where $\tilde \tau_1$ pair production, which offers the largest
overall reach in the $(m_{1/2}, m_0)$ plane, is accessible becomes
smaller (larger) with decreasing (increasing) values of
$\tb$. However, much of this changed reach is compensated by the
change of the region excluded by the requirement that \lsp\ is the
LSP. \s

If $A_0$ is small or $\tb$ is large no scalar quarks, not even $\tilde
t_1$ or $\tilde b_1$ squarks, are light enough to be produced at
$\sqrt{s} = 800$ GeV, given current experimental constraints. At small
$\tb$ the $M_h \gsim 113$ GeV constraint can only be satisfied if
squarks are fairly heavy, whereas at larger $\tb$ the $b \ra s \gamma$
constraint excludes scenarios with light squarks. Only for small $\tb$
and large (and negative) $A_0$ does the lightest $\tilde{t}_1$ state
become accessible in a narrow strip of parameter space with $m_{0}
\lsim 500$; see the top--left panel in Fig.~9. \s

For $\tb = 40$, Fig.~8, the heavy Higgs bosons $H,A$ and $H^\pm$ are
accessible only in the small corner $m_{1/2} \sim 250$ GeV, $m_0\sim
200$ GeV between the regions ruled out by the $M_h$ and the non \lsp\
LSP constraints. This region disappears for $\tb \lsim 30$ [Fig.~9],
but can be significantly extended for larger $\tb$ values [Fig.~10],
due to the reduction of $M_A$ with increasing $\tb$. On the other
hand, choosing $A_0 = -1$ TeV rather than 0 increases $|\mu|$ as
determined from EWSB, and hence also the Higgs boson masses. This
explains the absence of a heavy Higgs region in the top--left panel of
Fig.~10. \s

By comparing Figs.~1--3 with Figs.~8--10 we see that the entire 2
$\sigma$ overlap region, where one has at the same time a light Higgs
boson with $M_h \sim 115$ GeV, a SUSY contribution which accounts for
the $(g_\mu-2)$ deviation and the requirement of a neutralino being a
good Dark Matter candidate, can be covered by neutralino searches
already at an 800 GeV collider. Most of this region can also be
covered by chargino searches, and much of it can be probed in addition
by charged slepton (in particular, $\tilde \tau_1$) searches; and
sneutrinos are accessible in at least part of this region. \s

\begin{figure}[htbp]
\centerline{\large $\tan\beta=40\ , \ A_0 = 0 \ ,\  {\rm sign}(\mu)>0$}
\hspace*{-.4cm}{\large $m_0$}\\[-1.cm]
\begin{center}
\hspace*{-.2cm} \epsfig{figure=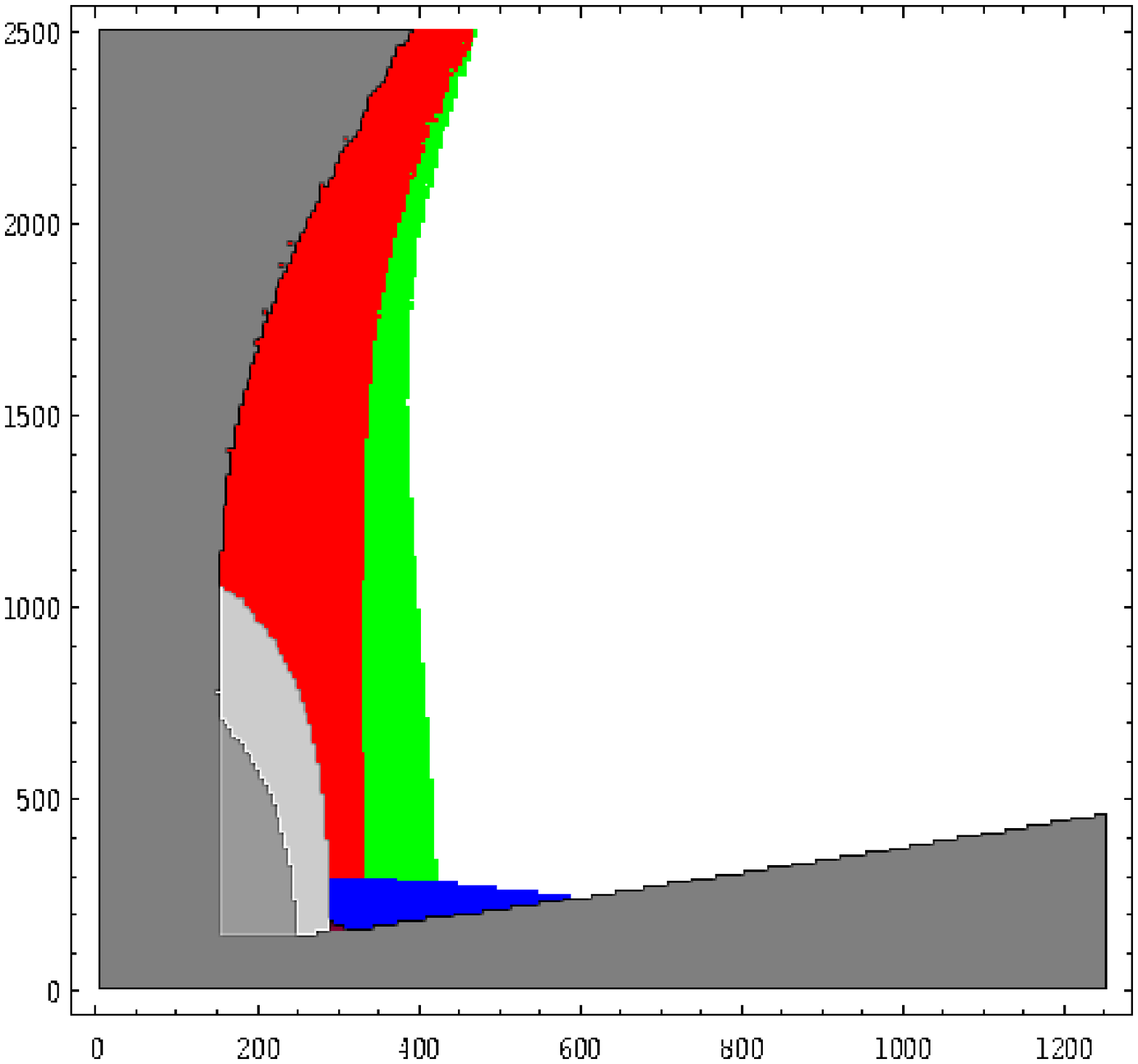,width=15cm}\\
\end{center}
\hspace*{14.5cm} {\large $m_{1/2}$}
\vspace*{-2mm}
\caption[]{The $(m_{1/2}, m_0)$ mSUGRA plane where SUSY and Higgs
particles can be produced at an $e^+e^-$ collider with a c.m. energy
$\sqrt{s}=500$ GeV. The grey areas are those excluded by theoretical
and experimental constraints. The colored regions are those where some
cross section is large enough for the particles to be produced:
$\tilde\chi_1^0 \tilde\chi_2^0$ (green), $\tilde\chi_1^+
\tilde\chi_1^-$ (red), $\tilde{\ell}^+ \tilde{\ell}^-$ (blue), $\tilde{\nu}
\bar{\tilde\nu}$ (purple), $\tilde{t}_1 \bar{\tilde t}_1$ (dark blue),
$\tilde{b}_1 \bar{\tilde b}_1$ (dark blue) and the heavy MSSM $H,A,H^\pm$
bosons (yellow). Note that some of these regions are overlapping.}
\end{figure}

\begin{figure}[htbp]
\noindent \hspace*{.9cm} $\tan\beta=5, \, A_0 = -1\, {\rm TeV}, \,  
{\rm sign}(\mu)>0$ \hspace*{1.5cm}
$\tan\beta=10, \, A_0 = 0, \,  {\rm sign}(\mu)>0$\\
\hspace*{-.7cm}{\large $m_0$}\\[16.2cm]
\hspace*{16.5cm} {\large $m_{1/2}$}\\[-17.9cm]
\begin{center}
\hspace*{-.2cm}\mbox{\epsfig{figure=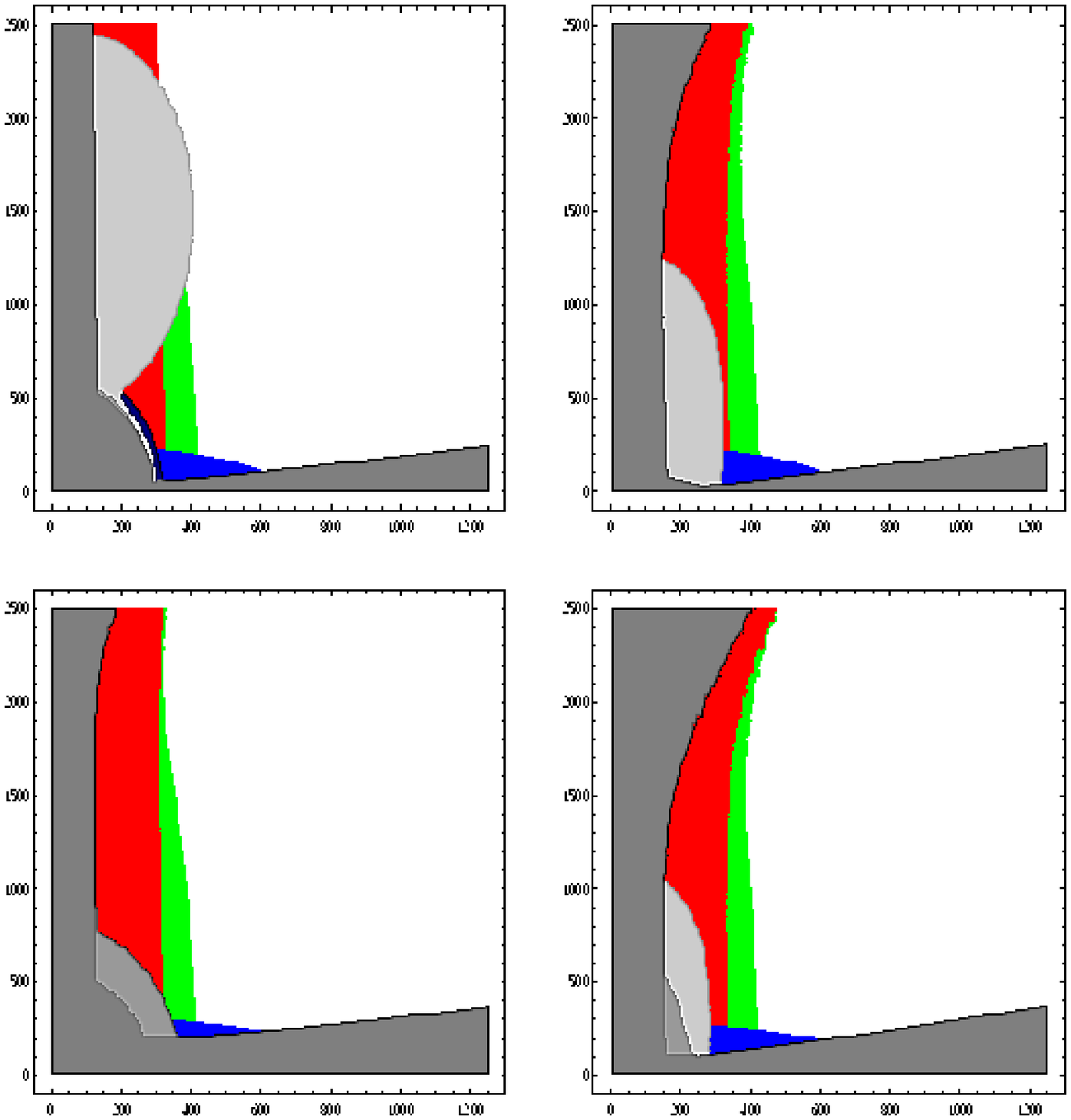,width=16.5cm}}
\end{center}
\noindent \hspace*{.9cm} $\tan\beta=20, \, A_0 = -1\, {\rm TeV}, \,  
{\rm sign}(\mu)>0$ 
\hspace*{1.5cm}
$\tan\beta=30, \, A_0 = 0 , \,  {\rm sign}(\mu)>0$\\[.3cm]
\caption[]{The $(m_{1/2}, m_0)$ mSUGRA plane where SUSY and Higgs particles
can be produced at an $e^+e^-$ collider with a c.m. energy $\sqrt{s}=500$ 
GeV for different values of $\tan \beta<40$ and $A_0$  with sign($\mu)
>0$. The notation is as in Fig.~11.}
\end{figure}

\begin{figure}[htbp]
\noindent \hspace*{.7cm} $\tan\beta=45, \, A_0 = -1\, {\rm TeV}, \,  
{\rm sign}(\mu)>0$ \hspace*{1.7cm}
$\tan\beta=50, \, A_0 = 0, \,  {\rm sign}(\mu)>0$\\
\hspace*{-.7cm}{\large $m_0$}\\[16.2cm]
\hspace*{16.5cm} {\large $m_{1/2}$}\\[-17.9cm]
\begin{center}
\hspace*{-.2cm}\mbox{\epsfig{figure=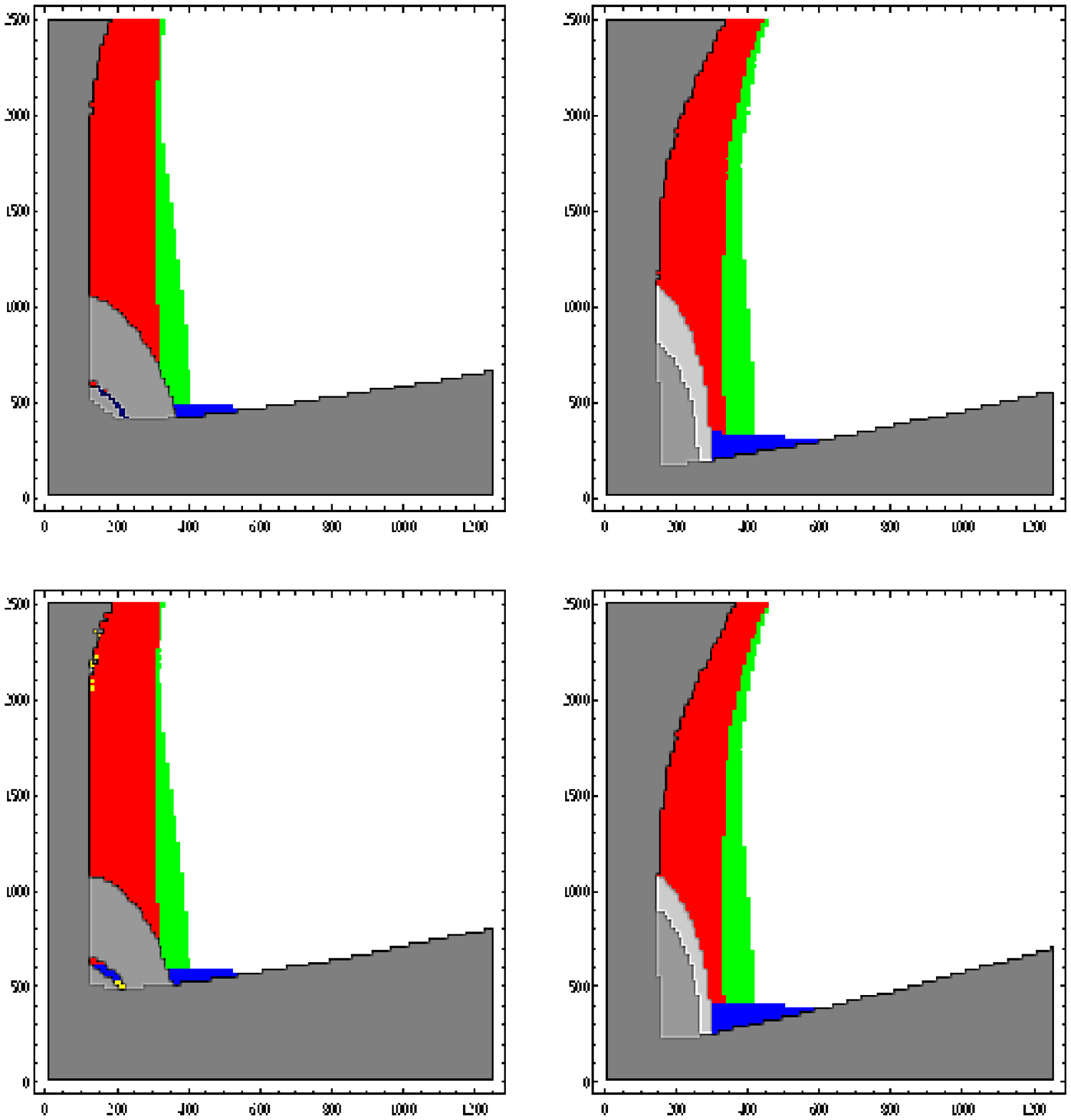,width=16.5cm}}
\end{center}
\noindent \hspace*{.7cm} $\tan\beta=55, \, A_0 = -1\, {\rm TeV}, \,
{\rm sign}(\mu)>0$ 
\hspace*{1.7cm}
$\tan\beta=60, \, A_0 = 0 , \,  {\rm sign}(\mu)>0$\\[.3cm]
\caption[]{The $(m_{1/2}, m_0)$ mSUGRA plane where SUSY and Higgs particles
can be produced at an $e^+e^-$ collider with a c.m. energy $\sqrt{s}=500$ 
GeV for different values of $\tan \beta>40$ and $A_0$  with sign($\mu)
>0$. The notation is as in Fig.~13.}
\end{figure}

\begin{figure}[htbp]
\centerline{\large $\tan\beta=40\ , \ A_0 = 0 \ ,\  {\rm sign}(\mu)>0$}
\hspace*{-.4cm}{\large $m_0$}\\[-1.cm]
\begin{center}
\hspace*{-.2cm} \epsfig{figure=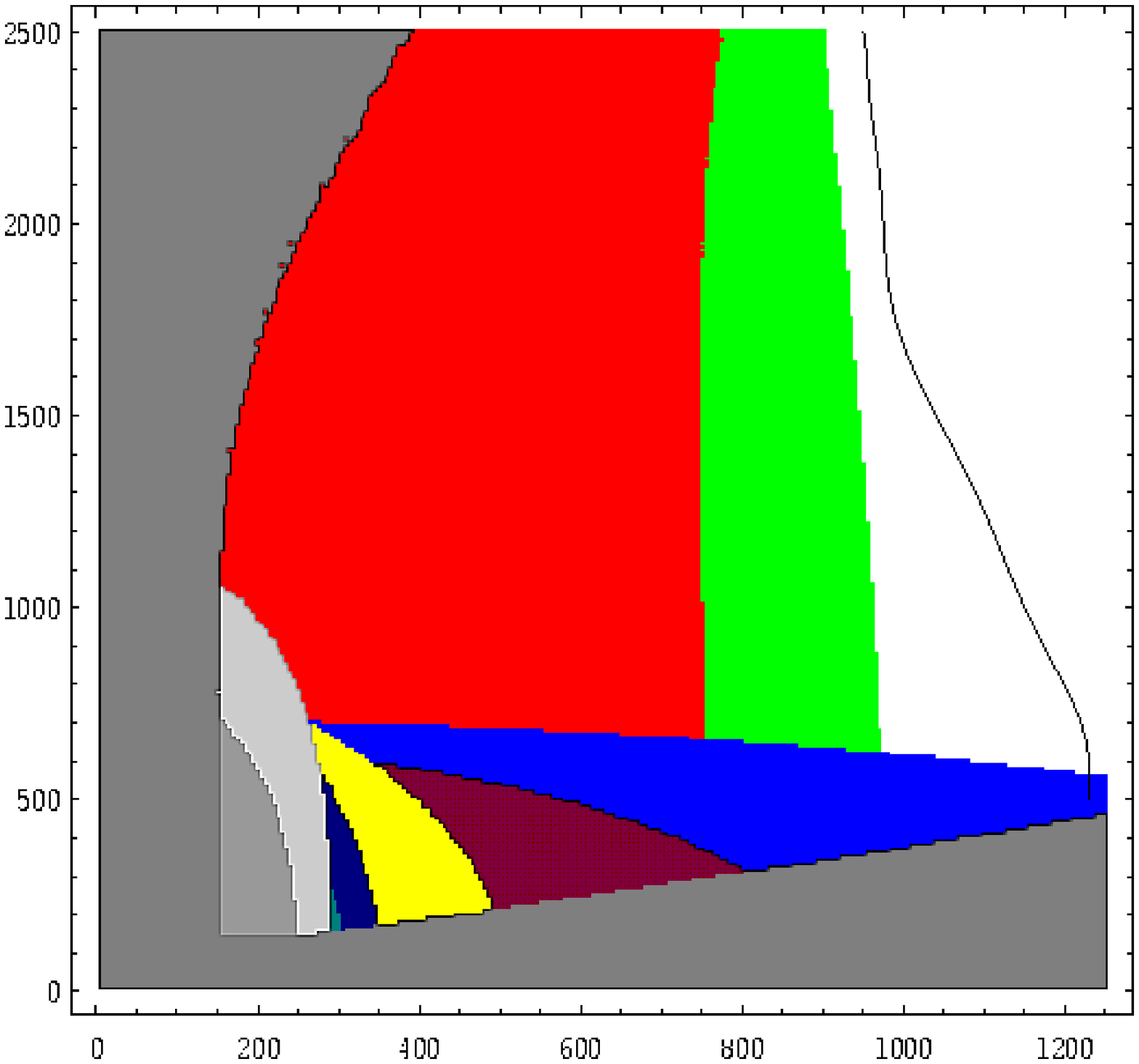,width=15cm}\\
\end{center}
\hspace*{14.5cm} {\large $m_{1/2}$}
\vspace*{-2mm}
\caption[]{The $(m_{1/2}, m_0)$ mSUGRA plane where SUSY and Higgs
particles can be produced at an $e^+e^-$ collider with a c.m. energy
$\sqrt{s}=1.2$ TeV. The grey areas are those excluded by theoretical
and experimental constraints. The colored regions are those where some
cross section is large enough for the particles to be produced:
$\tilde\chi_1^0 \tilde\chi_2^0$ (green), $\tilde\chi_1^+
\tilde\chi_1^-$ (red), $\tilde{\ell}^+ \tilde{\ell}^-$ (blue), $\tilde{\nu}
\bar{\tilde\nu}$ (purple), $\tilde{t}_1 \bar{\tilde t}_1$ (dark blue),
$\tilde{b}_1 \bar{\tilde b}_1$ (dark blue) and the heavy MSSM $H,A,H^\pm$
bosons (yellow). Note that some of these regions are overlapping. The black
line shows the $5\sigma$ reach contour for sparticles at the LHC in
the missing ${\rm E}_T$ channel with a luminosity $\int {\cal L}dt = 100$
fb$^{-1}$; adapted from Ref.~\cite{Charles}. }
\end{figure}

\begin{figure}[htbp]
\noindent \hspace*{.7cm} $\tan\beta=5, \, A_0 = -1\, {\rm TeV}, \,  
{\rm sign}(\mu)>0$ \hspace*{1.7cm}
$\tan\beta=10, \, A_0 = 0, \,  {\rm sign}(\mu)>0$\\
\hspace*{-.7cm}{\large $m_0$}\\[16.2cm]
\hspace*{16.5cm} {\large $m_{1/2}$}\\[-17.9cm]
\begin{center}
\hspace*{-.2cm}\mbox{\epsfig{figure=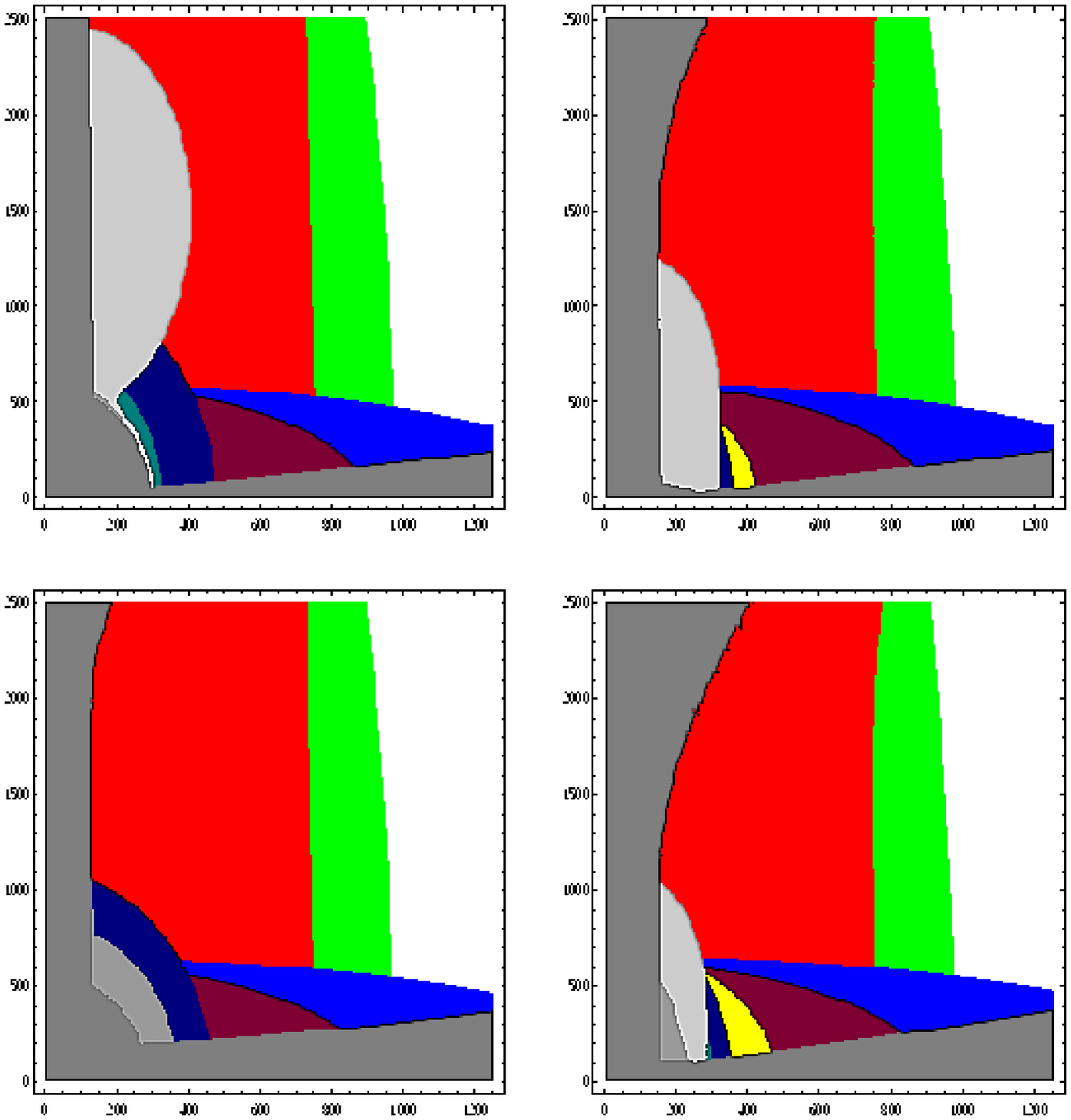,width=16.5cm}}
\end{center}
\noindent \hspace*{.7cm} $\tan\beta=20, \, A_0 = -1\, {\rm TeV}, \,  
{\rm sign}(\mu)>0$ 
\hspace*{1.7cm}
$\tan\beta=30, \, A_0 = 0 , \,  {\rm sign}(\mu)>0$\\[.3cm]
\caption[]{The $(m_{1/2}, m_0)$ mSUGRA plane where SUSY and Higgs particles
can be produced at an $e^+e^-$ collider with a c.m. energy $\sqrt{s}=1.2$ 
TeV for different values of $\tan \beta<40$ and $A_0$  with sign($\mu)
>0$. The notation is as in Fig.~14.}
\end{figure}

\begin{figure}[htbp]
\noindent \hspace*{.7cm} $\tan\beta=45, \, A_0 = -1\, {\rm TeV}, \,  
{\rm sign}(\mu)>0$ \hspace*{1.7cm}
$\tan\beta=50, \, A_0 = 0, \,  {\rm sign}(\mu)>0$\\
\hspace*{-.7cm}{\large $m_0$}\\[16.2cm]
\hspace*{16.5cm} {\large $m_{1/2}$}\\[-17.9cm]
\begin{center}
\hspace*{-.2cm}\mbox{\epsfig{figure=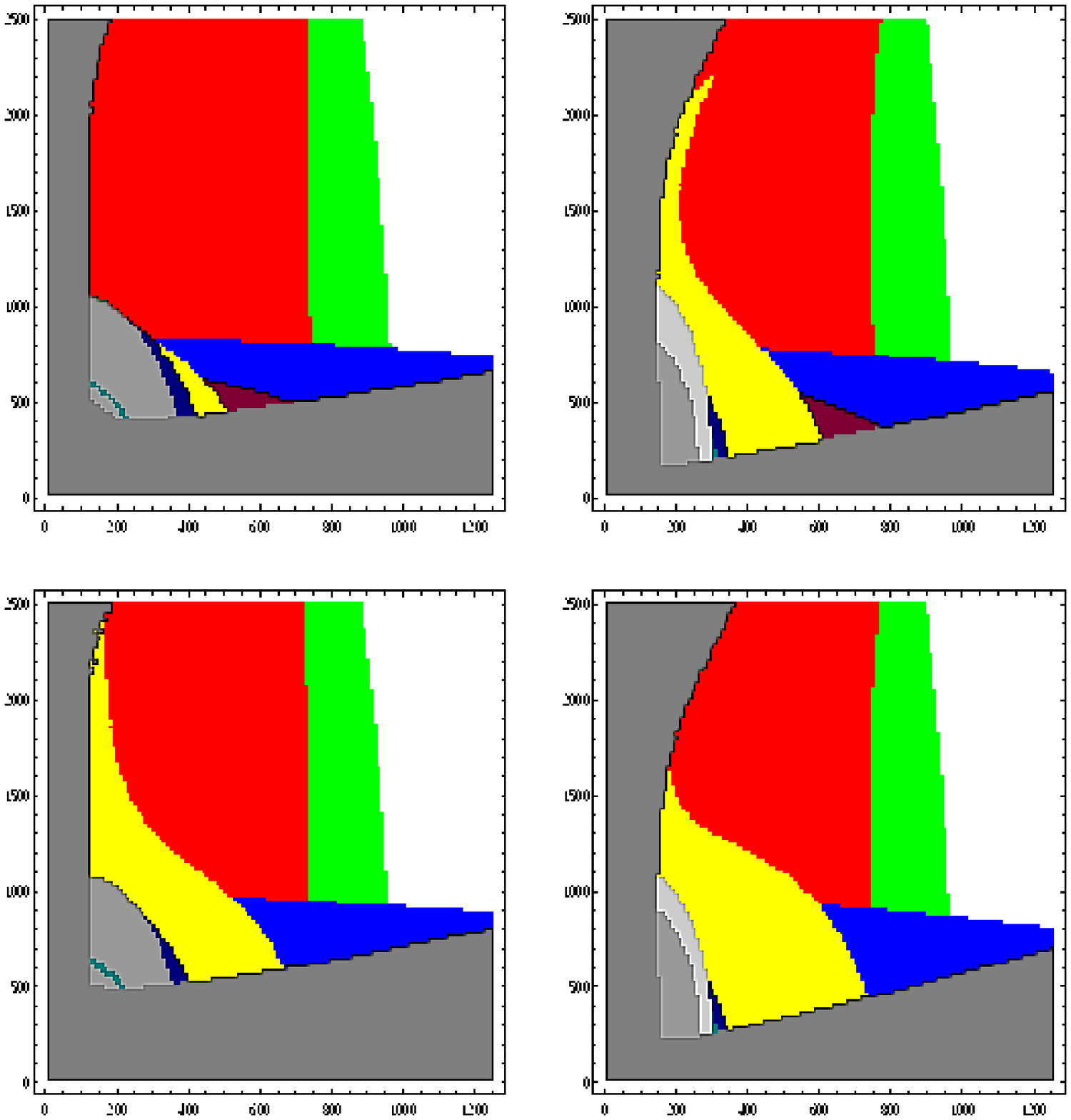,width=16.5cm}}
\end{center}
\noindent \hspace*{.7cm} $\tan\beta=55, \, A_0 = -1\, {\rm TeV}, \, 
{\rm sign}(\mu)>0$ 
\hspace*{1.7cm}
$\tan\beta=60, \, A_0 = 0 , \,  {\rm sign}(\mu)>0$\\[.3cm]
\caption[]{The $(m_{1/2}, m_0)$ mSUGRA plane where SUSY and Higgs particles
can be produced at an $e^+e^-$ collider with a c.m. energy $\sqrt{s}=1.2$ 
TeV for different values of $\tan \beta>40$ and $A_0$  with sign($\mu)
>0$. The notation is as in Fig.~14.}
\end{figure}

Figs.~11--13 show the SUSY reach for $\sqrt{s}=500$ GeV. The regions
where charginos, neutralinos and charged $(\tau)$ sleptons are
accessible [with the same requirement for cross section and luminosity
as before] can essentially be obtained from Figs.~8--10 by simply
rescaling $m_0$ and $m_{1/2}$. The reach is therefore reduced by a
factor $\sim 5/8$ in this case. Of course, the experimental
constraints remain the same, and thus exclude a relatively larger
region of the parameter space that would have been accessible. For
example, for $\tb =5$ and $A_0 = -1$ TeV [top--left frame in Fig.~12]
the $M_h \gsim 113$ GeV requirement has erased a sizeable region where
charginos or neutralinos can be produced at a 500 GeV $\ee$
collider. In this case, the stop region has also shrunk dramatically.
Moreover, sneutrino and heavy Higgs boson production is accessible
only in the tiny strip of parameter space at small $m_0$ and small
$m_{1/2}$ in the bottom--left frame of Fig.~13 where the new
contributions to $b \ra s \gamma$ decays are ``accidentally'' small;
see the discussion of Fig.~4. \s

On the other hand, Figs.~14--16 show that at $\sqrt{s}=1.2$ TeV the
prospects for SUSY particle and MSSM Higgs boson production are of
course much better. Again by simple rescaling, one reaches values of
$m_{1/2} \sim 750$ GeV and 1 TeV from chargino pair production and
mixed neutralino pair production, respectively, while $\tilde{\tau}$
pair production can be probed for $m_{0}\lsim 700$ GeV. The regions
where sneutrinos and top squarks can be produced also increase, and
there is now a little corner where even bottom squarks could be
accessible [top--left panel in Fig.~15].\footnote{At first sight it is
somewhat counter--intuitive that $\tilde b_1$ production is accessible
only at small $\tb$, given that $m_{\tilde b_1}$ decreases with
increasing $\tb$. The reason is that the increase of the region
excluded by the $b \ra s \gamma$ constraint is faster than the
increase of the region with small $m_{\tilde b_1}$.} For large values
of $\tb$, the Higgs domain increases dramatically and for $\tb \gsim
55$ [Fig.~16], the heavy $H,A$ and $H^\pm$ bosons can be produced in
very large areas of the parameter space. \s

In models with universal soft SUSY breaking gaugino masses the largest
value\footnote{Note that this value already requires severe fine-tuning
to reproduce the experimental values of the $W$ and $Z$ boson masses,
with typically $\mu^2 > 10^2 M_Z^2$.}  of $m_{1/2}$ which can be
probed at a 1.2 TeV $\ee$ collider through $\lsp \tilde \chi_2^0$
production, $m_{1/2} \simeq 1$ TeV, corresponds to squark and gluino
masses in the neighborhood of 2 TeV. This is only slightly lower than
the regions which are expected to be probed at the LHC with a high
luminosity, where $m_{\tilde{q}}$ and $m_{\tilde{g}} \lsim 2.5$ TeV
are accessible \cite{Charles}. This is illustrated in Fig.~14, where
we display, together with the reach of an $\ee$ collider with
$\sqrt{s}=1.2$ TeV, the domain that can be probed at the LHC with a
luminosity of $\int {\cal L} dt =100$ fb$^{-1}$ in the search for squarks
and gluinos in the jets plus missing transverse energy channel, shown
by the additional line that we adapted from the analysis of
Ref.~\cite{Charles}. In addition, there is of course the
complementarity between the two colliders: while LHC will primarily
probe the strongly interacting particles, the linear collider will be
more concerned with the weakly interacting neutralinos, charginos
sleptons and Higgs bosons.

\section*{5. Conclusions}

In this paper, we have performed a comprehensive analysis of all the
available constraints on the minimal Supergravity model. We used the
state of the art FORTRAN code {\tt SuSpect} for the calculation of the
SUSY and Higgs particle spectrum of the MSSM. We include all presently
known important effects [two--loop renormalization equations for
couplings and gaugino masses, the complete one--loop effective
potential for electroweak symmetry breaking, and the radiative
corrections to standard fermion, Higgs boson and sparticle
masses]. This allowed us to fairly reliably delineate the regions of
the mSUGRA parameter space which are still allowed by theoretical
constraints [from a proper EWSB breaking, neutralino LSP,
non--tachyonic Higgs and SUSY particles, etc..]  and by the present
experimental data [i.e., the bounds on SUSY particle and Higgs boson
masses from negative searches at LEP2 and the Tevatron, the
measurement of the $b\to s\gamma$ branching ratio, and the precision
electroweak measurements]. We have also indicated the regions of
parameter space where one obtains a light Higgs boson with a mass $M_h
\sim 115$ GeV, where the SUSY contributions can account for the
deviation of the experimental value of $(g_\mu-2)$ from its Standard
Model prediction, and where the lightest neutralino makes a good Dark
Matter candidate. \s

When taken at face value, our calculation shows that all constraints
[including the positive indications] can be satisfied only in fairly
narrow regions of mSUGRA parameter space. However, experimental errors
on the input parameters [in particular, on the mass of the top quark;
Fig.~5], as well as theoretical uncertainties, are still
significant. The latter lead to a residual scale dependence of our
results [Fig.~6], as well as to a significant uncertainty in the
prediction of the mass of the lightest Higgs boson. We also repeat our
cautionary remarks regarding the interpretation of the $b \ra s
\gamma$ and, in particular, the Dark Matter constraints, which can be
circumvented without significantly altering any collider signal, at
the cost of a somewhat more contrived model of particle physics and/or
cosmology. With these caveats in mind, the main outcome of our
analysis can be summarized as follows: \s

-- There are large areas of the $(m_{1/2}, m_0)$ parameter space which
are still allowed by present experimental constraints. In particular,
for large enough values of $\tb$, the bound on the lightest $h$ boson
mass, $M_h \gsim 113$ GeV, does not place too severe constraints. If
$\mu>0$, which is favored by the $(g_\mu-2)$ anomaly, the constraint
from the radiative decay $b \to s\gamma$ is not too restrictive [in
our conservative interpretation of experimental and theoretical
errors] even for large values of $\tb$. In fact, if $A_0 = 0$ it is
always superseded by the Higgs boson mass constraint, but for $A_0 =
-1$ TeV the $b \ra s \gamma$ constraint can be more severe [Figs.~3
and 4].  Precision electroweak measurements are easily accommodated in
the mSUGRA framework. \s

-- For $\tb \gsim 10$ and small values of the trilinear coupling
$A_0$, the requirement of a Higgs boson in the mass range 113 GeV
$\lsim M_h \lsim 117$ GeV favors moderate values of the common gaugino
mass parameter, $m_{1/2} \lsim 500$ GeV, leading to relatively light
chargino and neutralino states, $m_{\tilde\chi_1^\pm} \sim
m_{\tilde\chi_2^0} \sim 2 m_{\tilde\chi_1^0} \lsim 400$ GeV. For large
(and negative) values of $A_0$, which lead to a strong mixing in the
stop sector, a Higgs mass in this range can be accommodated in large
regions of the parameter space even for rather small $\tb (\sim 5)$
values. In this case $\tilde t_1$ squarks can be rather light, if the
soft breaking parameters $m_0$ and $m_{1/2}$ are not too high. The
range of $m_{1/2}$ favored by the LEP Higgs evidence strongly depends
on the exact value of $M_t$. We thus eagerly anticipate improved
measurements of this important parameter at the current run of the
Tevatron collider. \s

-- The $(g_\mu-2)$ excess, which can be accommodated in the MSSM only
if $\mu >0$, typically gives a stronger upper bound on $m_0$ than the
requirement $M_H = 115 \pm 2$ GeV. For $\tb\sim 40$, $m_0$ and
$m_{1/2}$ values below $\sim 600$ GeV [and slightly above $\sim 300$
GeV] are needed if the experimental value is to be reproduced within
$1\sigma$, implying again relatively light electroweak gaugino and
slepton states. However, the value of this upper bound increases
roughly proportional to $\tb$, so that at $\tb = 60$, $m_0$ as large
as 1.0 (1.6) TeV can be accommodated at the 1 (2) $\sigma$ level. \s

-- For small and moderate $\tb (\lsim 40)$ the requirement that the
density of the lightest neutralinos left over from the Big Bang is in
the right range to account for the Dark Matter density in the Universe
is very constraining indeed. In this case most of the region where
$\om$ is ``naturally'' in the interesting range is excluded by the
Higgs mass constraints, which requires SUSY breaking masses above
those preferred by Dark Matter calculations. Only a small band in the
region with a relatively light bino--like neutralino and relatively
light sleptons survives. In addition, there are ``exceptional''
regions: a narrow strip in the $\tilde \tau_1 \lsp$ co--annihilation
region near the boundary where the $\tilde{\tau}_1$ slepton is the
LSP, and a strip in the focus point region at large $m_0$ and small
$m_{1/2}$ values where neutralinos and charginos are relatively light
and have large higgsino components. Requiring in addition $M_h = 115
\pm 2$ GeV and a SUSY interpretation for the $(g_\mu - 2)$ anomaly
removes most of these ``exceptional'' regions with acceptable relic
density.  On the other hand, for large values of $\tb (\gsim 50)$, the
area of the $(m_0, m_{1/2})$ parameter space favored by cosmology
extends significantly due to the opening of the pseudoscalar
$A$--boson pole. This allows to fit all the requirements [$M_h$,
$(g_\mu-2)$ and the DM constraint] in a somewhat larger area of the
$(m_0, m_{1/2})$ parameter space. However, the Higgs mixing parameter
$\overline{m}_3^2$ in the scalar potential needs to be fine-tuned
quite severely to accommodate very large values of $\tb$
\cite{hightanbeta}. \s

-- In spite of the strong constraints on the mSUGRA parameter space
obtained by taking seriously all the positive indications for
supersymmetry it is still not possible to give tight limits on any one
single parameter. We found overlap regions with $5 \leq \tb \leq 60$,
0.1 TeV $\lsim m_0 \lsim$ 1.5 TeV, 160 GeV $\lsim m_{1/2} \lsim$ 550
GeV and for both $A_0=0$ and $A_0 = -1$ TeV. Allowing for a large
negative $A_0$ plays an important role in extending the allowed region
to smaller values of $\tb$. On the other hand, for larger $\tb$, $A_0
= 0$ is generally preferred, mostly due to the $b \ra s \gamma$
constraint. This implies that the allowed region in the $(m_{1/2},
m_0)$ plane could be further extended by considering more choices for
$A_0$, including positive values. \s

We have then analyzed the prospects for producing SUSY particles and
heavy Higgs bosons at high--energy and high--luminosity $\ee$
colliders, requiring a sample of 50 events per year to establish
discovery; this should be sufficient in the clean environment provided
by $\ee$ colliders. At c.m. energy $\sqrt{s} \sim 800$ GeV, we have
shown that charginos, neutralinos and sleptons [in particular $\tilde
\tau$ sleptons and sneutrinos] are accessible in rather large regions
of the parameter space. In particular, already at $\sqrt{s} = 800$ GeV
associated $\lsp \tilde \chi_2^0$ production is accessible in the
entire overlap region described above. Almost all of this region can
also be probed through $\tilde \chi_1^\pm$ pair production, and in
much of this region $\tilde \tau_1$ pair production can also be
studied.  In some areas of the parameter space, top squarks and even
bottom squarks can be produced. In the large $\tb$ regime, where the
present indications for SUSY can be accommodated in a larger fraction
of the $(m_{1/2}, m_0)$ plane, there is a large region of parameter
space where the heavier MSSM Higgs bosons $H,A$ and $H^\pm$ are
kinematically accessible. \s

Even for lower c.m. energies, $\sqrt{s} \sim 500$ GeV, charginos,
neutralinos and charged $(\tilde \tau)$ sleptons can be produced in a
significant region of parameter space not excluded by the present
constraints. However, discovery of sparticles can then no longer be
guaranteed [in the framework of mSUGRA] even if all positive
indications for SUSY hold up to further scrutiny. On the other hand,
if the c.m. energy of the collider is increased to $\sqrt{s}=1.2$ TeV,
the mSUGRA parameter space where SUSY and Higgs particles are
kinematically accessible and have sufficiently large cross sections to
be detected becomes very wide. The $\ee$ collider will then have a
search potential of SUSY particles that is comparable to the range
probed at the LHC. This is largely due to the fact that, thanks to the
high luminosities expected at future $\ee$ colliders, the process $\ee
\to \tilde\chi_1^0 \tilde\chi_2^0$ can probe large values of the
parameter $m_{1/2}$: only from kinematical arguments, values $m_{1/2}
\sim 1$ TeV can be probed at $\sqrt{s}=1.2$ TeV, corresponding to a
gluino mass of the order of 2 TeV. Heavy Higgs particles can be
searched if their masses are smaller than the beam energy. For large
values of $\tb$, this occurs in a large region of the mSUGRA parameter
space. \s

One might naively argue that the LHC would still have an edge over a
1.2 TeV $\ee$ collider, since at the LHC the heaviest sparticles
(squarks and gluinos) are produced directly, allowing access to
lighter sparticles (neutralinos, charginos and sleptons) through
cascade decays. In contrast, if $m_{1/2}$ is indeed near 1 TeV [which
seems highly unlikely, given the original motivation for
``weak--scale'' SUSY], an $\ee$ collider would ``only'' have access to
one or two production channels. However, in such a pessimistic
scenario the relatively small production cross section and the large
number of possible decay modes will make it very difficult, if not
impossible, to study any one decay mode in detail at the LHC. So far
at least the ability of LHC experiments to perform such studies has
only been demonstrated \cite{LHC} for significantly lower mass scales,
i.e. much higher production cross sections. \s

On the other hand, precision measurements at an $\ee$ collider could
reveal a great deal about the MSSM spectrum even if only the ``lower
end'' of the sparticle spectrum is accessible \cite{TESLA}. In
particular, thanks to the precise knowledge of the initial $e^+$ and
$e^-$ beam energies, threshold scans allow the measurement of some
sparticle masses at the permille level. Making use of the ability to
vary the beam polarization at will, various couplings appearing in the
production cross sections of SUSY and Higgs particles can be measured
with a high precision. Additional couplings can be determined through
the careful measurement of decay branching ratios. \s

The amount of information on the SUSY model can be maximized by
combining results from a high--luminosity $\ee$ collider with those
obtained at the LHC and elsewhere. In the near future measurements at
relatively low energies [at the $B-$factories, from searches for $\mu
\ra e \gamma$ decays etc.] will yield new information on the flavor
structure of the soft breaking terms. Any positive signal for large
flavor violation would rule out the mSUGRA model as commonly defined,
while absence of such a signal would strengthen the motivation for
assuming universal sparticle masses at some high scale. Of course, the
crucial test of SUSY will have to come from high--energy colliders. A
combination of the information on sleptons and electroweak gauginos
that one can obtain at $\ee$ colliders with the information on squark
and gluino production obtained at the LHC would allow very stringent
tests of the model. The flavor--conserving parts of the SUSY
Lagrangian at the low energy scale can then be reconstructed to a
considerable extent. Under the assumption of a Grand Desert, the
underlying structure of the theory at the GUT or Planck scale can then
be studied. For instance the combination of, on the one hand, the
measurement of the electroweak gaugino [chargino and neutralino]
masses at the linear collider with the measurement of the gluino mass
at the LHC allows to test gaugino mass unification.\footnote{A partial
test, of the unification of U$(1)_{\rm Y}$ and SU(2) gaugino masses,
is possible using $\ee$ data alone \cite{tsukamoto}. However, barring
a major disaster during construction, data from the LHC will become
available first, and should thus be included in this test.} Similarly,
the measurement of the slepton and Higgs boson masses at an $\ee$
collider together with the squark mass determination at the LHC would
allow to verify the second mSUGRA assumption that scalar soft SUSY
breaking mass parameters are also unified at the GUT scale
\cite{Peter}. A high--energy, high--luminosity $\ee$ linear collider
would thus be crucial for fully testing the mSUGRA model. \s

Of course it is quite possible, perhaps even likely, that mSUGRA in
its simplest version will already have been excluded before the next
$\ee$ linear collider commences operation. However, most of our
conclusions remain valid in a more general supersymmetric context. In
particular, the SUSY mass scales indicated by the positive evidence
for a 115 GeV Higgs boson and for a SUSY loop contribution to $(g_\mu
- 2)$ are not very sensitive to details of the SUSY model. Our
conclusions will then remain qualitatively the same, as long as there
are no large hierarchies between soft breaking parameters that are
assumed to be unified in mSUGRA [i.e. as long as all gaugino masses
and all scalar masses are similar at some high scale; a very large or
very small ratio of these two masses can be accommodated even in
mSUGRA]. In the absence of a compelling model of supersymmetry
breaking it becomes absolutely essential to collect as many
independent pieces of information about the soft breaking terms as
possible. By studying a highly constrained model we have thus chosen a
scenario which minimizes the advantages of $\ee$ colliders. As argued
above, even in this context the prospects for SUSY studies at such
colliders seem very bright. We are thus confident that linear high
energy, high luminosity $\ee$ colliders will play a crucial role in
revealing the secrets of Supersymmetry breaking, assuming that
Nature indeed makes use of the beautiful idea of weak scale
Supersymmetry.

\bigskip 

\nn {\bf Acknowledgments:} \s

\nn We thank Ben Allanach, Asesh Datta, Paolo Gambino, Jean--Francois Grivaz, 
Gilbert Moultaka and Margarete M\"uhlleitner for discussions. 
Special thanks go to Francois Richard and Peter Zerwas for their interest in 
this work and for helpful suggestions. This work is supported in part by the 
Euro--GDR Supersym\'etrie and by the European Union under contract 
HPRN-CT-2000-00149. The work of M.D. is supported in part by the
``Sonderforschungsbereich 375--95 f\"ur Astro--Teilchenphysik'' der
Deutschen Forschungsgemeinschaft.

\newpage
\setcounter{equation}{0}
\renewcommand{\theequation}{A.\arabic{equation}}
\newpage

\section*{Appendix: Production cross sections}

In this Appendix, we present for completeness expressions for the
production cross sections in $\ee$ collisions for diagonal and mixed
pairs of charginos \cite{Achi+}, neutralinos \cite{Achi0}, sleptons
\cite{Asleptons} and squarks \cite{Asquarks} as well as for the heavy
MSSM Higgs bosons \cite{Ahiggs,Ohmann}, using a unified notation.

\subsection*{1. Neutralino and Chargino production} 

The integrated cross section for mixed neutralino $\tilde\chi_i^0
\tilde\chi_j^0$ pair production can be written as
\beq
\sigma( \ee \to \tilde\chi^0_i \tilde\chi_j^0 ) &=& \frac{1}{1+ \delta_{ij} } 
\frac{2 \pi \alpha^2}{s_W^4 c_W^4} \, \frac{\lambda^{1/2}_{ij}}{s} 
(\sigma_{ZZ} + \sigma_{\tilde{e} \tilde{e}} + \sigma_{Z \tilde{e} }) ,
\eeq
with $\lambda^{1/2}_{ij}$ being the usual phase--space function with the 
reduced masses:
\beq
\lambda^{1/2}_{ij} = \frac{1}{2} \bigg[ (1- \mu_i^2 -\mu_j^2)^2 - 
4 \mu_i^4 \mu_j^4 \bigg]^{1/2} \ \ {\rm with} 
\ \mu_{i}^2 = m_{\tilde\chi_i^0}^2/s \ , \ \mu_{j}^2 =
m_{\tilde\chi_j^0}^2/s . 
\label{PS}
\eeq
The contribution of the $s$--channel $Z$--boson exchange, the
$t$--channel $\tilde{e}$ exchange and the $Z \tilde{e}$ interference
are expressed in terms of $\lambda_{ij}$ and the final particle
energies, $e_{i,j} =\sqrt{\lambda_{ij} +\mu^2_{i,j}}$ [note that the
neutralino masses here are the eigenvalues and the sign must be
included]:
\beq
\sigma_{ZZ} &=& \frac{(O^{''L}_{ji})^2 (L_e^2+ R_e^2)s^2}{|D_Z(s)|^2} 
\Bigg[ \frac{2}{3}\lambda_{ij} +2 e_i e_j -2 \mu_i \mu_j \Bigg], \non \\
\sigma_{Z \tilde{e}}&=& \frac{c_W^2 O^{''L}_{ji} s }{{\cal R}e D_Z(s)} 
\Bigg[ L_e f_{e_i}^L f_{e_j}^L [I_1(\mu^2_{\tilde{e}_L}) -
\mu_i \mu_j {\rm L} (\mu^2_{\tilde{e}_L}) ] -  R_e f_{e_i}^R f_{e_j}^R [I_1 
(\mu^2_{\tilde{e}_R})-\mu_i \mu_j {\rm L}(\mu^2_{\tilde{e}_R}) ]
\Bigg],  \non \\
\sigma_{\tilde{e} \tilde{e}}&=& \frac{c_W^4}{4} \Bigg[ 
  (f_{e_i}^L f_{e_j}^L)^2 [I_2(\mu^2_{\tilde{e}_L})-\mu_i \mu_j 
I_3(\mu^2_{\tilde{e}_L}) ] 
- (f_{e_i}^R f_{e_j}^R)^2 [I_2(\mu^2_{\tilde{e}_R}) -\mu_i \mu_j 
I_3(\mu^2_{\tilde{e}_R}) ], \Bigg]
\eeq
with $D_Z(s)= s-M_Z^2+i \Gamma_Z M_Z$. 
The various couplings are given, in terms of the weak isospin and electric 
charge of the electron and the elements of the matrix $Z$ which diagonalizes 
the neutralino mass matrix [which can be found in Ref.~\cite{GH} for 
instance] by: 
\beq
L_e= I_{3L}^e - Q_e s_W^2 \ \ ,  \ \ R_e= -Q_e s_W^2 \ \  , \ \ O^{''L}_{ij}
= - \frac{1}{2} Z_{i3} Z_{j3} +  \frac{1}{2} Z_{i4} Z_{j4}, \non \\
f_{e_i}^L = \sqrt{2} [ (I_{3L}^e -Q_e)\tan \theta_W Z_{i1} - I_{3L}^e 
Z_{i2} ] \ \  , \ \ f_{e_i}^R = \sqrt{2}  Q_e \tan \theta_W Z_{i1} .
\eeq
The kinematical functions $I_1, I_2$ and $I_3$, with
$\mu^2_{\tilde{e}_k} = m_{\tilde{e}_k}^2/s$ and $k=L,R$, read:
\beq
I_1 (\mu) &=&  \bigg[ (\mu_i^2+\mu_j^2 -2\mu^2)^2 - (e_i -e_j)^2 \bigg] 
\frac{ {\rm L}(\mu^2)} {4 \lambda_{ij}^{1/2}} 
- \bigg[\mu_i^2+\mu_j^2-2 \mu^2+ 1 \bigg] , \non \\
I_2 (\mu) &=& \frac{ (\mu^2_i+\mu^2_j-2\mu^2)(\mu_i^2+\mu_j^2-2\mu^2- 1)
-2(\lambda_{ij}-e_i e_j) } 
{ \frac{1}{4} (\mu_i^2+\mu_j^2-2\mu^2- 1)- \lambda_{ij} } 
- (\mu_i^2+\mu_j^2-2\mu^2) \frac{ {\rm L}(\mu^2)} {4
\lambda_{ij}^{1/2}} , \non \\
I_3 (\mu) &=& 
\frac{ {\rm L}(\mu^2)} {4 \lambda_{ij}^{1/2} (\mu_i^2+\mu_j^2-2\mu^2-
1)} ,
\eeq
and
\beq
{\rm L} (\mu^2) = \log \,  
\frac{ \mu_i^2 + \mu_j^2 -2 \mu^2 - 1 +  2 \lambda^{1/2}_{ij}  }
     { \mu_i^2 + \mu_j^2 -2 \mu^2 - 1 -  2 \lambda^{1/2}_{ij}  } .
\eeq

For chargino pair production, the cross section can be decomposed into the 
$s$--channel $\gamma, Z$--exchange contributions, the $t$--channel 
$\tilde{\nu}_e$ contribution and the interference terms: 
\beq
\sigma( \ee \to \tilde\chi^+_i \tilde\chi_j^- ) &=& 8 \pi \alpha^2
\frac{\lambda^{1/2}_{ij}}{s} (\sigma_{s} + \sigma_{t} + \sigma_{st}) .
\eeq
Using the same notation as previously [but now $\mu_i^2=
m_{\tilde\chi_i^\pm}^2/s$ and $\mu_j^2= m_{\tilde\chi_j^\pm}^2/s$] one
has for the different components:
\beq
\sigma_s &=& \delta_{ij} \left[ 1+ \frac{1}{2c_W^2 c_W^2} \frac{s}
{{\cal R}e D_Z(s)}  (L_e + R_e) (O^{'L}_{ij} + O^{'R}_{ij}) \right] 
\left( \frac{1}{3} \lambda_{ij}+ e_i e_j +\mu_i \mu_j \right) \non \\
&& + \frac{1}{4s_W^4 c_W^4} \frac{s^2} {|D_Z(s)|^2} (L_e^2+R_e^2) 
\bigg[ \bigg( (O^{'L}_{ij})^2+ (O^{'R}_{ij})^2 \bigg) \left( \frac{1}{3} 
\lambda_{ij}+ e_i e_j \right) + 2O^{'L}_{ij} O^{'R}_{ij} \mu_i \mu_j \bigg] ,
\non \\
\sigma_{st} &=& - \frac{V_{i1} V_{j1}}{8 s_W^2} \bigg\{ \delta_{ij} \bigg[ 
I_1(\mu^2_{\tilde{\nu}}) - \mu_i \mu_j {\rm L} (\mu^2_{\tilde{\nu}}) \bigg] 
+\frac{1}{s_W^2 c_W^2} \frac{s} {{\cal R}e D_Z(s)} L_e
\bigg[ O^{'L}_{ij} I_1(\mu^2_{\tilde{\nu}}) - O^{'R}_{ij} \mu_i \mu_j 
{\rm L} (\mu^2_{\tilde{\nu}}) \bigg] \bigg\} , \non \\
\sigma_{t} &=& \frac{ |V_{i1}|^2 |V_{j1}|^2}{16 s_W^4} \bigg[
I_2(\mu^2_{\tilde{\nu}})-\mu_i \mu_j I_3(\mu^2_{\tilde{\nu}}) \bigg],
\eeq
with the additional couplings $O^{'L}_{ij}, O^{'R}_{ij}$ expressed in
terms of the elements of the matrices $U$ and $V$ which diagonalize
the chargino mass matrix
\beq
O^{'L}_{ij}= \delta_{ij}s_W^2 -V_{i1} V_{j1} - \frac{1}{2} V_{i2} V_{j2} 
\ , \
O^{'R}_{ij}= \delta_{ij}s_W^2 -U_{i1} U_{j1} - \frac{1}{2} U_{i2} U_{j2}.
\eeq

\subsection*{2. Selectron and Sneutrino production} 

The integrated cross section for the pair production of left--handed
or right--handed selectrons, which occurs through the $s$--channel
$\gamma$ and $Z$ boson exchanges and the $t$--channel exchange of the
four neutralinos $\tilde\chi_l^0$, can be written as
\beq
\sigma(\ee \to \tilde{e}_i \tilde{e}_i^*) &=& 
\frac{\pi \alpha^2}{s } \Bigg\{ \frac{1}{3} \beta^{3}_i  
\Bigg[ Q_e^2 Q_{\tilde{e}}^2 + \frac{Q_e Q_{\tilde{e}} }{s_W^2 c_W^2} 
\tilde{a}_i L_e \frac{s} {{\cal R}e D_Z(s)} + \frac{L_e^2+R_e^2}{2c_W^4 s_W^4} 
\tilde{a}_i^2 \frac{s^2} {|D_Z(s)|^2} \Bigg] \non \\
&&+  4 \sum_{l=1}^{4} \sum_{k=1}^{4} |\lambda_{il}|^2 |\lambda_{ik}|^2 
H_{ilk} + 2 \sum_{l=1}^4 
|\lambda_{il}|^2 \bigg[ Q_e Q_{\tilde{e}}+ \frac{\tilde{a}_i
L_e}{c_W^2 s_W^2} \bigg] F_{il} \Bigg\} .
\eeq
The notation is as before with $\beta_i$ is the selectron velocity
$\beta_i^2=1- 4 \mu_i^2$ where in this case $\mu_i^2=
m_{\tilde{e}_i}^2/s$.  The couplings $\tilde{a}_i$ of the selectrons
to the $Z$ boson and the couplings $\lambda_{ik}$ between the
electron, the selectrons $\tilde{e}_i$ and the neutralinos
$\tilde\chi_k^0$ are given by
\beq
\tilde{a}_L = I_{3L}^{\tilde{e}} - Q_{\tilde{e}} s_W^2 \ \ \ , \ \ \
\tilde{a}_R = - Q_{\tilde{e}} s_W^2,  \hspace*{3cm} \non \\
\lambda_{Lk} = \frac{1}{2} \left( Z'_{k1} - \frac{\tilde{a}_L}{c_W s_W} Z'_{k2}
\right) \ \ , \ \
\lambda_{Rk} = - \frac{1}{2} \left( Z'_{k1} - \frac{\tilde{a}_R}{c_W s_W}
Z'_{k2} \right) \label{aiexp},
\eeq
with the rotated matrix elements: $Z'_{k1}=Z_{k1}c_W +Z_{k2} s_W$ and
$Z'_{k2}=-Z_{k1} s_W + Z_{k2} c_W$. The kinematical functions $F$ and
$H$, in the case where the two exchanged neutralinos $\tilde\chi_l^0$
and $\tilde\chi_k^0$ are different, is given by
\beq
H_{ilk} = + \frac{1}{2} \frac{ F_{il}- F_{ik} }{ \mu_l^2 - \mu_k^2} ,
\eeq
where $\mu_l^2 = m_{\tilde\chi_l^0}^2/s, \mu_k^2 =
m_{\tilde\chi_k^0}^2/s$ and the function $F_{ik}$ reads:
\beq
F_{ik} = \beta_i (-1+ 2\mu_i^2-2\mu_k^2) + 2[\mu_k^2+ (\mu_i^2- \mu_k^2)^2] \, 
\log \frac{ 2\mu_i^2 - 2\mu_k^2 - (1+\beta_i)}{2\mu_i^2 -2 \mu_k^2
-(1-\beta_i)} .
\eeq
In the case where the exchanged neutralinos are the same [i.e for the 
squared amplitudes], the function $H_{ilk}$ with $l=k$ reduces to:
\beq
H_{ikk} = - 2 \beta_i +(1- 2 \mu_i^2 + 2\mu_k^2) \,  
\log \frac{ 2\mu_i^2 - 2\mu_k^2 - (1+\beta_i)}{2\mu_i^2 -2 \mu_k^2
-(1-\beta_i)} .
\eeq
For the production of selectrons of different types, there is no
$s$--channel gauge boson exchange and the cross section simply  reads:
\beq
\sigma(\ee \to \tilde{e}_L \tilde{e}_R^*) &=& 
\frac{4\pi \alpha^2}{s} \sum_{l=1}^{4} \sum_{k=1}^{4} \lambda_{Ll} 
\lambda_{Rl} \lambda_{Lk}  \lambda_{Rk} H_{lk} ,
\eeq
where in terms of the phase space function defined in eq.~(\ref{PS}), 
\beq
l \neq k \ : \ \ H_{lk} &=& - \frac{\mu_l \mu_k}{\mu_l^2 - \mu_k^2} \bigg[
\log \frac{ \mu_{\tilde{e}_L}^2 + \mu_{\tilde{e}_R}^2 -2 \mu_l^2 
-1 - \lambda_{ \tilde{e}_L \tilde{e}_R}^{1/2} }
{\mu_{\tilde{e}_L}^2 + \mu_{\tilde{e}_R}^2 -2 \mu_l^2 -1+ \lambda_{
\tilde{e}_L \tilde{e}_R}^{1/2} } - (l \leftrightarrow k) \, \Bigg] ,
\non \\
l= k \ : \ \ H_{kk} &=& \frac{ 4 \lambda_{ \tilde{e}_L \tilde{e}_R}^{1/2}
\mu_l^2 } { (\mu_{\tilde{e}_L}^2 + \mu_{\tilde{e}_R}^2 -2 \mu_l^2 -1 - 
\lambda_{ \tilde{e}_L \tilde{e}_R}^{1/2})
(\mu_{\tilde{e}_L}^2 + \mu_{\tilde{e}_R}^2 -2 \mu_l^2 -1 + \lambda_{
\tilde{e}_L \tilde{e}_R}^{1/2})} .
\eeq
For the pair production of the electron sneutrino, the expression of
the total cross section is similar to the one of left--handed
selectrons except that the $s$--channel photon exchange is absent and
the couplings are different. It is given by:
\beq
\sigma(\ee \to \tilde{\nu}_e \tilde{\nu}_e^*) &=&  \frac{\pi \alpha^2}{s } 
\Bigg\{ \frac{1}{3} \beta^{3}_{\tilde{\nu}} \frac{L_e^2+R_e^2}{2c_W^4 s_W^4} 
\tilde{a}_\nu^2 \frac{s^2} {|D_Z(s)|^2} \Bigg] \non \\
&&+  4 \sum_{l=1}^{4} \sum_{k=1}^{4} |\lambda_{\tilde{\nu}l}|^2 
|\lambda_{\tilde{\nu} k}|^2 H_{\tilde{\nu} lk} + \frac{2 \tilde{a}_\nu^2 }
{s_W^2 c_W^2} \sum_{l=1}^4 |\lambda_{\tilde{\nu} l}|^2 \bigg] F_{\tilde{\nu}l} 
\Bigg\} ,
\eeq
where the functions $H$ and $F$ are given by the previous equations
with $\mu_i$ replaced by $\mu_{\tilde{\nu}_e}$ and
\beq
\tilde{a}_\nu=\frac{1}{2} \ \  , \ \ \lambda_{\tilde{\nu}l} = \frac{1}{2s_W}
V_{1l} . 
\eeq

\subsection*{3. Sfermion pair production} 

For sleptons of the second and third generation and for squarks, there is
only $s$ channel gauge boson exchange and the production cross sections,
in the absence of sfermion mixing, is simply given by: 
\beq
\sigma(\ee \to \tilde{f}_i \tilde{f}_i^*) = \frac{\pi \alpha^2 N_c}{3s} 
\beta_{i}^{3} \Bigg[ Q_e^2 Q_{\tilde{f}}^2 + \frac{Q_e 
Q_{\tilde{f}} }{s_W^2 c_W^2} \tilde{a}_i L_e \frac{s} {{\cal R}e D_Z(s)} 
+ \frac{L_e^2+R_e^2}{2c_W^4 s_W^4} \tilde{a}_i^2 \frac{s^2} {|D_Z(s)|^2} 
\Bigg], 
\eeq
with $N_c$ the color factor, $N_c=3(1)$ for squarks (sleptons), and
the $\tilde{a}_i$ are as in eq.~(\ref{aiexp}) for a given charge and
isospin.  However, in the case of third generation sfermions, the
mixing between the left--handed and right--handed states has to be
included. In this case the cross section becomes slightly more
involved and can be written as:
\beq
\sigma(\ee \to \tilde{f}_i \tilde{f}_j^*) = \frac{\pi \alpha^2 N_c}{3s} 
\lambda_{ij}^{3/2} \Bigg[ \delta_{ij} \bigg( Q_e^2 Q_{\tilde{f}}^2 + 
\frac{Q_e 
Q_{\tilde{f}} }{s_W^2 c_W^2} \frac{s  \tilde{a}_{ij} L_e} {{\cal R}e D_Z(s)} 
\bigg) + \frac{L_e^2+R_e^2}{2c_W^4 s_W^4} \frac{ \tilde{a}_{ij}^2 s^2} 
{|D_Z(s)|^2} \Bigg] ,
\eeq
with $\lambda_{ij}$ the phase space function eq.~(\ref{PS}) and the couplings 
$a_{ij}$ given by
\beq
a_{11}= I_{3L}^{\tilde{f}} \cos^2 \theta_{\tilde{f}}-Q_{\tilde{f}} s_W^2 , \,
a_{22}= I_{3L}^{\tilde{f}} \sin^2 \theta_{\tilde{f}}-Q_{\tilde{f}} s_W^2 , \,
a_{12}= a_{21}= - I_{3L}^{\tilde{f}} \sin \theta_{\tilde{f}} \cos
\theta_{\tilde{f}},
\eeq
with $\theta_{\tilde{f}}$ the angle of the unitary matrix which turns
the left-- and right--handed current eigenstates into the mass 
eigenstates: 
\beq
\tilde{f}_1 = \cos \theta_{\tilde{f}} \tilde{f}_L
+ \sin \theta_{\tilde{f}} \tilde{f}_R \ \ , \ \ 
\tilde{f}_2 = - \sin \theta_{\tilde{f}} \tilde{f}_L
+ \cos \theta_{\tilde{f}} \tilde{f}_R .
\eeq
In the case of squarks, one can include the QCD corrections which can
be rather important \cite{Schwinger,AQCD}. The standard corrections,
with virtual gluon exchange and gluon emission in the final state,
lead to an increase of the total cross section by $\sim 15\%$ far from
the kinematical threshold, with much bigger corrections closer to
threshold. In the case of diagonal pair production, they can be
included by using the Schwinger interpolation formulae
\cite{Schwinger}
\beq 
\sigma(\ee \to \tilde{q}_i \tilde{q}_i^*)&=& \sigma^{\rm Born}
\left[ 1+ \frac{4}{3} \frac{\alpha_s} {\pi} \left( \frac{\pi^2}{2
\beta_i} - \frac{1}{4} (1+\beta_i) (\pi^2 -6) \right) \right] ,
\eeq
which, up to an error of less than 2\%, reproduces the exact
results. The corrections for gluino exchange are in general smaller
and decouple for heavy gluinos. QED threshold corrections to slepton
pair production can also be of some importance \cite{freitas}.

\subsection*{4. Higgs boson production} 

The main production mechanisms of neutral Higgs bosons at $\ee$ 
colliders are the Higgs--strahlung process and pair production,
\begin{eqnarray}
 \ \ {\rm Higgs\mbox{-}strahlung:} \hspace{1cm} \ee & 
           \ra &  (Z) \ra Z+h/H ;
\hspace{5cm} \non \\
 \ \ {\rm pair \ production:} \hspace{1cm} \ee & \ra & (Z) \ra A+h/H ;
\non
\end{eqnarray}
as well as the $WW$ and $ZZ$ fusion processes, 
\begin{eqnarray}
\ \ {\rm fusion \ processes:} \hspace{0.8cm} \ \ee & \ra &  \bar{\nu} 
\nu \ (WW) \ra \bar{\nu} \nu \ + h/H \hspace{3.5cm}; \non \\
\ee & \ra &  \ee (ZZ) \ra \ee + h/H  \non .
\end{eqnarray}
The charged Higgs particle can be pair produced through virtual photon
and $Z$ boson exchange,
\begin{eqnarray}
\ \ {\rm charged \ Higgs: } \hspace{0.8cm} \ \ee & \ra &  \ (\gamma , 
Z^* ) \ \ra \ H^+ H^- \hspace*{3.93cm} \nonumber 
\end{eqnarray}
The production cross sections for the neutral Higgs bosons are
suppressed by mixing angle factors compared to the SM Higgs
production,
\begin{eqnarray}
\sigma(\ee \ra Zh) \ , \ \sigma(VV \ra h) \ , \ \sigma(\ee \ra AH) \ \ 
\sim \ \sin^2(\beta-\alpha) ; \\
\sigma(\ee \ra ZH) \ , \ \sigma(VV \ra H) \ , \ \sigma(\ee \ra Ah) \ \ 
\sim \ \cos^2(\beta-\alpha) ,
\end{eqnarray}
while the cross section for the charged Higgs particle does not depend
on any parameter other than $M_{H^\pm}$. In the decoupling limit,
$M_A \gg M_Z$, the $HVV$ couplings vanish, while the $hVV$ couplings
approach their SM Higgs values:
\beq
g_{HVV} & = & \cos(\beta-\alpha) \ra  \sin4\beta M_Z^2/2 M_A^2  \ra 0
; \\
g_{hVV} & = & \sin(\beta-\alpha)  \ra 1- {\cal O}(M_Z^4/M_A^4) \ \ \ra
1 . 
\eeq
Hence, the only relevant mechanisms for the production of the heavy Higgs 
bosons in this limit will be the associated pair production and the pair 
production of the charged Higgs particles. The cross sections, in the 
decoupling limit and for $\sqrt{s} \gg M_Z$, are given by [we use $M_H \sim 
M_A$]
\beq
\sigma (\ee \ra AH) &=& \frac{G_F^2 M_Z^4}{96 \pi s} (v_e^2+a_e^2)
\beta_A^3 , \\
\sigma (\ee \ra H^+H^-) &=& \frac{2G_F^2 M_W^4 s_W^4 }{3 \pi s } \left[
1+ \frac{v_e v_H}{8 s_W^2 c_W^2} + \frac{(a_e^2+ v_e^2)v_H^2}{
256 c_W^4 s_W^4} \right] \beta_{H^\pm}^3 ,
\eeq
where $\beta_j=(1-4M_j^2/s)^{1/2}$ is the velocity of Higgs bosons,
the $Z$ couplings to electrons are given by $a_e=-1, v_e=-1+4s_W^2$,
and to the charged Higgs boson by $v_H=-2+4s_W^2$. \s

The cross sections for $hA$ and $HZ$ production vanish in the
decoupling limit since they are proportional to $\cos^2 (\beta-
\alpha)$.  The cross section for the fusion process, $\ee \ra
\bar{\nu}_e \nu_e H$, is enhanced at high energies since it scales
like $M_W^{-2}\log s/M_H^2$.  This mechanism provides therefore a
useful channel for $H$ production in the mass range of a few hundred
GeV below the decoupling limit and small values of $\tb$, where
$\cos^2(\beta-\alpha)$ is not prohibitively small; the cross section,
though, becomes gradually less important for increasing $M_H$ and
vanishes in the decoupling limit. The cross section for the $ZZ$
fusion process is one order of magnitude smaller than that for $WW$
fusion.

\newpage

\end{document}